\documentclass[oldversion]{aa}  
\usepackage{graphicx}
\usepackage{txfonts}   
\usepackage{natbib}
\usepackage{times}
\usepackage{aalongtable}
\usepackage{rotating}
\usepackage{pdflscape}

\newcommand\msol{\textrm{M}_{\odot}}

\newcommand{\logg}{\ensuremath{\log g}}

\def\0{\phantom0}

\def\1{$^1$}
\def\2{$^2$}
\def\3{$^3$}
\def\4{$^4$}
\def\plm{$\pm$}

\def\Teff{T$_{eff}$}
\def\vmic{v$_{t}$}

\def\gsim{\;\lower.6ex\hbox{$\sim$}\kern-7.75pt\raise.65ex\hbox{$>$}\;}
\def\lsim{\;\lower.6ex\hbox{$\sim$}\kern-7.75pt\raise.65ex\hbox{$<$}\;}

\begin{document}

   \title{Extremely metal-poor stars in classical dwarf spheroidal galaxies : Fornax, Sculptor, and Sextans}

   \author{
M. Tafelmeyer\inst{1}
\and P. Jablonka\inst{1,2}
\and V. Hill\inst{3}
\and M. Shetrone\inst{4}
\and E. Tolstoy\inst{5}
\and M.J. Irwin\inst{6}
\and G. Battaglia\inst{7}
\and A. Helmi\inst{5}
\and E. Starkenburg\inst{5}
\and K.A. Venn\inst{8}
\and T. Abel\inst{9}
\and P. Francois\inst{2}
\and A. Kaufer\inst{10}
\and P. North\inst{1}
\and F. Primas\inst{7}
\and T. Szeifert\inst{10}
}

   \institute{
Laboratoire d'Astrophysique, Ecole Polytechnique F\'ed\'erale de Lausanne (EPFL), Observatoire, CH-1290 Sauverny, Switzerland
\and GEPI, Observatoire de Paris, CNRS UMR 8111, Universit\'e Paris Diderot,  F-92125, Meudon, Cedex, France
\and Department Cassiop\'ee, University of Nice Sophia-Antipolis, Observatoire de C\^{o}te d'Azur, CNRS, F-06304 Nice Cedex 4, France
\and McDonald Observatory, University of Texas, Fort Davis, TX 79734, USA
\and Kapteyn Astronomical Institute, University of Groningen, PO Box 800, 9700AV Groningen, the Netherlands
\and Institute of Astronomy, University of Cambridge, Madingley Road, Cambridge CB3 0HA, UK
\and European Southern Observatory, Karl-Schwarzschild-str. 2, D-85748, Garching bei M\"{u}nchen, Germany
\and Dept. of Physics \& Astronomy, University of Victoria, 3800 Finerty Road, Victoria, BC V8P 1A1, Canada
\and Kavli Institute for Particle-Astrophysics and Cosmology, Stanford University, SLAC National Accelerator Laboratory, Menlo Park 94025, USA
\and European Southern Observatory, Alonso de Cordova 3107, Santiago, Chile
}

   \date{Received ... ; accepted ...}

\abstract{ We present the results of a dedicated search for extremely
metal-poor stars in the Fornax, Sculptor, and Sextans dSphs. Five
stars were selected from two earlier VLT/Giraffe and HET/HRS surveys
and subsequently followed up at high spectroscopic resolution with
VLT/UVES.  All of them turned out to have [Fe/H] $\lsim -3$ and three
stars are below [Fe/H] $\sim -3.5$. This constitutes the first
evidence that the classical dSphs Fornax and Sextans join Sculptor in
containing extremely metal-poor stars and suggests that all of the
classical dSphs contain extremely metal-poor stars.  One giant in
Sculptor at [Fe/H]=$-3.96 \pm 0.06$ is the most metal-poor star ever
observed in an external galaxy. We carried out a detailed analysis of
the chemical abundances of the $\alpha$, iron peak, and the heavy
elements, and we performed a comparison with the Milky Way halo and
the ultra faint dwarf stellar populations. Carbon, barium, and
strontium show distinct features characterized by the early stages of
galaxy formation and can constrain the origin of their
nucleosynthesis.  }{ }{ }{ }{ }

\authorrunning{Tafelmeyer et al.}
\titlerunning{Extremely metal-poor stars in classical dSphs}
\maketitle

\section{Introduction}

Extremely metal-poor stars (EMPS), with [Fe/H] $< -3$, are eagerly
sought in the Milky Way and beyond, because they provide insight into
the earliest stages of chemical enrichment processes.  With only a few
exceptions, the interstellar gas out of which stars form is preserved
in their outer atmospheres even as they evolve up the red giant branch
(RGB).  High-resolution abundance analyses of RGB stars can provide
detailed information about the chemical enrichment of the galaxy at
the time the star was formed; this is a process we will refer to in
this context as chemical tagging.  Since RGB stars can have a wide
range of lifetimes, depending on their mass, this allows us to build up
a very accurate picture of how chemical evolution progressed with
time, from the very first stars until one Gyr ago.

The most elusive stars in this chemical tagging process are the
extremely metal-poor stars (EMPS), which sample the earliest epochs of
chemical enrichment in a galaxy. There have been numerous detailed
studies devoted to the search
\citep[e.g.][]{christlieb02,beers05} 
and characterization 
\citep[e.g.][]{mcwilliam95, ryan96, carretta02, cayrel04, cohen04,
francois07, lai08, bonifacio09} 
of EMPS in the Galaxy over the last two decades.
But similar studies in extra-galactic systems
only became feasible with the advent of the 8-10m class telescopes.  One
of the surprising results of spectroscopic studies with large
telescopes of individual RGB stars in nearby dwarf spheroidal galaxies (dSphs),
was the apparent lack of EMPS \citep[e.g.][]{helmi06, aoki09},
even though that the average metallicity of most of these systems
is very low.  However, recent results suggest that this lack of EMPS in dSphs
is merely an artifact of the method used to search for them
\citep[e.g.][]{kirby08, starkenburg10, frebel10a}.

The overarching motivation to find and study the most primitive stars
in a galaxy is the information they provide about the early chemical
enrichment processes and the speed and uniformity of this
environmental pollution.  A key question is whether chemical
enrichment progresses in the same way in dwarf galaxies as compared
with larger galaxies such as the Milky Way.  Probing the chemical
enrichment history in different environments leads to a better
understanding of which processes dominate in the early universe and
for how long. It is important to separate initial processes from
environmental effects, which can dominate the later development of
enrichment patterns. The studies of the metal-poor population in the
Galactic halo suggests that early enrichment occurs very quickly and
uniformly. But it is very hard to obtain a reliable overall picture,
given the large diffuse nature of the Galactic halo and the
overwhelming numbers of thin and thick disk stars that mask that
population.  Dwarf spheroidal galaxies are much smaller environments,
and it is much easier to study a significant fraction of their stellar
population in a uniform way. Although dSphs have almost certainly lost stars,
gas, and metals over time, most are unlikely to have gained any.  It is
therefore in those systems that we have the best chance to carefully
determine the detailed timescales of early chemical evolution and look
for commonalities with all chemical evolution processes in the early
universe.

The star formation histories of the three dwarf spheroidals (dSphs) we
are studying differ from each other.  Sculptor and Sextans appear to
be predominantly old and metal-poor systems, with no significant
evidence of any star formation occurring during the last 10 Gyr
\citep[e.g.][]{majewski99, hurley-keller99, monkiewicz99, lee03}.
Fornax, on the other hand, is larger, more metal-rich in
the mean, has five globular clusters, and a much more extended star
formation history.  Fornax also appears to have been forming stars quite
actively until a few 100Myr ago, and is dominated by a 4-7 Gyr old
intermediate-age population \citep[e.g.][]{stetson98, buonanno99,
coleman08}.  An ancient population is
unequivocally also present, as demonstrated by detection of a red
horizontal branch, and a weak, blue horizontal branch together with 
RR~Lyrae variable stars \citep{bersier02}. Despite their different
subsequent evolutionary pathways, there seems to be convincing evidence
that all dSphs had low mass star formation at the earliest times, but
it is still a point of contention if their early chemical evolution
differs from one dwarf galaxy to another.  (For a more in-depth
review of the properties of the classical dSphs and other dwarf
galaxies found in the local group, please see the 
review by Tolstoy, Hill \& Tosi 2009).

In this paper we present the results of a dedicated search for EMPS in
the Fornax, Sculptor, and Sextans dSphs.  In
Sect.~\ref{sect:sampleselect} we describe the sample selection,
observations, reduction, and basic measurements; in
Sect.~\ref{sec:atmo} we present the stellar atmospheric parameters; in
Sect.~\ref{sect:abundances} we describe the abundance analysis method;
and in Sect.~\ref{sect:results} we discuss our results and their
implications.  We summarize our conclusions in
Sect.~\ref{sect:conclusion}.

\section{Observations and data reduction}
\label{sect:sampleselect}

We have selected six extremely metal-poor candidate RGB stars with
V$<$18.7 mag, two targets per galaxy, in Fornax, Sculptor and
Sextans. Figure
 \ref{fig:cmd} shows the position in color and magnitude of our
 sample stars on the three dSph galaxy giant branches.
Five targets arise from our initial CaT surveys
\citep[Battaglia 2010, in prep.]{tolstoy04, battaglia06}, and have
metallicity estimates [Fe/H]$_{CaT}$ $\le -2.6$ in these studies.  

One star, Sex24-72, has no CaT estimate and was  instead
chosen  from the analysis of its spectrum obtained at a resolution
R=18,000 over 4814 to 6793\AA\ with the High Resolution Spectrograph
(HRS, \citet{tull98}) at the Hobby-Eberly Telescope (HET,
\citet{Ramsey98}).  This 1.3 hour spectrum was taken in February 2006
 after an initial short observation confirmed it as a radial velocity
 member of Sextans. This observation (program STA06-1-003) was
 conducted as part of a larger effort to search for the most
 metal-poor stars in northern dSph galaxies, but this star was the
 only very metal-poor radial velocity member found in Sextans.

\begin{figure}
\includegraphics[angle=-90, width=\hsize]{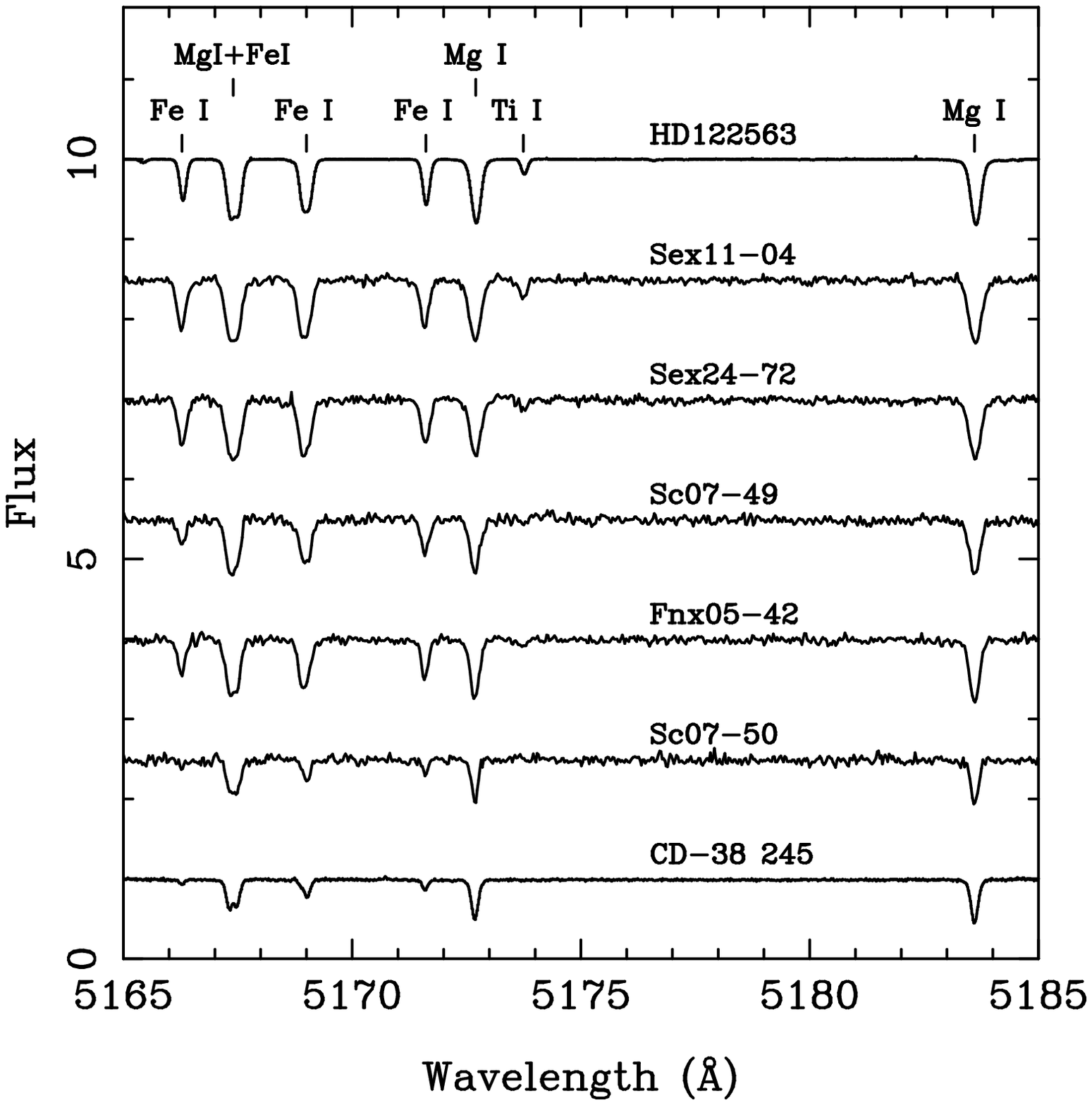}
\caption{Spectra of our five sample stars, ranked by decreasing value of [Fe/H],
  in the region of MgI. For comparison two Milky
  Way halo stars are included, HD122563 and CD $-$38 245, at [Fe/H]=$-$2.8 and
  [Fe/H]=$-$4.19, respectively.
\label{fig:allspectra}}
\end{figure}

\begin{figure}[!Hb]
\centering
\includegraphics[angle=-90, width=0.49\textwidth]{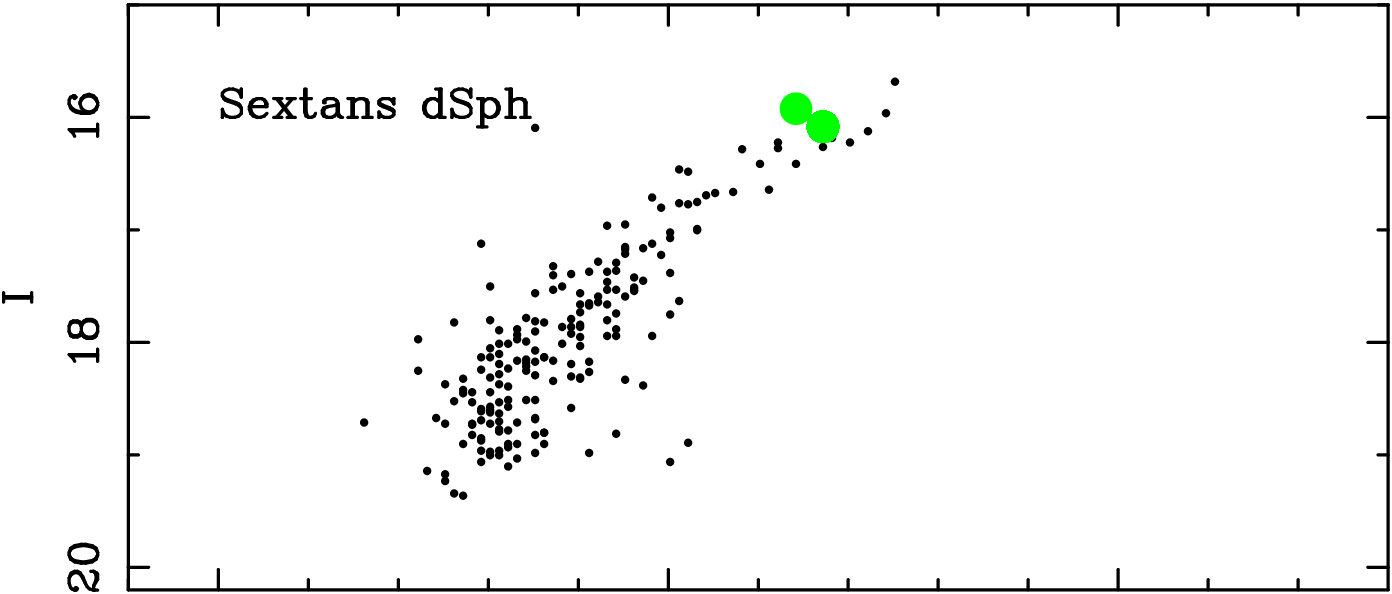}
\includegraphics[angle=-90, width=0.49\textwidth]{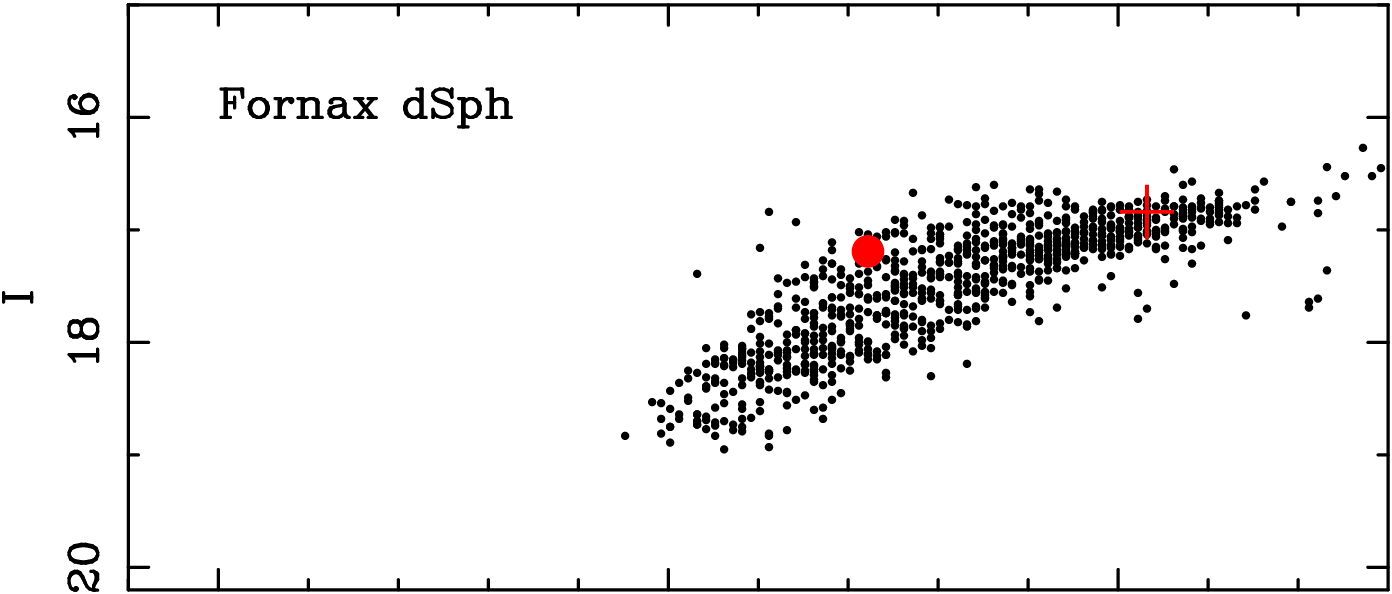}
\includegraphics[angle=-90, width=0.49\textwidth]{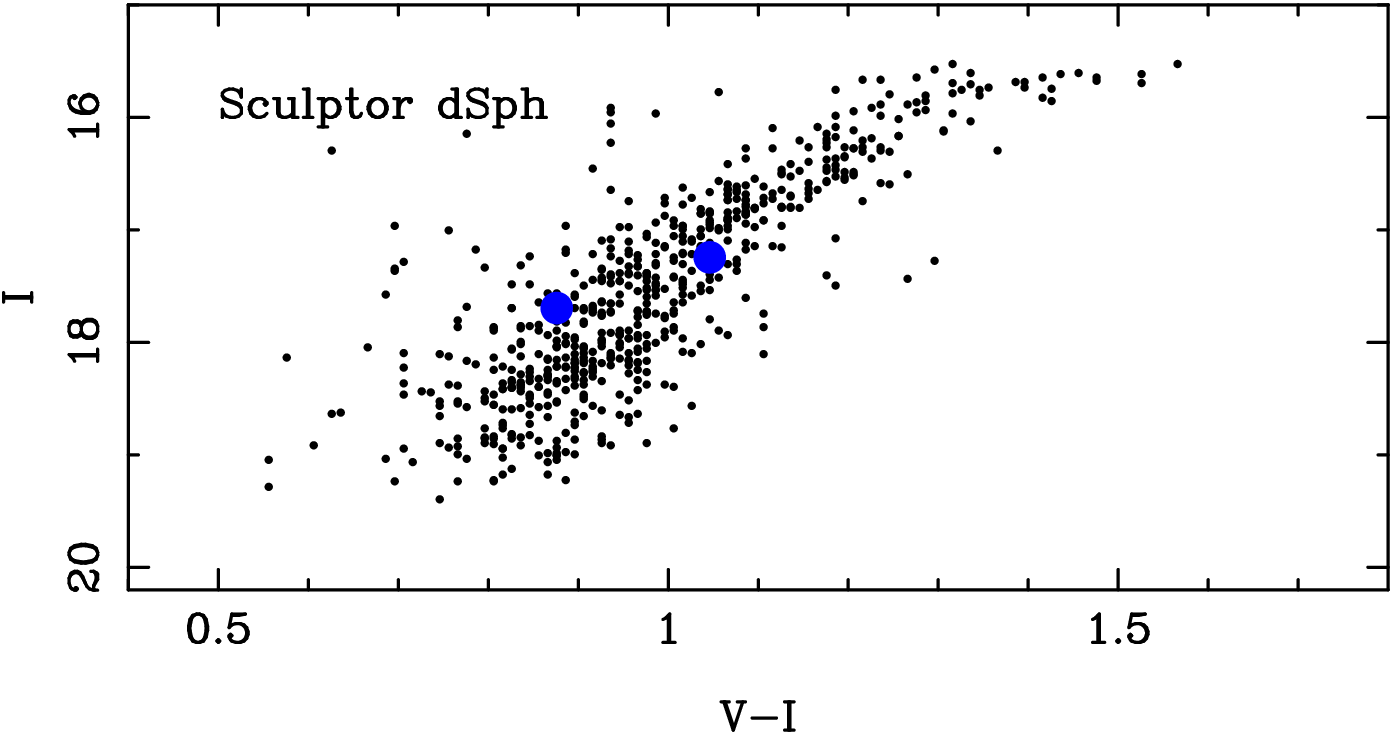}
\caption{Reddening corrected color-magnitude diagrams of Sextans, Fornax,
and Sculptor for our galaxy members selected from their spectra in the
CaT region. 
\label{fig:cmd}}
\end{figure}

High-resolution spectra were obtained with the UVES spectrograph
at VLT from June to September 2007 (Program ID 079.B-0672A) and from
April to September 2008 (Program IDs 081.B-0620A and 281.B-5022A). Our
(service) observing program was originally planned for the red arm
of UVES, centered on 5800\AA, covering 4700\AA-6800\AA\ at
R$\sim$40000.  From 4700\AA\ to 5800\AA\, the dispersion is
$\sim$0.028 \AA/pix and $\sim$0.034 \AA/pix from 5700\AA\ to
6800\AA. In the following sections we will refer to this wavelength
range as ``red''. The stars Scl07-49 and Scl07-50 were the first  to
be observed in our sample (Period 79).  As soon as we realized that
our sample stars contained extremely metal-poor stars, we extended the
wavelength domain blue ward in order to get as many lines as possible
for each element.  The spectrum of Scl07-50  was subsequently completed by the
blue arm of UVES, from $\sim$3700\AA\ to $\sim$5000\AA, with
$\sim$0.030 \AA/pix. In the following sections, we will refer to this
wavelength range as ``blue''. The other stars in Sextans and
Fornax were observed in dicroid mode (Period 81), the blue arm
centered on 3900\AA ($\sim$0.028 \AA/pix) and the red arm centered on
5800\AA ($\sim$0.034 \AA/pix). The total coverage is $\sim$ 3200\AA\ -
6800\AA\, with effective usable spectral information starting from
$\sim$ 3800\AA. In summary, all stars have spectra covering 3800\AA\
to 6800\AA, except Scl07-49, for which we have spectra covering
4700\AA-6800\AA. Our observations are summarized in
Table~\ref{tab:log}. One star in Fornax turned out to be an M dwarf
and was not included in this work.  The signal-to-noise ratios are
given per pixel.  Figure \ref{fig:allspectra} presents the
spectra in the region of MgI of our five sample stars, ranked by
decreasing value of [Fe/H], together with the two Milky Way halo
stars, HD122563 (\Teff=4600K, \logg=1.1, \vmic=2.0, [Fe/H]=$-$2.8) and
CD$-$38 245 (\Teff=4800K, \logg=1.5, \vmic=2.2, [Fe/H]=$-$4.19) from
the ESO large program ``First Stars'' \citep{cayrel04}, bracketing
the metallicity range of the present work.

\begin{table*}[h!]
\begin{center}
\label{tab:log}
\begin{tabular}{lcccccc}
\hline
ID & $\alpha$ (J2000) & $\delta$ (J2000)    & t$_{blue}$(s) & t$_{red}$(s) & S/N            &  [Fe/H]$_{CaT}$\\
     &                  &                   &              &             & 4000\AA\ 5300\AA\  6300\AA\ & \\
\hline

Sex24-72  & 10 15 02.65 & $-$01 29 55.9 &  7215  &  7215  &~~~~11~~~~~~~~    36~~~~~~~~   49   & $-2.70$ \\
Sex11-04  & 10 13 41.77 & $-$02 11 24.1 &  5644  &  8049  &~~~~11~~~~~~~~    38~~~~~~~~   50   & $-2.56$ \\
Fnx05-42  & 02 41 30.96 & $-$33 55 44.9 & 21035  & 21035  &~~~~13~~~~~~~~    34~~~~~~~~   45   & $-2.76$ \\
Fnx M dwarf& 02 40 20.64 & $-$34 12 42.7 &        &        &                                    &                 \\
Scl07-49  & 01 00 05.02 & $-$34 01 16.6 &         & 20730  &~~~~~~~~~~~~~~~~   36~~~~~~~~   41  & $-2.77$  \\
Scl07-50  & 01 00 01.14 & $-$33 59 21.4 & 36060  & 27045  &~~~~27$^{*}$~~~~~~~~   30~~~~~~~~ 37 & $-2.83$  \\
\hline
\end{tabular}			
\end{center}
\caption[]{Log of observations. The exposure times are given for 
the blue and red arms of UVES. The signal-to-noise ratios are 
measured at 4000\AA\ (with the exception of Scl07-050, for which 
S/N per pixel was measured at~4500\AA), 5300\AA\ , and 6300\AA\ .  The
metallicity estimates  are derived from CaT or from an HET/HRS spectrum in the case of Sex24-72.}

\end{table*}\begin{table*}[h!]
\begin{center}
\label{tab:log}
\begin{tabular}{lcccccc}
\hline
ID & $\alpha$ (J2000) & $\delta$ (J2000)    & t$_{blue}$(s) & t$_{red}$(s) & S/N            &  [Fe/H]$_{CaT}$\\
     &                  &                   &              &             & 4000\AA\ 5300\AA\  6300\AA\ & \\
\hline

Sex24-72  & 10 15 02.65 & $-$01 29 55.9 &  7215  &  7215  &~~~~11~~~~~~~~    36~~~~~~~~   49   & $-2.70$ \\
Sex11-04  & 10 13 41.77 & $-$02 11 24.1 &  5644  &  8049  &~~~~11~~~~~~~~    38~~~~~~~~   50   & $-2.56$ \\
Fnx05-42  & 02 41 30.96 & $-$33 55 44.9 & 21035  & 21035  &~~~~13~~~~~~~~    34~~~~~~~~   45   & $-2.76$ \\
Fnx M dwarf& 02 40 20.64 & $-$34 12 42.7 &        &        &                                    &                 \\
Scl07-49  & 01 00 05.02 & $-$34 01 16.6 &         & 20730  &~~~~~~~~~~~~~~~~   36~~~~~~~~   41  & $-2.77$  \\
Scl07-50  & 01 00 01.14 & $-$33 59 21.4 & 36060  & 27045  &~~~~27$^{*}$~~~~~~~~   30~~~~~~~~ 37 & $-2.83$  \\
\hline
\end{tabular}			
\end{center}
\caption[]{Log of observations. The exposure times are given for 
the blue and red arms of UVES. The signal-to-noise ratios are 
measured at 4000\AA\ (with the exception of Scl07-050, for which 
S/N per pixel was measured at~4500\AA), 5300\AA\ , and 6300\AA\ .  The
metallicity estimates  are derived from CaT or from an HET/HRS spectrum in the case of Sex24-72.}

\end{table*}

The spectra were reduced with the standard VLT UVES reduction
pipeline. Individual sub-exposures of 1500-3000~s were combined with
the IRAF task scombine. The final spectra were normalized with
\texttt{DAOSPEC}\footnote{\texttt{DAOSPEC} has been written by
  P.B. Stetson for the Dominion Astrophysical Observatory of the
  Herzberg Institute of Astrophysics, National Research Council,
  Canada.}, and although \texttt{DAOSPEC} also measures equivalent
widths, we remeasured the equivalent widths by hand with the
\texttt{IRAF} task \texttt{splot} from Gaussian fits.  Indeed, at low
metallicities, a large fraction of lines are only slightly above the
noise level and need to be checked individually to prevent spurious
detections.  Moreover, the continuum level around each line was
further checked and adjusted individually, yielding more accurate
equivalent widths.

The expected uncertainty in the equivalent widths are estimated
with the formula of \citet{cayrel88}: $$ \sigma=1.5 \times (S/N)^{-1}
\times \sqrt{FWHM \times \delta x }$$, where S/N is the signal-to-noise
ratio per pixel $FWHM$ is the line full width at half maximum and
$\delta x $ the pixel size, i.e, $\sim$9m\AA, $\sim$2.6m\AA,
$\sim$1.8m\AA, for our typical S/N, of 10, 35 and 50, respectively. 
This formulation however does not include continuum placement 
uncertainties, and therefore provides a lower limit to the uncertainties. As
also reported in Section \ref{sec:errors}, we conservatively 
considered only equivalent widths larger than $\sim$20m\AA\ in the red
and $\sim$30m\AA\ in the blue, and performed synthesis for the smaller
lines.

\section{Stellar parameters}
\label{sec:atmo}

\subsection{Stellar models}

Photospheric 1-D models for the sample giants were extracted from the
new MARCS\footnote{{http://marcs.astro.uu.se/}} spherical model
atmosphere grids \citep{gustafsson03, gustafsson08}. The abundance
analysis and the spectral synthesis calculations were performed with 
the code \texttt{calrai}, first developed by
\citet{spite67} (see also the atomic part description in \citet{cayrel91}), 
and continuously updated over the years. This code assumes local
thermodynamic equilibrium (LTE), and performs the 
radiative transfer in a a plane-parallel geometry.

The program \texttt{calrai} is used to analyze all DART datasets and in
particular the VLT/giraffe spectroscopy of Fornax, Sculptor, and
Sextans dSphs (Hill et al. in prep; Letarte et al., 2010; Tafelmeyer et
al., in prep). The results of these works are partly summarized in
\citet{tolstoy09}. The homogeneity of the analyses allows secure
comparisons of the chemical patterns in all metallicity ranges and
between galaxies.

As we discuss in a following subsection, we had to combine the results
provided by \texttt{calrai} with those of
\texttt{turbospectrum} \citep{alvarez98} in order to properly take  into account the
continuum scattering in the stellar atmosphere for the abundances
derived from lines in the blue part of the spectra.  We have conducted
a series of tests on HD122563 and CD$-$38 245, sampling the two
extremes of our metallicity range, to verify the compatibility of the
two codes. In all these tests, the continuum scattering is
treated as absorption in order to put both codes on same
ground.  Both \texttt{calrai} and the version of
\texttt{turbospectrum} adopted for our calculations use spherical
model atmospheres and plane-parallel transfer for the line formation,
also referred to {\it sp} computations in the terminology of
\citet{heiter06}. These authors recommend the use of spherical model
atmospheres in abundance analyses for stars with \logg $\leq 2$ and
4000 $\leq$\Teff$\leq$ 6500K, and our target stars fall into this
range. As they point out, geometry has a small effect on line
formation and a plane-parallel transfer in a spherical model
atmosphere gives excellent results with systematics below 0.1 dex
compared to a fully spherical treatment.

Because the original abundances of HD122563 and CD$-$38 245 were derived
by \citet{cayrel04} using MARCS plane-parallel models ({\it pp}
computations) and an earlier version of \texttt{turbospectrum}, we
proceeded to reproduce their results with our techniques. We first
checked that we could reproduce the abundances of \citet{cayrel04}
with {\it pp} computation of the current version of
\texttt{turbospectrum}: they agree within 1/1000th of a dex. Moving
from {\it pp} \texttt{turbospectrum} computations to {\it sp} induces
a mean $-0.056 \pm 0.04$dex shift for CD$-$38 245 and $-0.12 \pm 0.07$
dex for HD122563 for all iron lines (iron is taken here as reference,
as it provides the largest number of lines).  Considering only lines
with excitation potential above 1.4 eV, as we do in the analysis of
our dSph star sample, the mean difference in iron abundance between
the {\it sp} versions of \texttt{turbospectrum} and \texttt{calrai} is
$-0.003$ and $-0.08$ dex for CD$-$38 245 and HD122563,
respectively. Restricting our analysis to FeI and to lines with
excitation potential larger than 1.4eV, we get a difference of $-0.038
\pm 0.04$dex and $-0.108 \pm 0.04$ for CD$-$38 245 and HD122563,
respectively. These numbers perfectly agree with the work of
\citet{heiter06}, which describes the difference between {\it sp} and  fully spherical 
      {\it ss} models increasing with decreasing stellar gravity and
      increasing effective temperature. The variation in stellar
      gravity is the primary factor of systematics between spherical
      and parallel models.  CD$-$38 245 is hotter than HD122563 by
      only 200K, but its gravity is 0.4 dex larger.

If we consider all chemical elements and lines for our sample stars
and keep the same atmospheric parameters, the mean of the difference
in abundances between the programs \texttt{calrai} and
\texttt{turbospectrum} for lines with an excitation potential larger
than 1.4eV are 0.03 (std=0.05) for Fn05-42, $-$0.02 (std=0.05) for
Scl07-49, $-$0.01 (std=0.03) for Scl07-50, 0.02 ( std=0.08) for
Sex11-04, and $-$0.05 (std=0.06) for Sex24-72.  These values are
reached when both codes consider scattering as absorption.

In conclusion, \texttt{calrai} and
\texttt{turbospectrum} lead to highly compatible results, well within
our observational error bars, and their results can be combined with
confidence.

\subsection{Photometric parameters}

Optical photometry ($V$, $I$) was available for our sample stars from
the ESO 2.2m WFI \citep[Battaglia 2010, in prep]{tolstoy04,
battaglia06}.  This was supplemented by near-infrared $J$, $H$, $K_s$
photometry from 2MASS \citep{skrutskie06} and from $J$, $K$ photometry
made available from VISTA commissioning data, which were also calibrated
onto the 2MASS photometric system. Table \ref{tab:photometry} presents
the corresponding stellar magnitudes.

With the exception of the Sculptor stars, for which we have no $H$-band
data, we considered four different colors $V-I$, $V-J$, $V-H$ and
$V-K_s$ and the CaT metallicity estimates to get initial values of
\Teff, following the calibration of \citet{ramirez05}, with a
reddening law (A(V)/E(B-V)=3.24) and extinction of E(B-V)=0.05, 0.017
and 0.02 for Sextans, Sculptor, and Fornax, respectively
\citep{schlegel98}.  The calibration of \citet{ramirez05} is our  
 temperature calibration for
all DART samples. It has the
advantage of using the same photometric system as our observations,
avoiding  precarious Johnson-Cousin system conversions and thereby providing
a more robust \Teff\ scale.

The photometric surface gravities were calculated using
the standard relation  

\noindent $\logg_{\star} = \logg_{\odot} +
\log{\frac{\mathcal{M}_{\star}}{\mathcal{M}_{\odot}}} + 
4 \times \log \frac{T_{\mathrm{eff}\star}}{T_{\mathrm{eff}\odot}} + \\
~~~~~~~~~~~~0.4 \times (M_{\mathrm{Bol}\star} - M_{\mathrm{Bol}\odot})$

\noindent with $\logg_{\odot} =4.44$, $T_{\mathrm{eff}\odot}$=5790K and 
$M_{\mathrm{Bol}\star}$ the absolute bolometric magnitude calculated
from the $V$-band magnitude using the calibration for the bolometric
correction from \citet{alonso99}, assuming distances of 140~kpc for
Fornax and 90~kpc for Sextans and Sculptor \citep{karachentsev04}. The
mass of the RGB stars was assumed to be $\mathcal{M}_{\star}$ = 0.8
$\mathcal{M}_{\odot}$.

\subsection{Spectroscopic metallicities, temperatures, and microturbulence velocities}

The effective temperatures and microturbulence velocities (v$_{mic}$)
were determined spectroscopically from lines of FeI by requiring no
trend of abundance with either excitation potential or equivalent
width.  According to \citet{magain84}, random errors on equivalent
widths can cause a systematic bias in the derived microturbulence
velocities. This effect is caused by the correlation of the error on
each equivalent width and the abundance deduced from it, and can be
avoided by using predicted equivalent widths instead of measured ones
in the plot of iron abundances vs. equivalent widths. The former are
calculated using the line's log gf value and excitation potential and
the stellar atmospheric parameters. They are thus free of random
errors and yield unbiased microturbulence velocities. We estimated the
change in velocities that would compensate for the differences of
slopes when changing to predicted equivalent widths, and found that
much less than 0.1 km/s was sufficient in all cases. This is small
compared to the uncertainties given in Sect.~\ref{sec:finalatmo}.
We nevertheless used the predicted equivalent widths instead of
the measured ones for the Sextans stars with [Fe/H]$\geq -3$.
Extrapolation of the model atmospheres, which was necessary for the
metallicity domain below $-$3, can cause errors in the predicted
equivalent widths, therefore they were not used for the remaining
more metal poor stars of our sample.

Surface gravities are often determined spectroscopically by requiring
iron to have ionization equilibrium.  For our sample stars, [FeI/H]
abundances were determined from 20-70~FeI lines per star, depending on
S/N and metallicity. However, the FeII abundances were deduced from a
small number (2 to 4) of weak lines and were consequently fairly
uncertain. Moreover, the atmospheres of metal-poor stars are
characterized by low electron number densities and low
opacities. Hence local thermodynamical equilibrium (LTE) is not
fulfilled in the atmospheric layers where these lines are formed.  The
main non-LTE effect on iron is over-ionization, which is caused by
photo-ionization of excited FeI lines from a super-thermal radiation
field in the UV \citep[see e.g.][ and references therein]{asplund05}.
Therefore we did not use spectroscopic gravities but rather kept the
photometric ones.  Note that all our stars, except Scl07-50, have
higher iron abundances when derived from ionized species instead of  
neutral ones. Even though the differences are typically within the
errors, we see this as  clear evidence for over-ionization owing to
NLTE effects. The situation is even clearer for titanium, with
differences between [TiI/H] and [TiII/H] of $\sim$0.4~dex. The first
ionization potential of titanium is slightly lower than that of iron,
while the second ionization potential is roughly the same. Therefore
one expects the overionization caused by the UV radiation field to be more
prominent for titanium than for iron.  Moreover, almost all
\ion{Ti}{I} lines measured here arise from low excitation levels, and
are therefore also more prone to NLTE effects on the level population
than the higher excitation \ion{Fe}{I} lines.

For Scl07-049, we count $\approx$20 FeI lines ($\approx$50\% of the
total number of FeI lines) with excitation potentials below
1.4~eV. They may well be affected by non-LTE effects, as these are
particularly important for low excitation lines.  This influences the
atmospheric parameters by biasing the diagnostic slopes and the
metallicities. Hence, these lines were not considered, neither for the
determination of the mean metallicity nor for the effective
temperature.  {For Fnx05-042 the influence of
these lines was negligible, thus they were kept in the analysis. The
fraction of low excitation lines is much smaller in the case of the
two Sextans stars and Scl07-050 (10 to 20\% of the total). Hence, all
lines were included.

\subsection{Final atmospheric parameters and error estimates}
\label{sec:finalatmo}

The convergence to our final atmospheric parameters, presented in
Table \ref{tab:atmo}, was achieved iteratively, as a trade-off between
minimizing the trends of metallicity with excitation potentials and
equivalent widths on the one hand and minimizing the difference
between photometric and spectroscopic temperatures on the other hand.
We started from the photometric parameters and adjusted \Teff\ and
\vmic\ by minimizing the slopes of the diagnostic plots, allowing for
deviation by no more than 2 $\sigma$ of the slopes. This yielded new
metallicities, which were then again fed into the photometric calibration
to derive new photometric temperatures and gravities.  In this
way, after no more than two or three iterations, we converged on our final
atmospheric parameters. The errors in atmospheric parameters,
+150~K for \Teff\, +0.2~km/s for \vmic\, and +0.3~dex log(g), also
reported in Table~\ref{tab:atmoerr}, reflect the range of parameters
over which we can fulfill our conditions of both minimal 
slopes in the diagnostic plots and small difference between 
photometric and spectroscopic temperatures.

However, there is a strong correlation between
excitation potential and equivalent width, in the sense that the
majority of the weak lines have high excitation potential, and most of
the strong lines have low excitation potentials.  Therefore, \Teff\ and
\vmic\ cannot be determined independently from each other.  To a
certain degree, it is possible to compensate for an uncertainty in one
parameter by modifying the other.  $\Delta$\Teff =150K is the maximum
variation in effective temperature that could still lead to an
acceptable solution, and $\Delta$\vmic= 0.2km/s is the corresponding
change in microturbulence velocity keeping the slopes between
abundance and equivalent width or excitation potential to zero within
2~$\sigma$.

\begin{table*}[ht!]
\begin{center}
\caption[]{Visual and near-IR photometry. $J,H,Ks$ for Sextans and Fornax from 2MASS, for Sculptor from  VISTA ; $V,I$ from ESO 2.2m WFI.}
\label{tab:photometry}
\begin{tabular}{lccccc}
\hline 
 ID       &  $V $   &    $I$   &  $J$            & $H$                 & $Ks$                \\
\hline                                                                           
Sex24-72  & 17.35 $\pm 0.02$ & 16.03 $\pm 0.02$ & 15.13$\pm 0.04$& 14.53 $\pm 0.04$  & 14.42$\pm 0.06 $ \\
Sex11-04  & 17.23 $\pm 0.02$ & 16.02 $\pm 0.02$ & 14.92$\pm 0.04$& 14.30 $\pm 0.04$  & 14.13$\pm 0.09 $ \\
Fnx05-42  & 18.48 $\pm 0.02$ & 17.23 $\pm 0.02$ & 16.15$\pm 0.09$& 15.72 $\pm 0.15$  & 15.31$\pm 0.19 $ \\
Scl07-49  & 18.35 $\pm 0.02$ & 17.27 $\pm 0.02$ & 16.44$\pm 0.02$&                   & 15.76$\pm 0.02 $ \\
Scl07-50  & 18.63 $\pm 0.02$ & 17.73 $\pm 0.02$ & 16.81$\pm 0.02$&                   & 16.13$\pm 0.03 $ \\
\hline
\end{tabular}			
\end{center}
\end{table*}

\begin{table*}[ht!]
\begin{center}
\caption[]{Photometric and spectroscopic stellar parameters: 
The stellar photometric temperatures in Sextans and Fornax 
are determined from $V-I$, $V-J$, $V-H$, and $V-Ks$, those 
in Sculptor from $V-I$, $V-J$, and $V-Ks$; phot and spec refer to
the finally adopted photometric and spectroscopic temperatures. The
last column indicates the spectroscopically derived microturbulence
velocities.}
\label{tab:atmo}
\begin{tabular}{c|cccccc|c|c}
\hline 
& \multicolumn{6}{c|}{T$_{eff}$ (K)} & log(g) & v$_{mic}$ (km/s) \\
ID & $V-I$ & $V-J$ & $V-H$ & $V-Ks$ & phot & spec & phot &\\
\hline
Sex24-72 & 4409 & 4393 & 4345 & 4432 & 4395 &   4430 & 0.75 & 2.2\\
Sex11-04 & 4547 & 4321 & 4263 & 4377 & 4377 &  4320 & 0.57 & 2.2\\
Fnx05-42 & 4446 & 4270 & 4331 & 4364 & 4353 &   4325 & 0.70 & 2.3\\
Scl07-49 & 4688 & 4625 &      & 4579 & 4631 &   4450 & 1.28 & 2.4\\
Scl07-50 & 5008 & 4723 &      & 4647 & 4793 &  4775 & 1.56 & 2.2\\
\hline
\end{tabular}			
\end{center}
\end{table*}

\begin{table*}[ht!]
\begin{center}
\caption[]{ Errors owing to uncertainties in stellar atmospheric
  parameters, where $\Delta$(T,v) stands for a simultaneous shift - consequence
  of their correlated impact on abundances - of \Teff\ by +150~K and
  \vmic\ by +0.2~km/s and $\Delta_g$ for a shift of log(g) by
  +0.3~dex. The uncertainties are roughly symmetric, therefore
  negative shifts of the atmospheric parameters yield the same errors
  with opposite signs. }
\label{tab:atmoerr}
\begin{tabular}{l|rr|rr|rr|rr|rr}
\hline
&\multicolumn{2}{|c|}{\textbf{Sex24-72}}&\multicolumn{2}{|c|}{\textbf{Sex11-04}}&\multicolumn{2}{|c|}{\textbf{Fnx05-42}}&\multicolumn{2}{|c|}{\textbf{Scl07-49}}&\multicolumn{2}{|c}{\textbf{Scl07-50}}\\
&$\Delta$(T,v)&$\Delta_g$&$\Delta$(T,v)&$\Delta_g$&$\Delta$(T,v)&$\Delta_g$&$\Delta$(T,v)&$\Delta_g$&$\Delta$(T,v)&$\Delta_g$\\
\hline
FeI &   0.16&$-$0.07&   0.13&  0.04  &   0.23& 0.00  &   0.17& 0.04  &   0.15&$-$0.04\\
FeII&$-$0.16&$-$0.10&$-$0.29&  0.02  &$-$0.28& 0.01  &$-$0.09& 0.08  &$-$0.00& 0.07  \\
Al  &       &       &$-$0.06& $-$0.18&$-$0.13&$-$0.08&       &       &$-$0.04&$-$0.03\\
Ba  &$-$0.07& 0.04  &$-$0.10&  0.02  &$-$0.19& 0.11  &       &       &       &       \\
C   &   0.30&$-$0.40& 0.30  &$-$0.35 &0.37   &$-$0.30&       &       &0.40   &$-$0.30\\
Ca  &$-$0.05& 0.04  &   0.02& $-$0.02&$-$0.07&$-$0.04&$-$0.02& 0.00  &$-$0.09&$-$0.08\\
Co  &$-$0.04&$-$0.07&$-$0.17&  0.00  &$-$0.05&$-$0.02&       &       &   0.04&$-$0.01\\
Cr  &   0.04& 0.01  &   0.08&  0.05  &   0.04&$-$0.02&   0.03& 0.01  &   0.03&$-$0.02\\
Mg  &$-$0.10&$-$0.01&$-$0.08& $-$0.06&$-$0.13&$-$0.05&   0.00&$-$0.06&$-$0.10&$-$0.04\\
Mn  &$-$0.03& 0.02  &   0.03& $-$0.03&   0.07& 0.06  &       &       &   0.05&$-$0.01\\
Na  &$-$0.09& 0.06  &   0.05&  0.05  &$-$0.01& 0.01  &   0.01& 0.03  &$-$0.01& 0.01  \\
Ni  &$-$0.04&$-$0.05&$-$0.05&  0.01  &$-$0.08&$-$0.02&   0.02& 0.02  &$-$0.00&$-$0.03\\
O   &       &       &$-$0.03&  0.03  &       &       &       &       &$-$0.03& 0.13  \\
Sc  &$-$0.15& 0.05  &$-$0.17& $-$0.01&$-$0.28& 0.05  &$-$0.14& 0.11  &$-$0.11& 0.08  \\
Si  &$-$0.12& 0.06  &$-$0.29& $-$0.21&$-$0.22&$-$0.09&       &       &$-$0.12&$-$0.07\\
SrII&$-$0.28&       &$-$0.29&        &       &       &       &       &       &       \\
TiI &   0.09& 0.05  &   0.19&  0.07  &   0.07&$-$0.03&   0.10& 0.01  &       &       \\
TiII&$-$0.18& 0.04  &$-$0.19&  0.00  &$-$0.28& 0.08  &$-$0.17& 0.11  &$-$0.11& 0.10  \\
Y   &$-$0.12&       &$-$0.14&  0.00  &$-$0.23&       &       &       &       &       \\
\hline
\end{tabular}           
\end{center}
\end{table*}

\section{Determination of abundances}
\label{sect:abundances} 

Our line list (shown in Table~\ref{tab:linelist}) combines the
compilation of \citet{shetrone03}, \cite{francois07} and
\citet{cayrel04}.  The solar abundances of \citet{anders89} are
adopted, with the exception of C, Ti, and Fe \citep{grevesse98}.  In
the following all abundances, including those of the comparison
samples, are scaled to these solar values.  All abundances are listed
in Table~\ref{tab:abundances} and they were derived from the
equivalent widths, with the exception of those marked in
Table~\ref{tab:linelist}.

\label{sec:abundances}

\subsection{The influence of scattering}

The color-magnitude diagrams in Fig.~\ref{fig:cmd} show that the
stars from our sample are located close to the tip of the RGB. Thus, at low
metallicity and low temperature, a proper treatment of continuum
scattering in the stellar atmosphere is mandatory.

Our standard line formation code, \texttt{calrai}, treats
continuum scattering as if it were absorption in the source function,
i.e., $S_{\nu}= B_{\nu}$, an approximation that is valid at long
wavelengths. In the blue, however, the scattering term must be
explicitly taken into account: $S_{\nu}=(\kappa_{\nu} \times B_{\nu} +
\sigma_{\nu} \times J_{\nu})/(\kappa_{\nu} + \sigma_{\nu})$.
\cite{cayrel04} have shown that the ratio $\sigma$/$\kappa$ in the
continuum is 5.2 at $\lambda$ = 350 nm, whereas it is only 0.08 at
$\lambda$ = 500 nm, for $\tau_{\nu}$=1 and a giant star with \Teff =
4600 K, log g = 1.0, and [Fe/H] = $-$3.  Therefore, we used the code
\texttt{turbospectrum} \citep{alvarez98}, which includes the proper
treatment of scattering in the source function, to correct our
original \texttt{calrai} abundances.

Figure~\ref{fig:scattering} illustrates how scattering influences the
abundances.  All calculations are done with \texttt{turbospectrum}. In
the lower panel, we plot the differences in abundances between those
derived assuming a pure absorption source function and those
explicitly including a scattering term.  Abundances tend to be
overestimated when scattering is not properly taken into account, and
the main driver of the scattering effect is the line wavelength. In
addition, the lower the gravity, the larger the effect of continuum
scattering. There is also mild evidence that the higher the excitation
potential of the lines, the larger the differences between the
abundance estimates. This is a consequence of the increase in
formation depth of the lines. In conclusion, the corrections for some
of our stars can be large because of the combined effects of being
very metal poor and high up on the RGB (luminous and cool).  The upper
panel in Fig.~\ref{fig:scattering} shows, with the example of [FeI/H],
that when scattering is properly treated, no spurious wavelength
dependence remains.  The larger dispersion at bluer wavelengths arises
from lower signal-to-noise ratio spectra in this domain.

{\it To avoid any bias in the effective temperatures, we did
not consider the iron lines at short wavelengths, because they have
large scattering corrections, but instead restricted our analyses to
iron lines at $\lambda >$ 4800\AA.} The only exception to this rule is
Scl07-050, for which we could detect only around 10 lines with
$\lambda >$ 4800\AA. Therefore we used the blue lines as well for the
sake of better statistics. However, as Fig.~\ref{fig:scattering}
shows, Scl07-050 has the smallest influence of scattering, never
exceeding 0.2 dex, even at the lowest wavelengths.  We estimated that
the bias in the slopes of iron abundances vs. excitation potentials
causes errors in effective temperatures of no more than 30K.  This is
negligible compared to the uncertainties given in
Table~\ref{tab:atmoerr}.

 In summary, our method has been the following: \texttt{calrai}
  was used to determine the stellar atmospheric parameters. These
  parameters were kept unchanged.  The abundances were corrected
  line-by-line using \texttt{turbospectrum} in two modes: (i)
  treating continuum scattering as if it were absorption in the source
  function, (ii) with a proper scattering term in the source function.
  The difference between the two outputs was then applied to the
  \texttt{calrai} abundances. The corrections are listed in Table
  \ref{tab:linelist}.

\begin{figure}
\begin{center}
\includegraphics[angle=-90, width=\hsize]{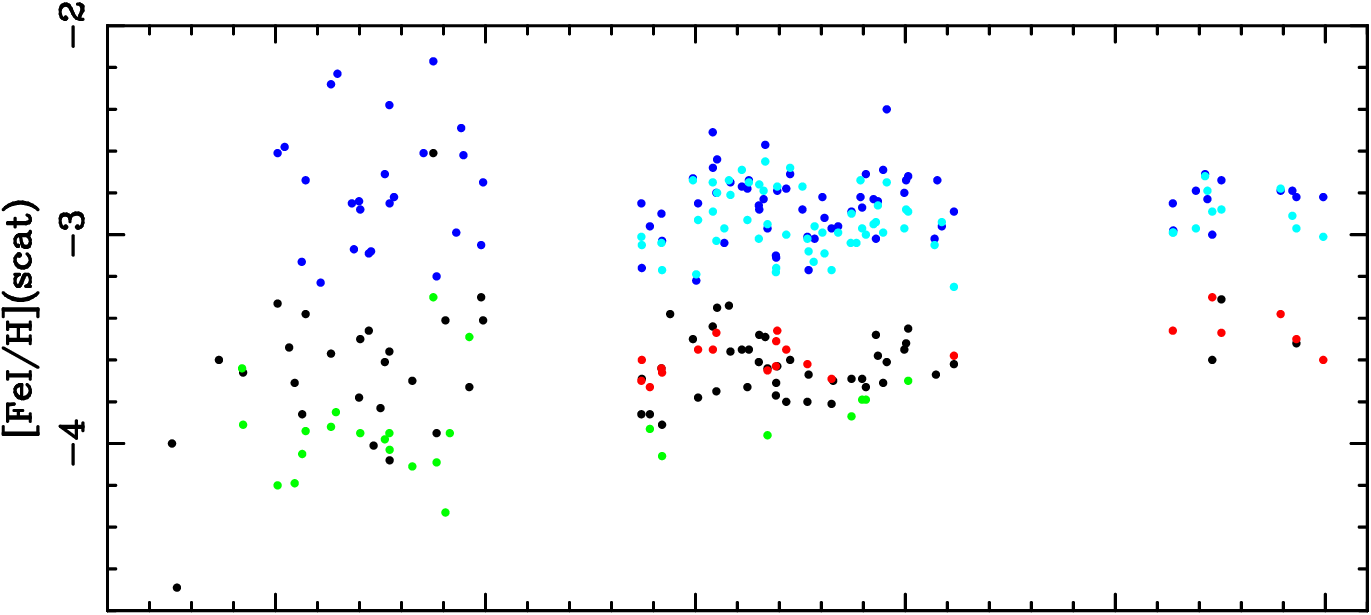}
\includegraphics[angle=-90, width=\hsize]{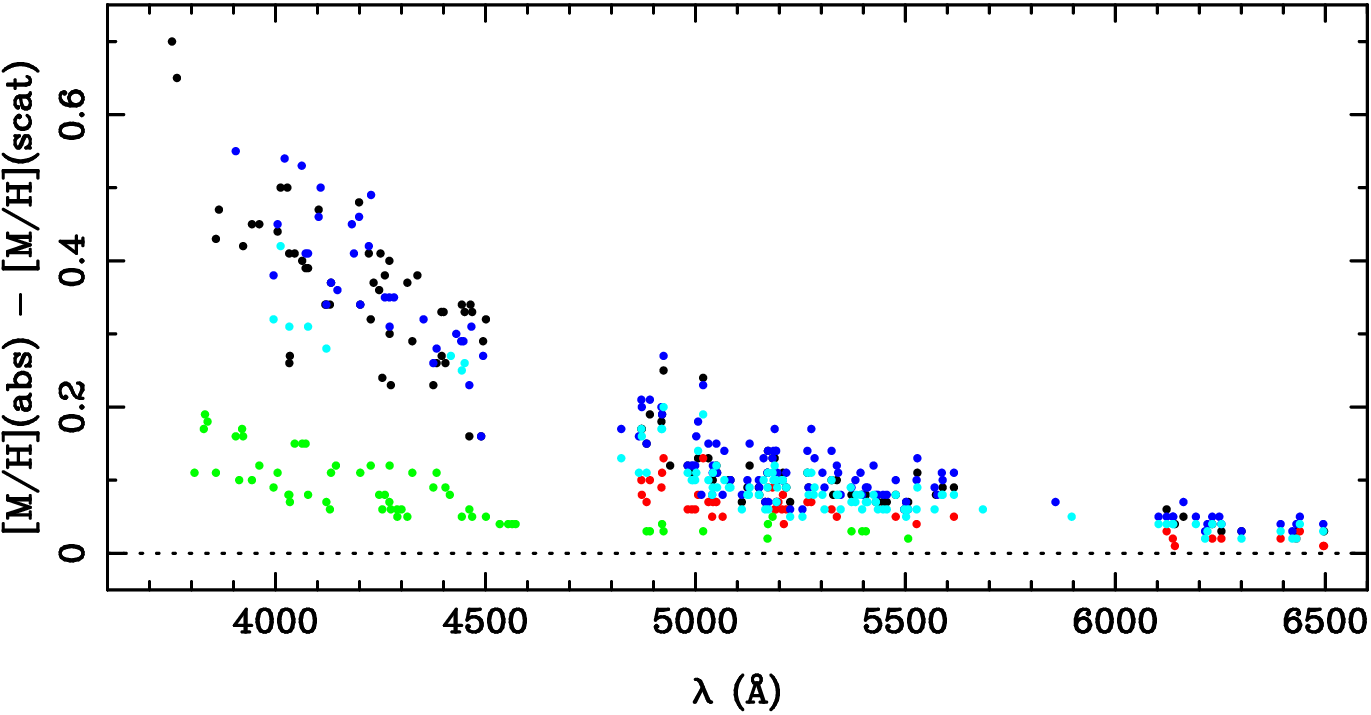}
\caption{{\bf Lower panel:} The differences in abundances derived 
assuming a pure absorption source function and when a scattering term
is introduced. All calculations are done with
\texttt{turbospectrum}. All lines are considered and are shown at
their wavelengths. Black points trace Fnx05-42 , blue stands for the 2
Sextans stars, red for Scl07-49 and green for Scl07-50.  {\bf Upper
panel:} [FeI/H] when scattering is properly taken into account.  There
is no trend of abundance with wavelength, while there was a clear
increase at bluer wavelengths before correction. }
\label{fig:scattering}
\end{center}
\end{figure}

\subsection{Errors}
\label{sec:errors}
The errors given in Table~\ref{tab:abundances} and shown in the
figures were derived in the following way: Although we used
\texttt{splot} to measure the line equivalent widths, we used the
error estimates of \texttt{DAOSPEC}, which provide the uncertainty in
equivalent width, $\Delta$EW, derived from the residual of the
Gaussian fit.  This includes the uncertainties in the placement of
the continuum.  We propagated this $\Delta$EW throughout the abundance
determination process, thus providing for each line $\delta_{DAO}$,
the abundance uncertainty corresponding to EW~$\pm~\Delta$EW. The
abundance uncertainty is not necessarily symmetric, and in all cases 
we adopted the largest one. 
For the lines for which abundances were determined from
synthetic spectra, $\delta_{DAO}$ was estimated from the range of
abundances for which a good fit of the observed line profile could be
achieved. 
For the abundance of elements measured from more than one line, the
mean abundance and dispersion
($\sigma (X)$) were computed, weighting the lines by
1/$\delta_{DAO}^2$. The corresponding error on the mean, $\delta_{\sigma}$, is
given as $\frac{\sigma (X)}{\sqrt{N_X}}$, where N$_X$ is the number of
lines used to determine the abundance of element X.  In order to avoid
artificially small errors due to low number statistics when too few
lines were available to measure a robust $\sigma (X)$, we took as a
minimum for $\sigma (X)$, the dispersion of Fe abundances and assumed
$\delta_{Fe} = \frac{\sigma (Fe)}{\sqrt{N_X}}$ as the smallest
possible error value. The final error in [X/H] is then max
($\delta_{DAO}$, $\delta_{\sigma}$, $\delta_{Fe}$). To get the error
in [X/Fe], the uncertainty of [Fe/H] was added in quadrature.

The considerations above do not contain errors due to uncertainties
in the atmospheric parameters.  As explained in Sect.~\ref{sec:atmo}, 
there is a strong correlation between micro-turbulence velocity and
effective temperature. Therefore we give errors as a combination
of $\Delta$T and $\Delta v_{mic}$. The error boundaries reflect the
parameter range for which the slopes of both diagnostic plots
(excitation potential and equivalent widths versus iron abundance) 
can be brought close to zero.

\subsection{Abundances not based on equivalent widths} 
\label{sec:noeqw}

Some of the lines, primarily in the blue, gave uncertainties of
0.5~dex or more. The reasons for this are manifold: extremely strong
lines ($\sim$ 300m\AA ), large residuals of the Gaussian fit because of
low S/N, problems of continuum placement (caused by low S/N or broad
absorption bands in the vicinity), very strong blends.  Whenever
possible, we did not use these lines and used instead those in the red
part of the spectrum. For the elements for which this strategy could
not be adopted, the abundances were determined from synthetic spectra:

\textbf{(i) Hyperfine structure:} 

The abundances of barium, cobalt, manganese, and scandium from the red lines
were determined from spectral synthesis, in order to consider hyperfine
splitting of the lines. The atomic data of the hyperfine components
were taken from \cite{prochaska00} and, if no data were given there,
from Kurucz database\footnote{http://kurucz.harvard.edu/atoms.html}.
For the three blue lines of scandium at 4246.82\AA, 4314.08\AA, and
4400.39\AA\ no hfs data were available, hence we did not correct the
abundances.

\textbf{(ii) Blends:} 

If there were significant blends, the abundances were derived
from synthesis to correct for the contribution of these blends. The silicon
and aluminum lines of Sex24-72 were blended by molecular bands, the
Al line at 3944.01\AA\ and the Si line at 3905.52\AA\ were not
used at all, since the blends were too strong. The chromium abundance
of the line at 5208.42\AA\ was larger than the abundances derived from
the other lines by 0.3-0.5~dex in all stars. The synthetic spectra
revealed the presence of an iron blend. After taking this into
consideration, the abundance of the 5208.42\AA\ line
agreed well with the other chromium lines.  The abundances of two magnesium lines,
which lie close to a strong hydrogen line at 3835.38\AA\ , 
were determined with synthesis, to make sure that
their abundances were not affected by the H line wings.

\textbf{(iii) Very weak or strong lines:}

The equivalent widths were determined from Gaussian fits to the line
profile. However, for strong lines ($\gsim$250~m\AA ), this
might not give correct results due to the presence of significant 
wings. Similarly, if lines are too weak and narrow ($\lsim$20~m\AA ),
the line profile may deviate from a Gaussian shape and therefore a 
Gaussian fit can be incorrect. Thus, for  elements for which
only very weak or very strong lines were available, abundances were
determined from spectral synthesis: the two sodium resonance lines of
Sex24-72 at 5889.97\AA\ and 5895.92\AA\ (the uncertainty in [Na/H]
derived from the line at 5889.97\AA\ was much larger than 0.5~dex,
therefore only the other line is used here) and  the very weak yttrium
line in Sex11-04.

\section{Results}
\label{sect:results}

\subsection{Comparison samples}
\label{sec:comp}
We compare our results with high-resolution spectroscopic analyses of
RGBs with [Fe/H] $\le -2.5$ in the Milky Way halo 
\citep{cayrel04, cohen08, aoki05, aoki07, cohen06, lai08, honda04} 
and in faint (Ursa MajorII and Coma Berenices, Bo\"otes I,  Leo IV)
and classical (Sextans, Draco, Fornax, Carina, Sculptor) dSphs 
\citep[][]{letarte10, shetrone01, fulbright04, koch08,  
cohen09, aoki09, norris10, simon10, frebel10, frebel10a}.  We distinguish
between unmixed and mixed RGBs in order to be able to draw a reliable
comparison sample for the elements that are sensitive to mixing.
The term ``mixed'' refers to stars for which the material from deep
layers, where carbon is converted into nitrogen, has been brought to
the surface during previous mixing episodes.  Mixing has occurred
between the atmosphere of RGB stars and the H-burning layer where C is
converted into N by the CNO cycle \citep{cayrel04}.

Following \citet{spite05} and \citet{spite06}, we divide stars into
these two categories according to their carbon and nitrogen abundances
or the ratio between the carbon isotopes $^{12}$C and $^{13}$C, when
possible.  The criteria are [C/N]$\le -0.6$ or
log($^{12}$C/$^{13}$C)$\le 1$ for mixed stars and [C/N]$> -0.6$ or
log($^{12}$C/$^{13}$C)$> 1$ for unmixed ones.  Figure \ref{fig:TLog}
presents the \logg\ vs \Teff\ diagram for our comparison sample and
the stars analyzed in this paper.  Mixed stars are generally more
evolved and are located on the upper RGB above the RGB bump luminosity
\citep{gratton00}.  The luminosity of the RGB bump decreases with
increasing metallicity \citep{fusipecci90}. At [Fe/H]$<-2.5$ it is
predicted around
$\log{\frac{\mathcal{L}_{\star}}{\mathcal{L}_{\odot}}} \sim 2.6$
\citep[N. Lagarde \& C. Charbonnel, private communication]{spite06}.
None of our dSph stars have N abundances, therefore, we rely on
their location at the top of the \logg\ vs \Teff\ diagram to conclude
that they are indeed mixed stars.

 The abundances of our comparison sample have been derived using
  plane-parallel atmospheric models.  We have run \texttt{calrai} with
  plane-parallel atmospheric models \citep{gustafsson75} on our sample
  stars. The effective temperatures were changed by less than
  100K. There was no systematic shift in abundances. The mean
  differences in abundances ([M/H]) derived in {\it pp} and {\it sp}
  models are well within the observational errors: 0.04 dex (std
  =0.04) for Sex24-72, 0.02 dex (std=0.09) for Sex11-04, 0.005 dex
  (std =0.13) for Fnx05-42, $-$0.038 (std=0.05) for Scl07-49, and
  0.005 (std=0.017) for Scl07-50. Similarly, the abundance ratios were
  modified by 0.005 dex (std =0.04) for Sex24-72, $-$0.05 dex
  (std=0.09) for Sex11-04, $-0.03$ dex (std =0.13) for Fnx05-42,
  $-$0.002 (std=0.05) for Scl07-49, and $-$0.003 (std=0.018) for
  Scl07-50. As a conclusion, samples can be safely inter-compared.

\begin{figure}
\includegraphics[angle=-90, width=\hsize]{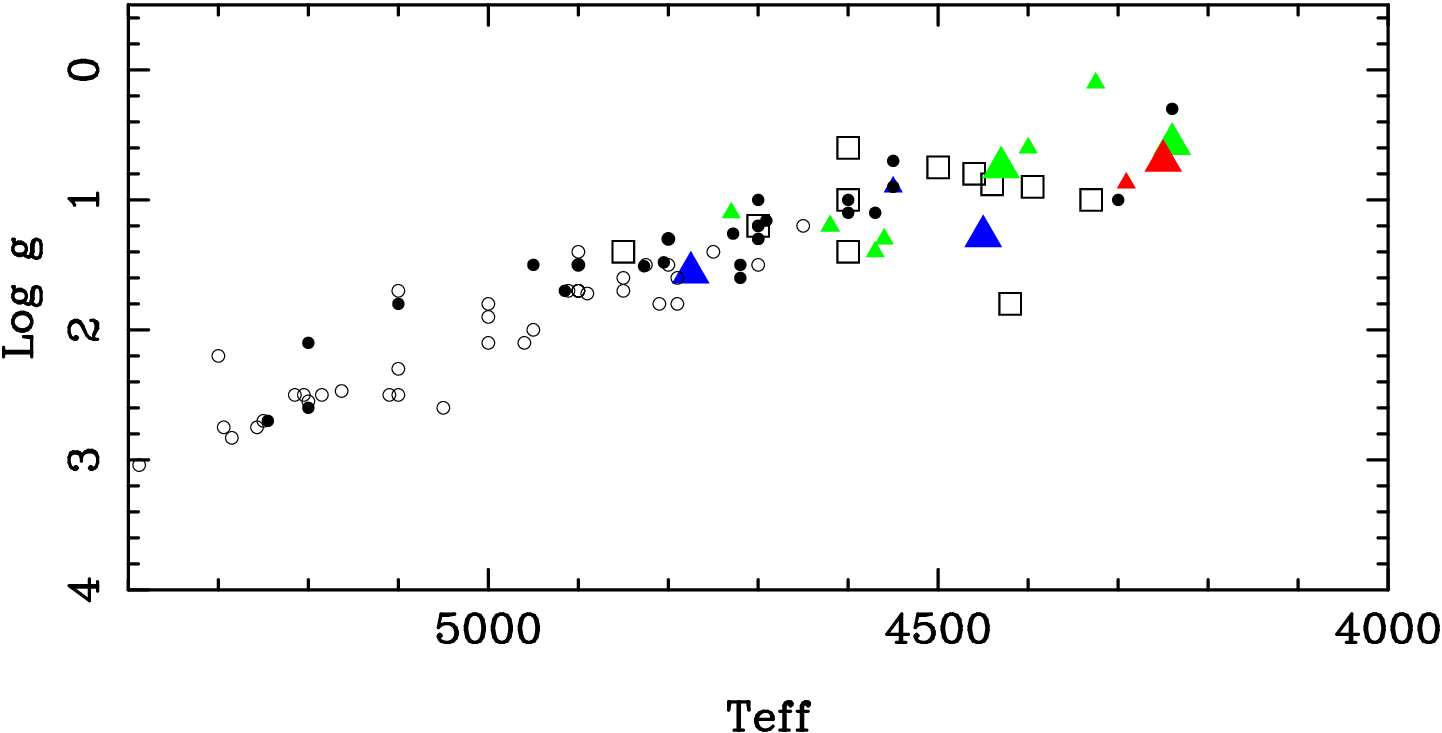}
\caption{DART dSph targets are in color: Sextans(Green) Scl(Blue) Fnx (Red). Large triangles
identify the 5 stars analyzed here, while small triangles indicate stars
in earlier publications \citep[][]{letarte10, aoki09, shetrone01, frebel10a}.  Open squares : UFDs, Carina and Draco
\citep{fulbright04, koch08, koch08b, frebel10, cohen09, norris10, simon10}. 
The Milky Way halo stars are represented by
circles; open circles are unmixed RGBs
and filled circles mixed RGBs \citep{cayrel04, cohen08,
aoki05, spite05, aoki07, cohen06, lai08, honda04}.
\label{fig:TLog}}
\end{figure}

\subsection{Iron}

Our results place all our sample stars at [Fe/H] $\lsim -3$ and three
stars below [Fe/H] $\sim -3.5$. This constitutes the first clear
evidence that likely all classical dSphs contain extremely
metal-poor stars.

\begin{figure}
\begin{center} 
\includegraphics[angle=-90, width=6cm]{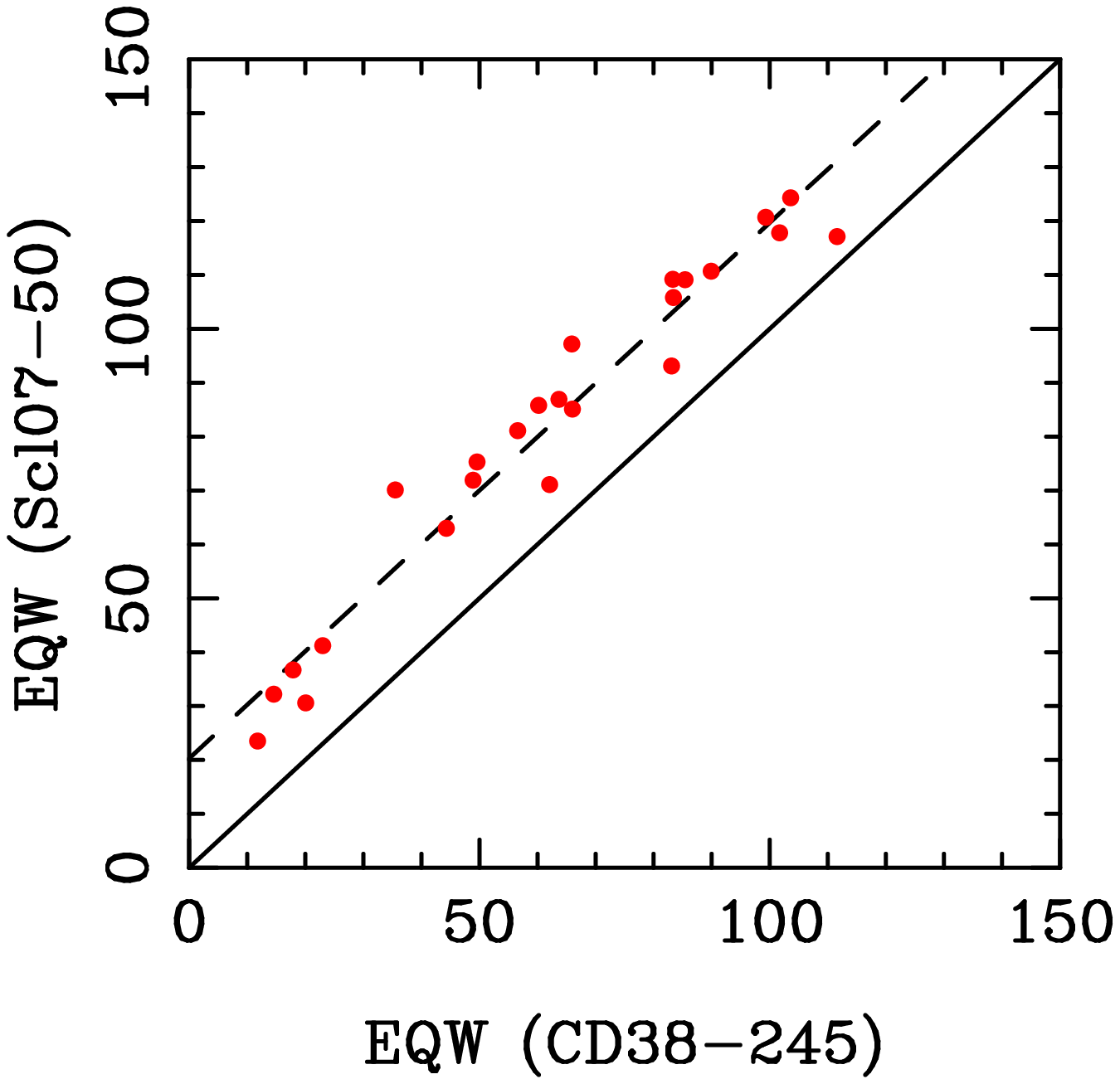} 
\caption{The comparison of the equivalent widths of Scl05-50 ([Fe/H]=$-3.96$)
and the Milky Way halo star CD$-$38 245 ([Fe/H]=$-4.19$). The dashed
lined follows the mean $\sim$ 20m\AA\ difference between the
equivalent widths of the two stars.}
\label{fig:cd38eqw}
\end{center}
\end{figure}

While the detailed abundances of our sample stars 
are
	 presented here after full analysis, the existence of
	 Scl07-50 and other EMPS in classical dSphs was already reported earlier
	 \citep{hill10a, tolstoy10}.
Scl07-50 at [Fe/H]=$-3.96 \pm 0.10$ is the most metal-poor star ever
observed in an external galaxy, but more fundamentally, it
considerably revises the metallicity floor of dSphs, setting it at
comparable level with the Milky-Way. Figure \ref{fig:cd38eqw}
compares the equivalent widths of CD$-$38 245 and Scl07-50, which have
very similar atmospheric parameters.  The lines
of CD$-$38 245 are on average  $\sim$~20m\AA\ weaker, corresponding
to the $\sim$0.2dex difference in metallicity between the two stars.

In Fig.~\ref{fig:compareFe} our UVES high-resolution [Fe/H] values 
are compared to
the FLAMES LR [Fe/H] estimates derived from the CaT ( \citep[with the
calibration of][]{battaglia08} for Sex11-04, Fnx05-42, Scl07-49
and Scl07-50 and from R$\sim$18000 spectroscopy at the HET for
Sex24-72. 

It is obvious that for 
[Fe/H] $\lsim -2.5$, the traditional
method based on the CaT absorption features
predicts  a metallicity that is too high. The underlying physical reasons for
this has recently been explained by Starkenburg et
al. (2010), who show that it reflects the
change of the profile of the CaT lines, from wing-dominated to core
dominated as the metallicity drops. They provide a new calibration
which is valid down to [Fe/H]=$-4$. Figure~\ref{fig:compareFe} indicate
with arrows the corresponding new CaT metallicities for stars of the
present sample.

\begin{figure}
\begin{center} 
\includegraphics[angle=-90, width=\hsize]{compareFe.ps} 
\caption{Comparison for our sample stars between [Fe/H] from high
  resolution UVES measurements ([Fe/H]$_{HR}$) and that from the CaT
  with the calibration of \citet{battaglia08} ([Fe/H]$_{LR}$), as
  filled circles.  For Sex24-72 HET/HRS spectroscopy
  at R$\sim 18000$ is used because CaT data are not available.  Blue stands
  for the Sculptor's stars, red for the Fornax' star, and green is for
  Sextans's ones.  The CaT metallicity determined with the new
  calibration of Starkenburg et al. (2010) for low metallicities are
  shown as open circles.  Crosses provide the same comparison at [Fe/H]
  $>-$2.5, where the classical (old) CaT calibration is robust.}
\label{fig:compareFe}
\end{center}
\end{figure}

\begin{figure}
  \begin{center}
    \includegraphics[width=8.5cm]{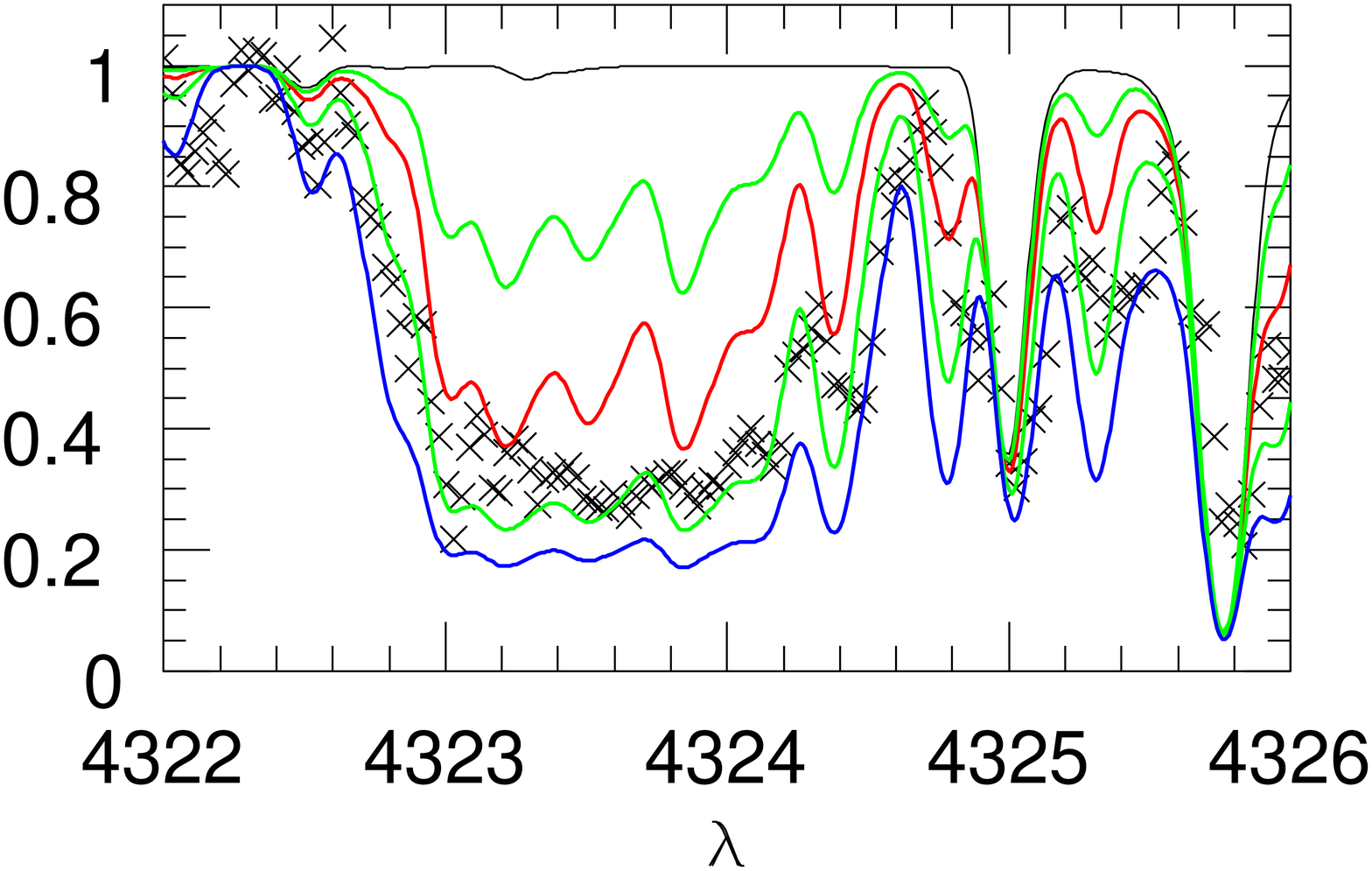}
    \caption{Synthesis of the molecular CH band at 4323\AA\ of Sex24-72. The crosses are the observed spectrum, the red line corresponds to the synthetic spectrum with [C/Fe]~=~0.0, the green lines are for [C/Fe]~=~$\pm$0.5, the blue line is for [C/Fe]~=~1.0, and the black line for the complete absence of C.}
    \label{fig:ch}
  \end{center}
\end{figure}

\subsection{Carbon}

Carbon abundances were determined from synthesis of the CH molecular
bands in the blue part of the spectrum. As an example,
Fig.~\ref{fig:ch} shows the synthesis of the CH band at 4323\AA\ for
Sex24-72. Because \texttt{calrai} does not include molecular bands, the
synthetic spectra were calculated with
\texttt{turbospectrum} using the same CH line list as in
\citet{cayrel04}.  The carbon abundances are sensitive to the assumed
oxygen abundances through the locking of C into the CO molecule. 
We only have upper limits for oxygen, hence we used Mg as an indicator of
the oxygen abundance, assuming [O/H]=[Mg/H]. Increasing by $+0.2$ dex
the adopted [O/Fe] translates into a maximum increase of $+0.05$ of
the derived C abundance (the exact dependence is a function of the
absolute C/O ratio in the star). This uncertainty is negligible with
respect to the errors on the fits to the data in combination with a
poor signal-to-noise ratio and  uncertainties associated with
the molecular lines' oscillator strengths.

\begin{figure}
\includegraphics[angle=-90, width=\hsize]{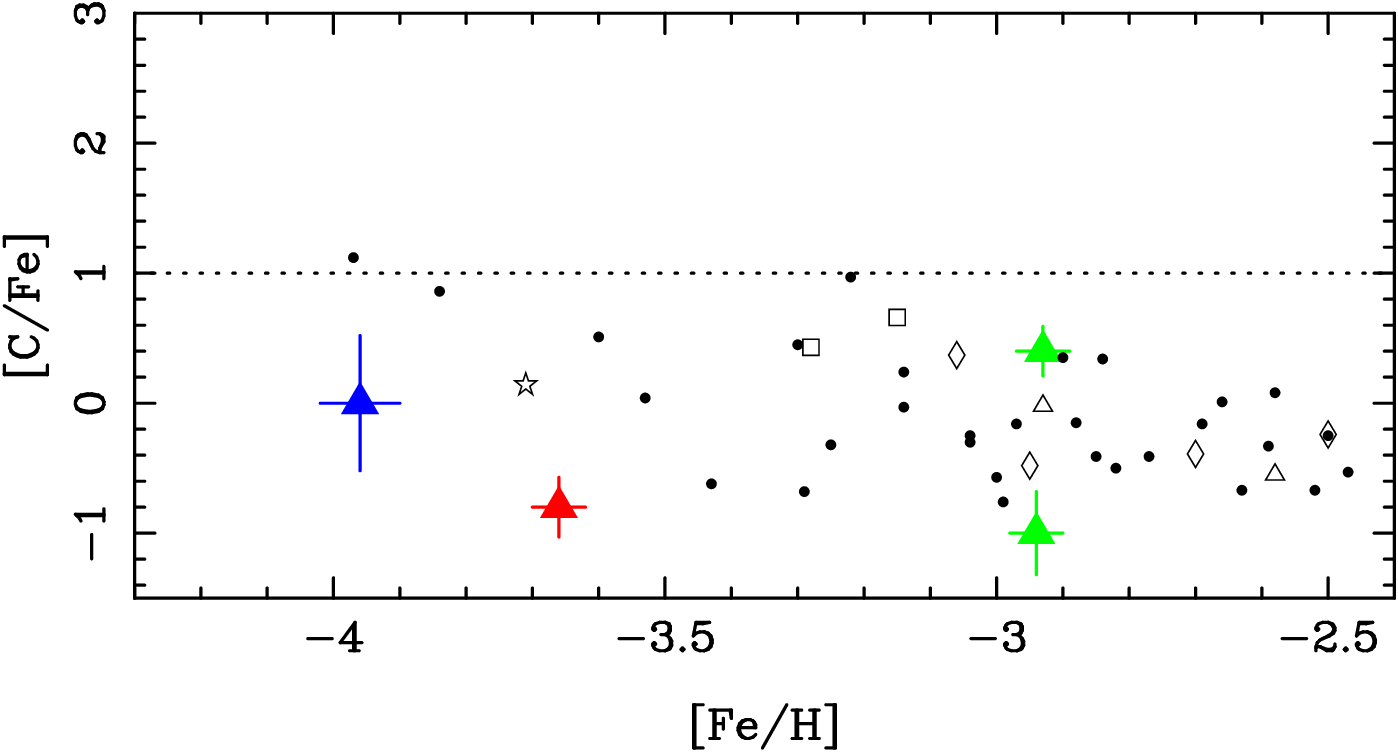}
\includegraphics[angle=-90, width=\hsize]{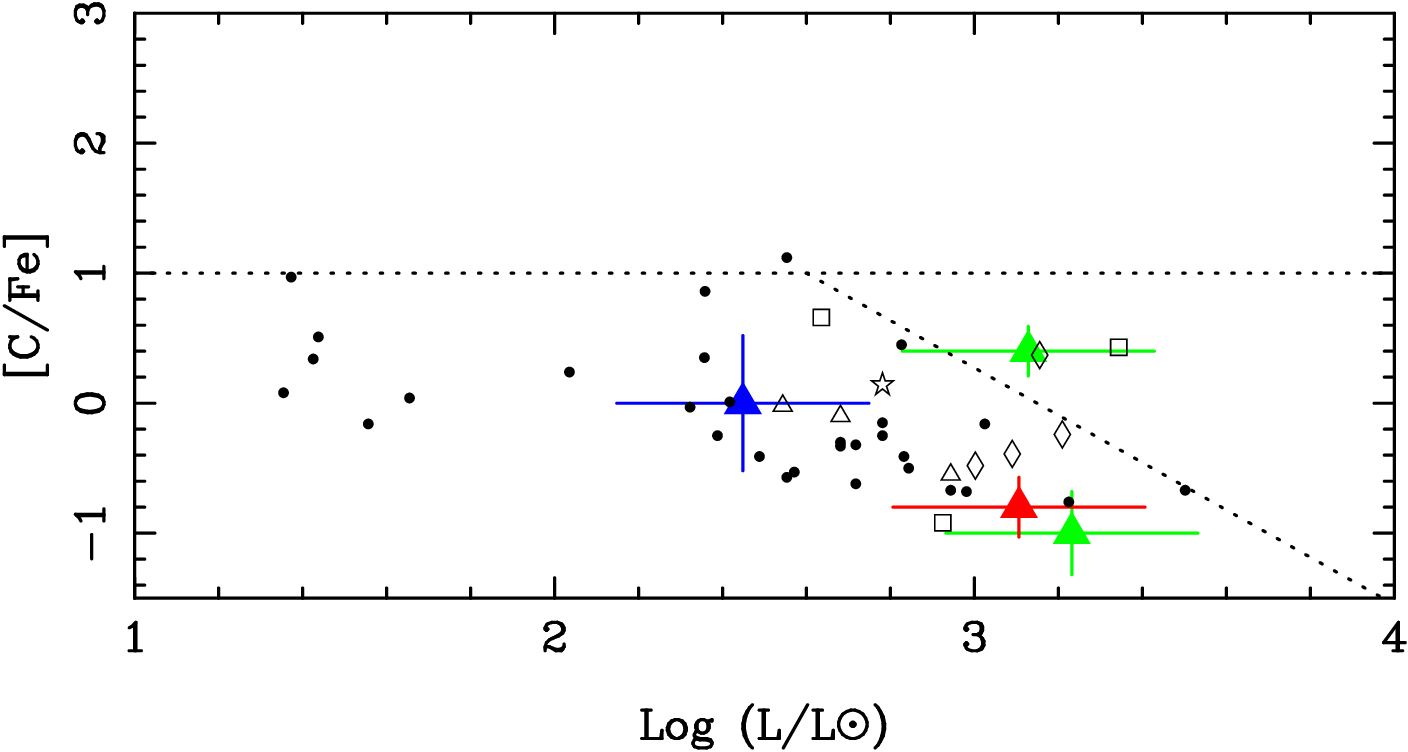}
\caption{DART dSph targets are in colored triangles : Sextans(Green) Scl(Blue) Fnx (Red). Other symbols are squares (Ursa MajorII), triangles (Coma Berenices), diamonds (Draco), star (Bo\"otes)  \citep{fulbright04, koch08, frebel10, cohen09, norris10}. Only mixed Milky Way halo stars (black dots) are considered \citep{cayrel04, cohen08,honda04, aoki05, aoki07, cohen06, lai08}. The dashed line delineates the regions in which they are observed.
\label{fig:CLFe}}
\end{figure}

Figure \ref{fig:CLFe} shows [C/Fe] as a function of the stellar
metallicity and luminosity for our sample stars, drawing a comparison
with the Milky Way halo, Draco, Bo\"otes I, Ursa Major II and Coma
Berenices EMPS. The [C/Fe] does not depend on metallicity. The
bolometric luminosities are derived from the gravities and effective
temperatures, assuming a mass of $\mathcal{M}= 0.8
\mathcal{M}_{\odot}$ for all stars.  The first clear result is that
none of our stars is C-rich in the classical sense ([C/Fe]$>$1).  All
but one star, Sex24-72, have [C/Fe] $\leq$ 0 as expected for their
tip-RGB luminosities \citep{spite05}.  The onset of the
extra mixing ($\delta\mu$) lies at
$\log{\frac{\mathcal{L}_{\star}}{\mathcal{L}_{\odot}}}= 1.2$ for a 0.8
$\msol$ star. \citet{eggleton08} also show a metallicity dependence of
the luminosity at the onset of the mixing, increasing from
$\log{\frac{\mathcal{L}_{\star}}{\mathcal{L}_{\odot}}}= 1.4$ to 2.4
for metallicity passing from 0.02 (solar) to 0.0001 ($-2.3$).  An
independent evidence confirming that our sample stars have experienced
thermohaline convection comes from the low ratio
$^{12}$C/$^{13}$C=6$^{+2}_{-1}$, that we measure in Sex24-72, the most C-rich
star of our sample, in which this can be done \citep{spite06}.

\citet{aoki07}  argued that given that mixing on the RGB and extra mixing 
at the tip of the RGB lower the surface abundance of carbon in
late-type stars compared to their earlier stage of evolution, the
definition of carbon-enriched EMP stars (CEMPS) should depend on the
stellar luminosity.  Hence, for
$\log{\frac{\mathcal{L}_{\star}}{\mathcal{L}_{\odot}}} > 3 $, a giant
star would be considered CEMPS for [C/Fe]$>0$. Applying this
criterion, one Sextans star of our sample, Sex24-72 with [C/Fe]$=0.4
\pm 0.19$ , is carbon enhanced.

Both \citet{frebel10} and \citet{cohen09} reported similar cases
in Ursa Major II and in Draco. We note that \citet{frebel10} measure
[C/Fe]=$-0.07 \pm 0.15$ for HD122563, which they use as reference, while
\citet{spite05} calculated [C/Fe]=$-0.47 \pm 0.11$ for the same star.
Part of this apparent discrepancy results from differences in the C
and Fe solar abundances, which induces a 0.08 dex shift. Frebel et al
[C/Fe]= $-0.15 \pm 0.15$ in Spite et al.'s scale.  The effective
temperature difference between the two studies (100K) cannot explain
the difference.  Indeed, increasing \citet{frebel10}'s \Teff\ by this
amount would actually increase [C/Fe] by $+$0.07 dex and lead to an
even larger discrepancy.  Taking the same atmospheric parameters as
\citet{spite05} for HD122563, but using spherical models as we do for
the dSph stars, we derived [C/Fe]=$-0.6 \pm 0.13$.  Taking into
account the 0.03dex difference in the C and Fe solar abundances
between the two works, our abundance ratio is $-0.57 \pm 0.13$ in
Spite et al. scale, i.e.,. fully compatible. Moreover, we  checked that this
0.1dex difference was entirely caused by passing from {\it pp} 
to {\it sp} models.

  An immediate consequence of
the checking result on HD122563 is that Sex24-72 is definitely
carbon-rich (since we seem to tend to have low carbon
abundances). Another consequence is to attribute a maximum uncertainty
of $\sim$0.3 dex in the carbon abundances of Frebel et
al. (2010). Shifting all of their [C/Fe] values downward by this
amount does not change their identification of a C-rich star in
Ursa Major II.

The origin and nature of these moderately enhanced carbon stars in
dSphs and ultra faint dwarfs (UFDs) is intriguing, because they are 
not observed in our
Galaxy.  Mass transfer by a companion AGB is an unlikely hypothesis:
we would expect a higher [C/Fe] than is observed at these very low
metallicities, because low metallicity AGBs tend to produce more carbon
than metal-rich ones \citep{cristallo09, bisterzo10, suda10,
karakas10}, and because the efficiency of carbon-depletion is significantly
reduced in these carbon-rich stars
\citep{denissenkov08,stancliffe09}. This is often accompanied by an increase 
in magnesium abundance, while Sex24-72 [Mg/Fe] is not high.  All this
suggests a pristine carbon enrichment.  Another piece of evidence
is the very low [Ba/H] of the stars with [C/Fe] $>$ 0.  A vast
majority of Milky Way carbon-rich stars are also over-abundant in
barium, hence one would expect that the C-rich stars would also be
Ba-rich \citep[e.g.][]{beers05,aoki07}.

Our two stars in Sextans are located very close to
each other in the HR diagram, share the same metallicity,
and for most of the other elements have the same abundance
ratios.  C appears to be a significant exception and
probably requires primordial C inhomogeneities. We
note that this dispersion in [C/Fe] has so far only been
seen in low-luminosity dSphs. Simulations of 
\cite{revaz09} suggest that these small
systems are prone to significant
dispersion in elemental abundances, as a consequence of
their sensitivity to feedback/cooling processes during 
their star formation histories. \cite{meynet06} show that rotating
massive stars can loose a large amount of carbon enhanced
material. When these ejecta are diluted with supernovae ejecta,
the [C/Fe] abundance ratios are very similar to those observed in
CEMPs. This mechanism for the very early 
production of significant amounts of carbon together
with classical abundances for the other elements, 
should be more easy to detect in low-mass dSphs than 
in higher mass systems, owing to poor mixing in the early phases of
star formation.  Clearly larger samples of stars are needed 
to cover the entire mass range of the dSph galaxies and properly
test this hypothesis.

We note that the dispersion in [C/Fe] (1.4 dex) among Sextans
stars with similar luminosities echoes the spread in abundances of
other elements for this galaxy. This does not imply that this
dispersion is the rule in other dSphs, in particular when they
experience very different star formation histories. We note however
that star-to-star variation in carbon is also seen in Draco.  This is
in contrast to the homogeneity seen in other elements, implying a
specific process leading to the spread in carbon.

\subsection{Oxygen}

None of the forbidden lines at 6300\AA\ and 6363\AA\ could be detected,
hence only upper limits could be derived. For the two
Sculptor stars these limits were clearly above [O/Fe] $\gsim -2.0$ and
consequently disregarded.  The upper limits for the other stars can be
found in Table~\ref{tab:abundances}.

\subsection{The even-Z elements}

\subsubsection{Abundances}
\label{sec:MgAbun}
The Mg abundances were generally determined from the lines in the red
part of the spectrum, at 5172\AA , 5183\AA , and 5528\AA . In the case
of Scl07-050, two blue lines at 3829\AA\ and 3832\AA\ were available
as well. These lines are in the region that is affected by continuum
scattering and close to the strong H 3835\AA\ line, whose wings
influence the continuum. Therefore we did spectral 
synthesis for these two blue lines.

For Si, two blue lines at 3905.5\AA\ and 4102.9\AA\ could be used in
principle for all stars. However, for the most C-rich star in our
sample, Sex24-72, synthesis showed that both lines are blended by
molecular bands. The 3905.5\AA\ even vanishes completely within a
strong molecular line and can therefore not be used. Hence, for the
two Sextans stars and Fnx05-42, we only used the 4102.9\AA\ line.  For
Scl07-050, only the line at 3905\AA\ is used, since the
other one is too weak to be detected. We miss the blue part of
Scl07-049 spectrum, therefore we could only get an upper limit for from the
extremely weak red line at 5684.52\AA\ .

The Ca abundance is generally derived from the six lines at 5589\AA ,
5857\AA , 6103\AA , 5122\AA , 6162\AA , and 6439\AA, or from a subset of
these six lines for the stars with lower Ca abundances.  However, for
Scl07-050, none of these six lines was detectable. Therefore its Ca
abundance is based on only one blue line at 4227\AA.  This line is not
considered for the other stars, since it is in the very low S/N
region.  This line is a resonance line and very sensitive to NLTE
effect as reflected by the suspiciously low [Ca/Fe] of Scl07-050,
which does not reflect the abundance ratios of the other
$\alpha$-elements for this star.

The TiI abundances were determined from three to seven lines in the
red part of the spectra for all stars with the exception of Scl07-050,
for which no TiI line could be detected at all. For this star an upper
limit for [TiI/Fe] of +0.35 dex is derived from the line at 4981\AA
. Generally all TiI lines are very weak, with a large fraction of
lines having equivalent widths between 20m\AA\ and 40m\AA.  The TiII
abundances were mainly derived from three to seven red lines. Most of
the blue lines had to be disregarded because of  too low  S/N. The few
blue lines that were not removed gave abundances very close to the
ones determined from the red lines.  Again, Scl07-50 is an exception,
since no red line of TiII could be detected. Hence, its TiII abundance
is based on its nine blue lines.

\begin{figure}
\includegraphics[angle=-90, width=\hsize]{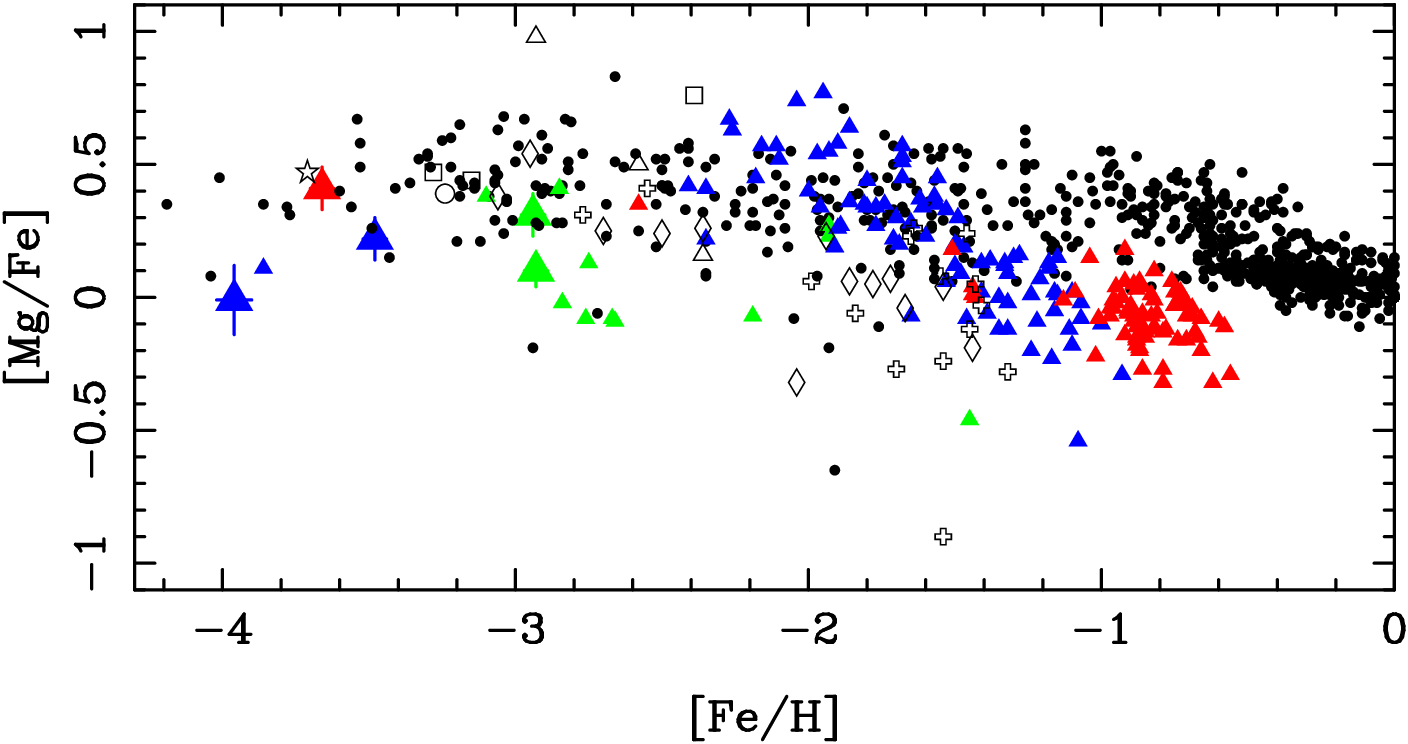}
\caption{DART dSph targets are in colored triangles : Sextans(Green)
  Scl(Blue) Fnx (Red). Large symbols stand for the sample stars of
  this work, small ones stand for earlier works and publications
  \citep[Hill et al., in prep; ][]{letarte10, frebel10a,aoki09,
    shetrone01}. Other symbols are squares (Ursa MajorII,
  \citet{frebel10}), triangles (Coma Berenices, \citet{frebel10}),
  diamonds (Draco, \citet{shetrone01, fulbright04, cohen09}), star
  (Bo\"otes, \citet{norris10}), cross (Carina,
  \citet{shetrone03,koch08}), and circle (Leo IV, \citet{simon10}). All Milky Way halo stars from our
  comparison sample are considered (black dots)
  \citet{cayrel04,honda04, aoki05, aoki07, cohen06,
    cohen08,lai08}. For the purpose of this figure, \citet{Venn04}'s
  compilation was added at [Fe/H]$\ge -2.5$.
\label{fig:MgFe}}
\end{figure}

\begin{figure}[!Hb]
\centering
\includegraphics[angle=-90, width=0.49\textwidth]{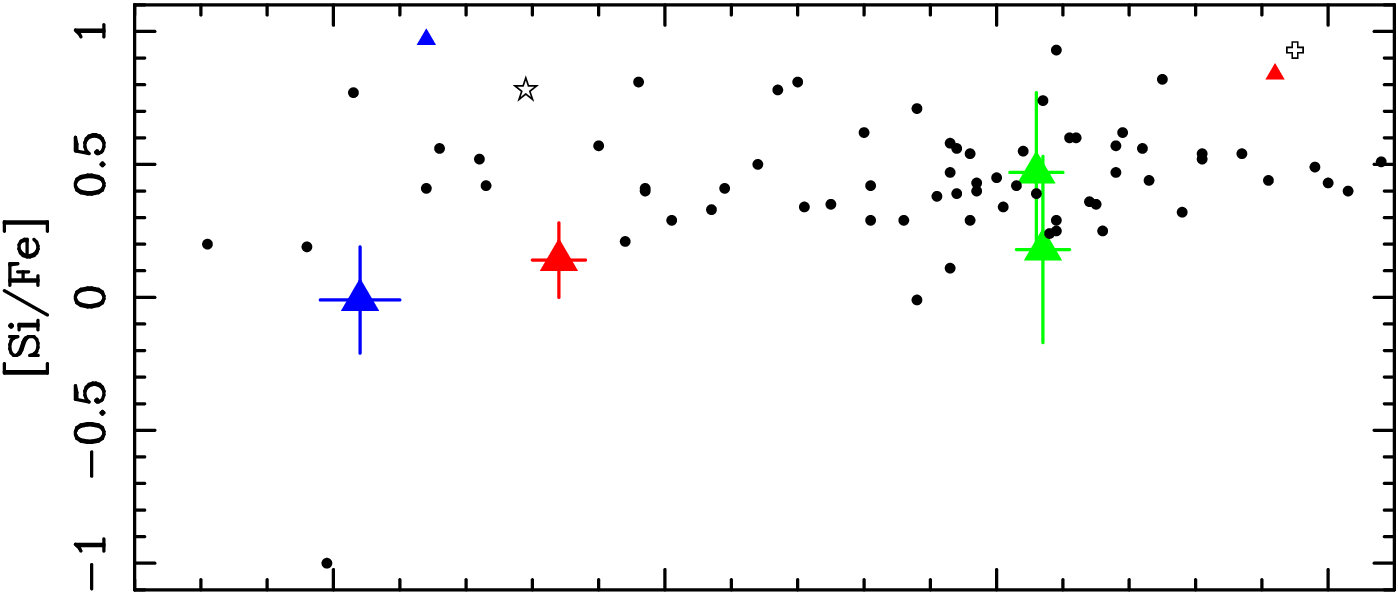}
\includegraphics[angle=-90, width=0.49\textwidth]{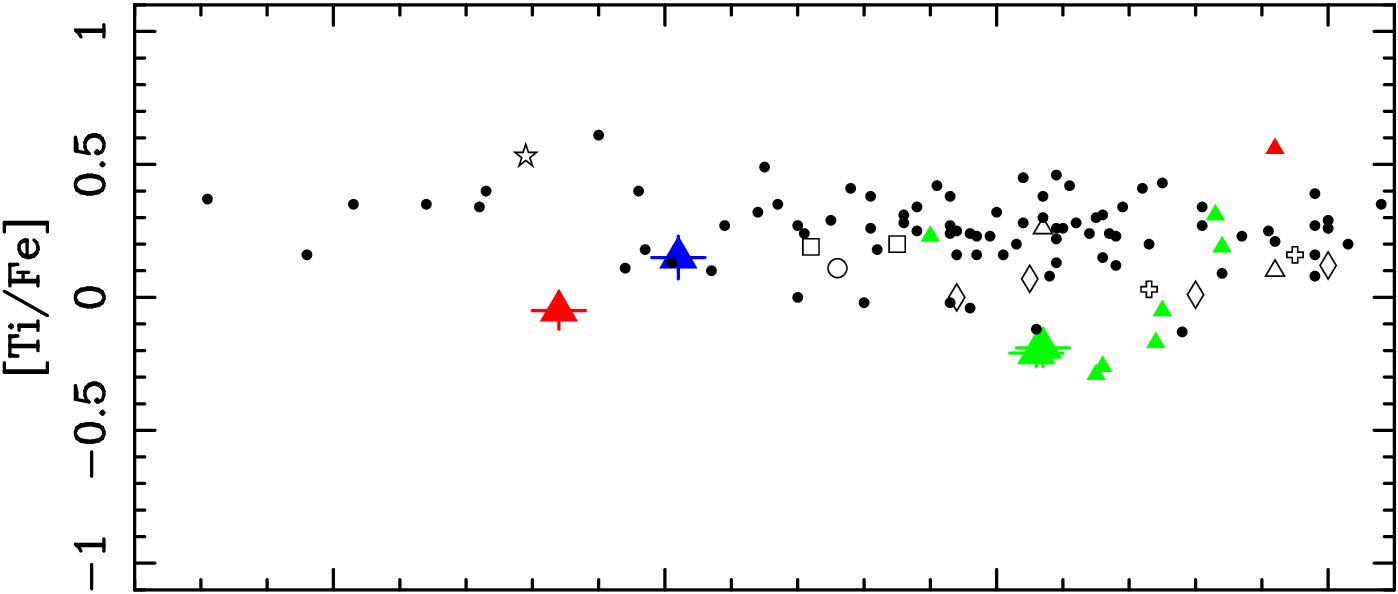}
\includegraphics[angle=-90, width=0.49\textwidth]{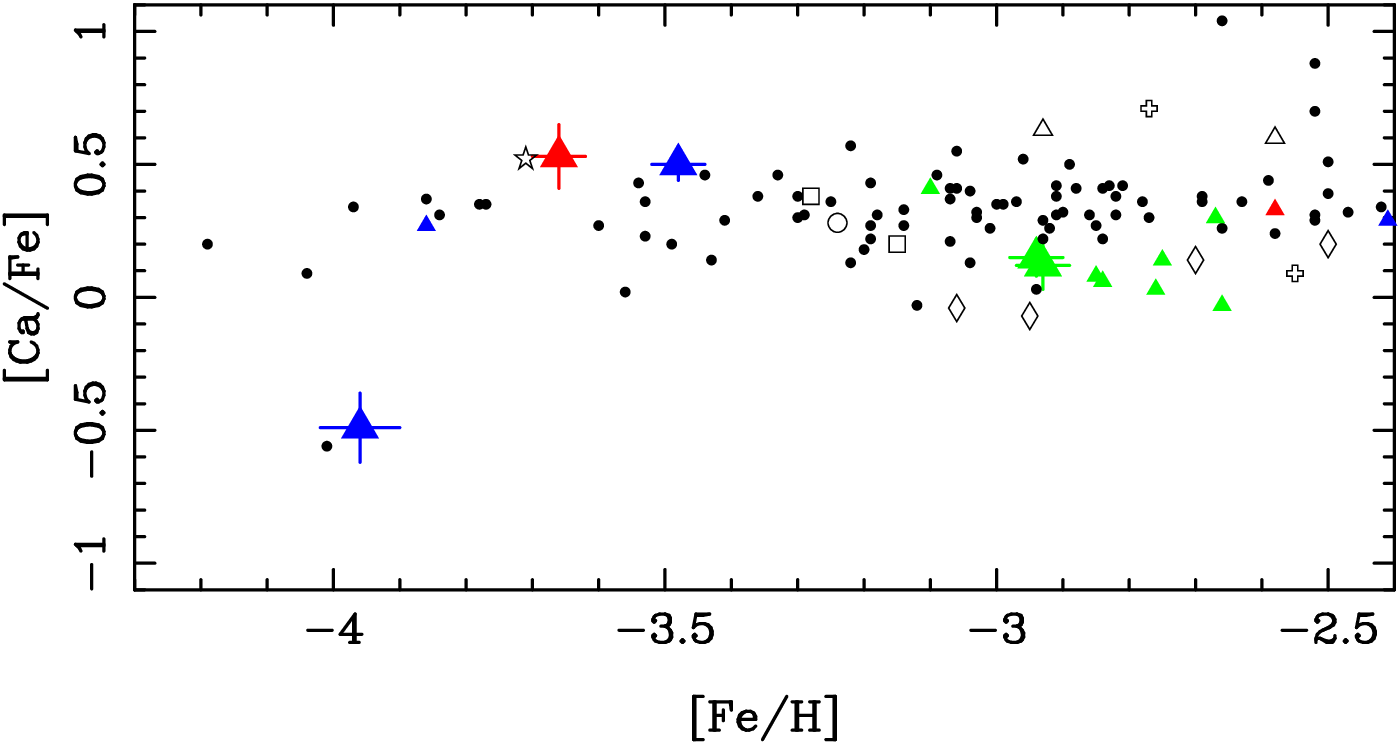}
\caption{Variation of the $\alpha$ elements with [Fe/H]. Symbols
are as in Fig. \ref{fig:MgFe}
\label{fig:alphaFe}}
\end{figure}

\subsubsection{Analysis}

Figure \ref{fig:MgFe} presents [Mg/Fe] as a function of [Fe/H] for all
dSph galaxies for which very metal poor and/or extremely metal poor
stars have been found and abundances measured based on high-resolution
spectroscopy.  It extends the metallicity range covered up to [Fe/H]=0
in order to include the full chemical evolution of the dSph galaxies
and make a clear comparison with the Milky Way. Our choice of
magnesium was driven because it is the most extensively
measured $\alpha$-element.  We included the abundance ratios derived
by \citet{shetrone03} in Carina at [Fe/H]$\ga -2$ and those compiled by
\citet{Venn04} in the Galactic halo for [Fe/H]$\ga -2.5$.

 From the faintest UFDs to the most luminous of the classical dSphs
and to the Milky Way,  Figure \ref{fig:MgFe} samples nearly the full
range of dwarf galaxy masses, allowing common as well as distinct
features to be uncovered.  Below [Fe/H]$\sim -3$, galaxies are
essentially indistinguishable. This means that the very first stages
of star formation in galaxies have universal properties, both in terms
of nucleosynthesis in massive stars and physical conditions triggering
star formation.  This was also foreseen by
\citet{frebel10,frebel10a,simon10}. Conversely, at [Fe/H]$\ga -2$
galaxies have clearly imprinted the peculiarities of their star
formation histories in their abundance ratios: the explosions of Type
Ia supernovae (SNe Ia) occur at lower global enrichment in iron for
lower star formation efficiency
\citep[see also][]{tolstoy09}.  In between these two extremes, at
least one galaxy, Sextans, shows signs of inhomogeneous interstellar
medium, revealed by the dispersion in [Mg/Fe] for stars of similar
[Fe/H]. Although solar [Mg/Fe] are observed in Milky Way halo very
metal-poor stars, their fraction over the total number of observed
stars is significantly larger in Sextans. \citet{aoki09} attributes
the origin of low [Mg/Fe] as likely due to low-mass type II
supernovae, but \citet{revaz09} show that even as low as [Fe/H]$\sim
-2.7$, SNe Ia would have time to explode. Only larger samples will be
able to solve this matter.

While magnesium is produced in hydrostatic nuclear burning phase in
Type II SN progenitors, the synthesis of silicon to calcium comes
partly from the presupernova explosion but is also augmented by an
important contribution from explosive oxygen burning in the shock
\citep{woosley86}. [Si/Fe], [Ti/Fe], [Ca/Fe] are shown in 
Fig. \ref{fig:alphaFe} and hardly display any differences from [Mg/Fe]
in the same metallicity range, once the uncertainties discussed in
Section \ref{sec:MgAbun} are considered.  This confirms the homogeneity in massive
star products of the extremely metal poor stars in all galaxies.

\subsection{The odd-Z elements}

\begin{figure}[!Hb]
\centering
\includegraphics[angle=-90, width=0.49\textwidth]{NaFe.ps}
\includegraphics[angle=-90, width=0.49\textwidth]{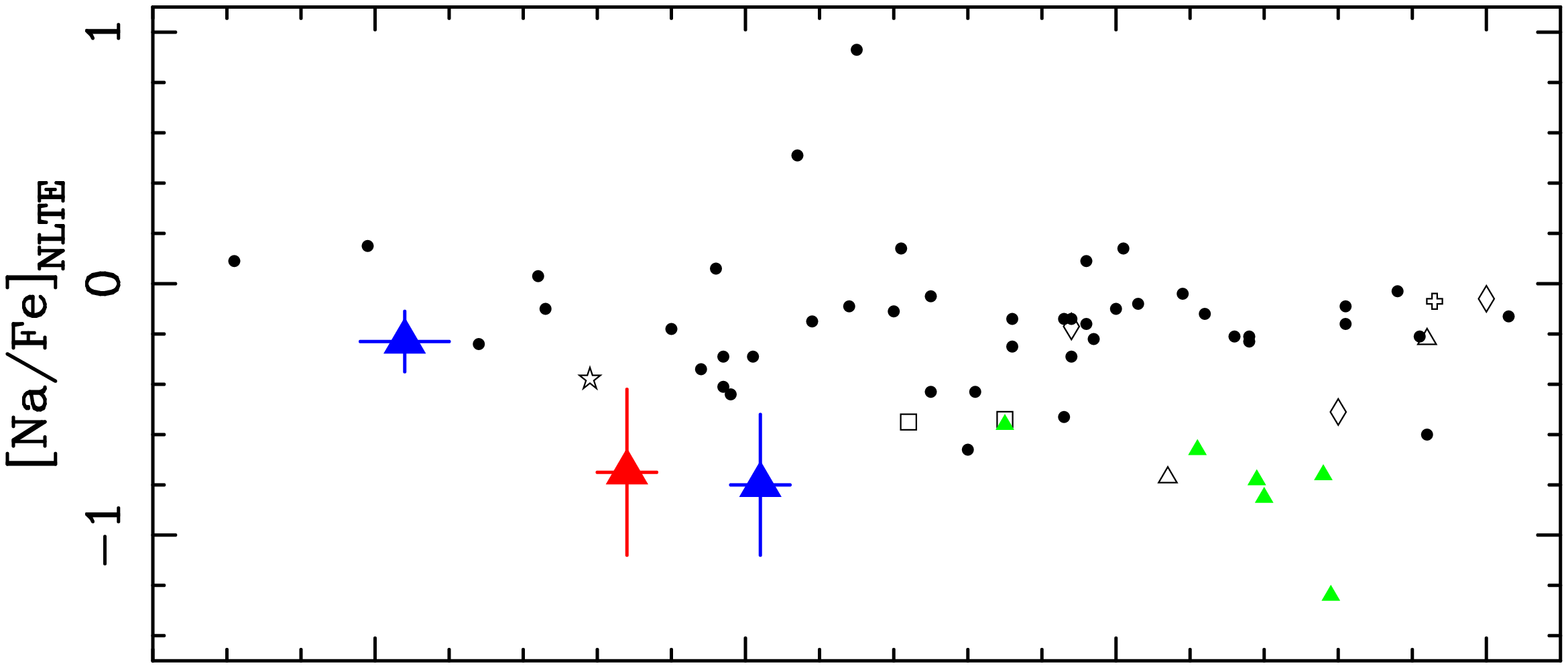}
\includegraphics[angle=-90, width=0.49\textwidth]{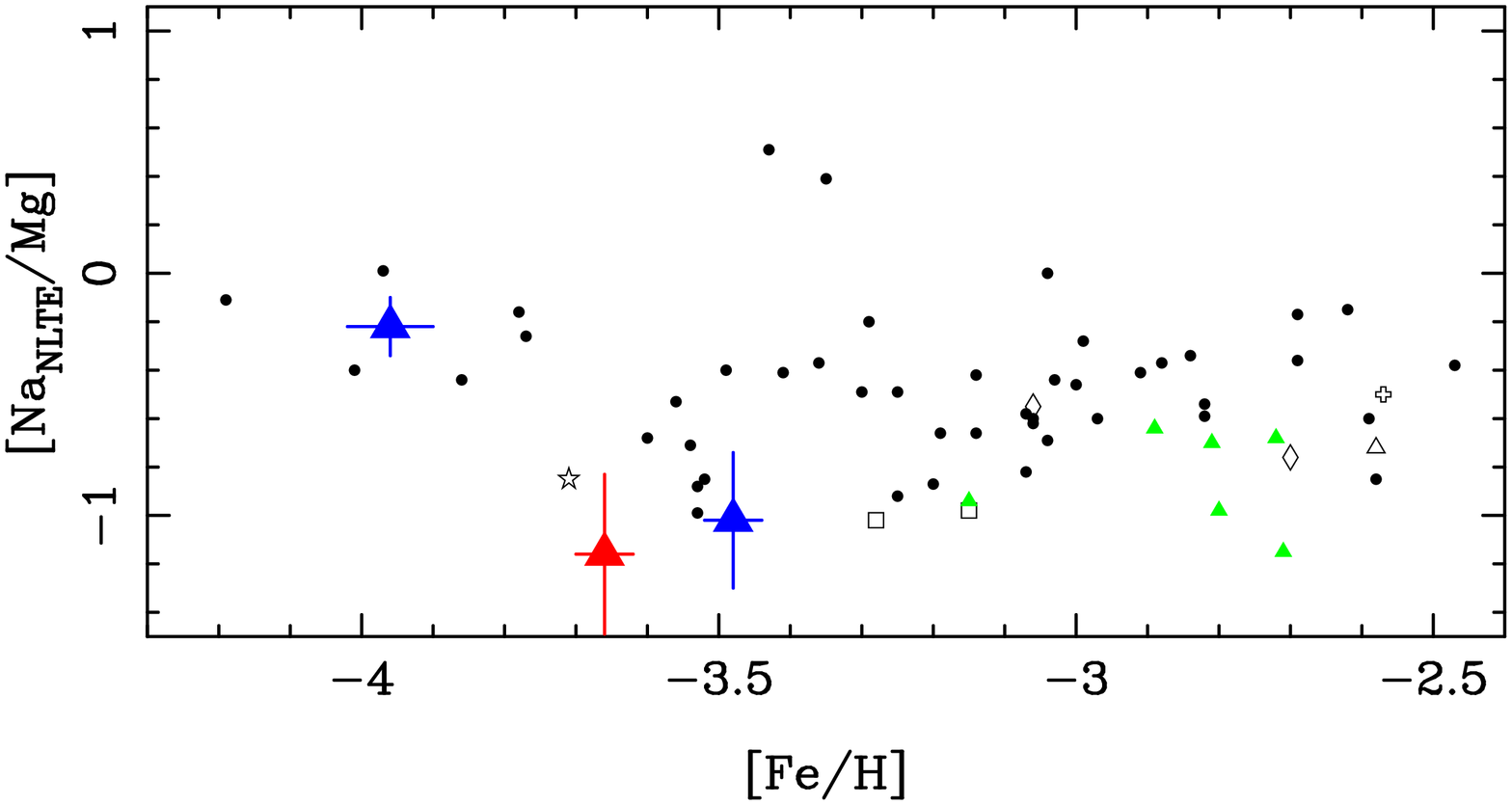}
\caption{ {\it Upper panel:} The [Na/Fe] versus [Fe/H] relation in LTE calculations.
{\it Middle panel:} The relation between sodium, corrected for NLTE
effects, and iron abundances. {\it Lower panel:} The relation between
sodium corrected for NLTE effects, and magnesium as function of
[Fe/H]. In the three panels, the symbols are as in Fig.\ref{fig:MgFe}.
\label{fig:NaNLTE}}
\end{figure}

\begin{figure}[!Hb]
\centering
\includegraphics[angle=-90, width=0.49\textwidth]{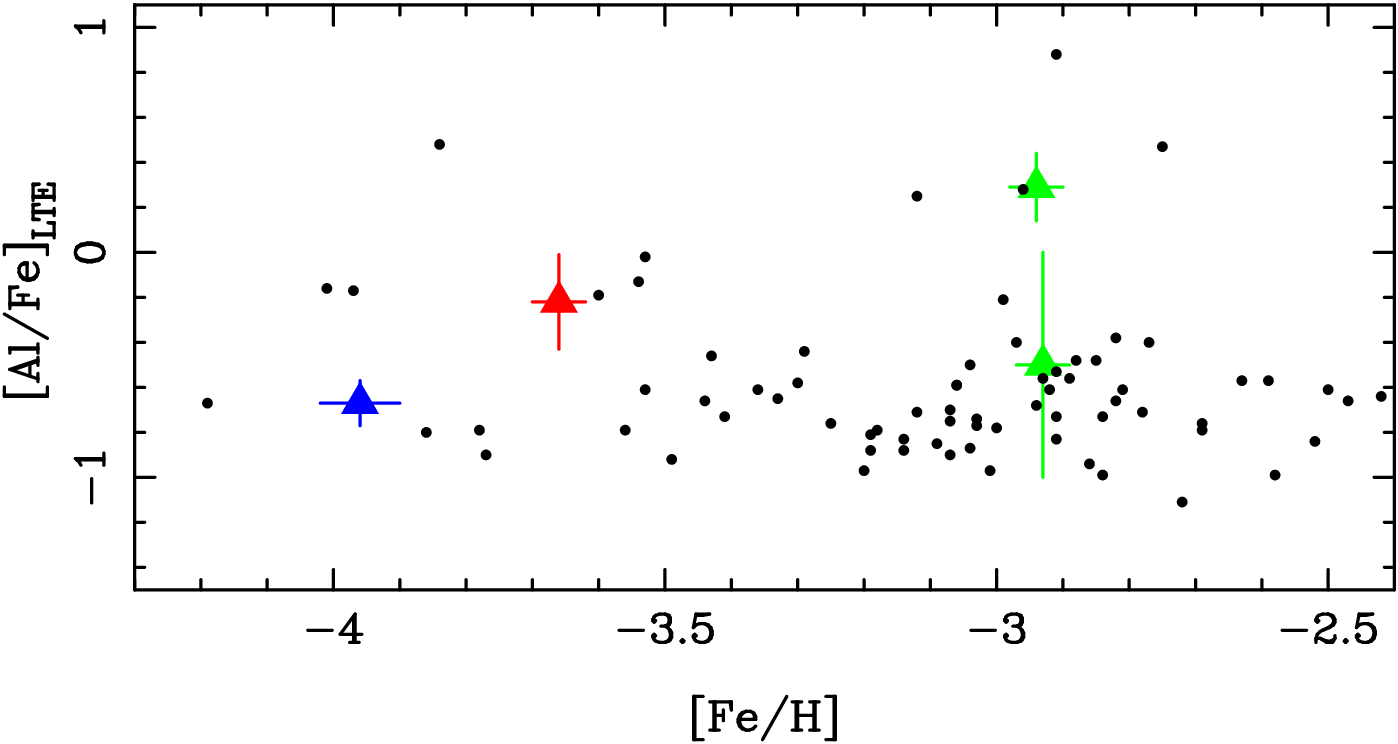}
\caption{Relation between aluminum and iron. Symbols as in Fig.\ref{fig:MgFe}.
\label{fig:AlFe}}
\end{figure}

\subsubsection{Abundances}

Both sodium and aluminum abundances are based on strong resonance
lines. The D lines at at 5890\AA\ and 5896~\AA\ were used for Na, the
two lines at 3944\AA\ and 3961\AA\ for Al.  The equivalent widths of
Table~\ref{tab:linelist} were used for all stars, except Sex24-72 (see
Section~\ref{sec:noeqw}).  
The scandium abundance of Scl07-50
was derived from the two lines at 4247\AA\ and 4314\AA, since none of
the other lines of Table~\ref{tab:linelist} could be detected for this
star. For the other stars the two lines at 5031\AA\ and 5527\AA\ were
used, except for Fnx05-42, where only the line at 5031~\AA\ could be
seen.  We also used the lines from the blue part with lower S/N for
confirmation.  After hyperfine structure and scattering corrections
the scandium abundances derived from the blue lines and the 5031~\AA\
line respectively, agreed within the errors.

\subsubsection{NLTE Effects}

Both Na lines used here are resonance lines, they are consequently very
sensitive to NLTE effects. Following \citet{baumueller98},
\citet{cayrel04} applied a uniform correction of $-$0.5~dex to all 
stars of their sample. But this did not account for a possible 
dependence of NLTE corrections on atmospheric parameters.
\citet{andrievsky07} calculated NLTE corrections for a grid of
parameters ranging from $-$2.5 to $-$4.0 in metallicity, 4500~K to
6350~K in effective temperature and 0.8 to 4.1 in logg (see their
Table 2). They applied their results to the samples of giant and turnoff
stars of \citet{cayrel04} and \citet{bonifacio06}.  We employed their grid
using linear interpolations in temperatures and metallicities to
estimate the NLTE corrections both for our sample stars and our
comparison sample. We calculated the errors on these interpolated
corrections as the difference between our values and the ones
provided by the closest points in the grid.

The NLTE corrections for a given set of atmospheric parameters also
depend on the line strength. \citet{andrievsky07} values are therefore
only accurate for the given reference equivalent widths. For
measured equivalent widths that are significantly different from the reference
values, we modified our NLTE corrections according to Fig. 3 of
\citet{andrievsky07}. The Na equivalent widths of Sex11-04 and Sex24-72
are more than 50~m\AA\ stronger than the highest value covered by 
\citet{andrievsky07}, leading to widely uncertain corrections. 
Therefore we did not consider them. For the same reason, some
of the stars in our comparison sample had to be excluded.

Figure~\ref{fig:NaNLTE} displays [Na/Fe]$_{NLTE}$ and [Na$_{NLTE}$/Mg]
as functions of metallicity. The error bars correspond to the
quadratic sum of the errors given in Table~\ref{tab:abundances} and
the errors in the NLTE corrections mentioned above.

Just as in the analysis of sodium, only the resonance lines of aluminum can be
detected at low metallicities, and they are affected by NLTE effects.
\citet{andrievsky08} calculated NLTE corrections for Al in giants,
but for temperatures above 4700~K, implying an extrapolation by up
to almost 500~K for some of our stars. Furthermore while they
mentioned the dependence of the NLTE corrections on line strength,
unfortunately they did not give explicit values.  Because
NLTE corrections of our Al abundances are too uncertain, we kept the
LTE values in Figure~\ref{fig:AlFe}.

\subsubsection{Analysis}

The production of aluminum and sodium is assumed to be bimodal. At
low metallicities, in the absence of heavy elements, aluminum is the
product of Ne burning in massive stars during their RGB phase, with a
very small contribution of C burning, whereas sodium is mainly
produced through C burning. These primary processes are only based on
the $^{12}$C production during the He-burning phase \citep{woosley95}.
It is therefore expected that both [Al/Fe] and [Na/Fe] are constant
with [Fe/H]. In metal-rich environments however, with a
sufficient amount of neutron rich elements that might act as neutron
donors, the yields of Al and Na depend on the neutron excess and their
abundances are expected to be metallicity dependent \citep{woosley95,
gehren06}. \citet{andrievsky07, andrievsky08} found constant [Na/Fe] and [Al/Fe]
values of $-$0.21$\pm$0.13 and $-$0.06$\pm$0.10,  respectively.

In the upper panel of Fig.~\ref{fig:NaNLTE},
we present the results of our original LTE calculations for Na.
In the middle panel of Fig.~\ref{fig:NaNLTE} we show the NLTE values
of [Na/Fe] for our EMP stars and the comparison sample.   The NLTE
corrections together with their uncertainties are reported in Table \ref{tab:NLTEcorr}.
Within the errors, dSph and Galactic halo stars agree quite well,
showing no slope with [Fe/H], as expected at these low metallicities.
The largest discrepancies correspond to the largest uncertainties in
NLTE corrections and for the Sextans stars, they
have low and dispersed abundance ratios, as was also noticed for Mg. 
Figure~\ref{fig:NaNLTE} shows that the
dispersion in sodium abundances perfectly reflects the dispersion in
magnesium in Sextans and confirms that the site of production of Na is
the same as that of Mg, in dSphs just as in our Galaxy halo.

The LTE abundances of aluminum agree with the halo for three of our
stars. Only Sex11-04 seems above the bulk of the halo distribution.
This is spurious however, given its atmospheric parameters, i.e.
[Fe/H]~$\sim -$3 and very low temperature, the non-LTE corrections of
this star are expected to be much smaller than for the other stars,
which should be substantially moved upward in [Al/Fe] after NLTE
corrections \citep[see Fig.~2 of][]{andrievsky08}.

\begin{table}[ht!]
\begin{center}
\caption[]{NLTE corrections to the sodium D lines according to \citet{andrievsky07}.}
\label{tab:NLTEcorr}
\begin{tabular}{lcc}
\hline
Star      &      D1 (5889.97\AA)      &      D2 (5895.92\AA) \\
\hline
&{\bf  MW Halo}&\\
BD+23-3130    &  $-$0.56\plm0.03 &  $-$0.47\plm0.03\\
BS16080-054   &  $-$0.65\plm0.2  &  $-$0.60\plm0.2 \\
BS16084-160   &  $-$0.55\plm0.14 &  $-$0.33\plm0.14\\
BS16550-087   &  $-$0.55\plm0.21 &  $-$0.30\plm0.09\\
BS16928-053   &  $-$0.60\plm0.20 &  $-$0.53\plm0.20\\
CS30312-059   &  $-$0.53\plm0.18 &  $-$0.32\plm0.12\\
HE0132-2429   &  $-$0.13\plm0.07 &  $-$0.11\plm0.06\\
HE1347-1025   &  $-$0.13\plm0.07 &  $-$0.11\plm0.06\\
HE1356-0622   &  $-$0.56\plm0.21 &  $-$0.45\plm0.21\\
HE1424-0241   &  $-$0.06\plm0.02 &  $-$0.06\plm0.02\\
BS16467-062   &  $-$0.11\plm0.07 &  $-$0.08\plm0.04\\
BS16929-005   &  $-$0.31\plm0.10 &  $-$0.22\plm0.08\\
& {\bf Comparison dwarf galaxies}  &\\
UMa II-S1     &  $-$0.49\plm0.11 &  $-$0.36\plm0.11\\
UMa II-S2     &  $-$0.50\plm0.17 &  $-$0.37\plm0.17\\
ComBer-S2     &  $-$0.52\plm0.13 &  $-$0.49\plm0.13\\
ComBer-S3     &  $-$0.30\plm0.25 &  $-$0.30\plm0.26\\
Boo-1137      &  $-$0.27\plm0.11 &  $-$0.16\plm0.08  \\
Draco-3157    &  $-$0.10\plm0.40 &  $-$0.15\plm0.40    \\
Draco-19219   &  $-$0.75\plm0.30 &  $-$0.67\plm0.15    \\
Draco-19629   &  $-$0.20\plm0.36 &  $-$0.34\plm0.15    \\
S10-14        &  $-$0.52\plm0.10 &  $-$0.30\plm0.20 \\
S11-13        &  $-$0.55\plm0.11 &  $-$0.53\plm0.06   \\
S11-37        &  $-$0.58\plm0.12 &  $-$0.52\plm0.07    \\
S12-28        &  $-$0.56\plm0.09 &  $-$0.53\plm0.09\\
S14-98        &  $-$0.61\plm0.09 &  $-$0.60\plm0.13  \\
S15-19        &  $-$0.68\plm0.11 &  $-$0.59\plm0.14 \\
& {\bf This work} &\\
Scl07-50      &  $-$0.21\plm0.08 &  $-$0.13\plm0.05\\
Fnx05-42      &  $-$0.80\plm0.32 &  $-$0.60\plm0.32\\
Scl07-49      &  $-$0.65\plm0.26 &  $-$0.50\plm0.26\\
\hline
\end{tabular}
\end{center}
\end{table}

\subsection{The iron peak elements}

\begin{figure}[!Hb]
\centering
\includegraphics[angle=-90, width=0.49\textwidth]{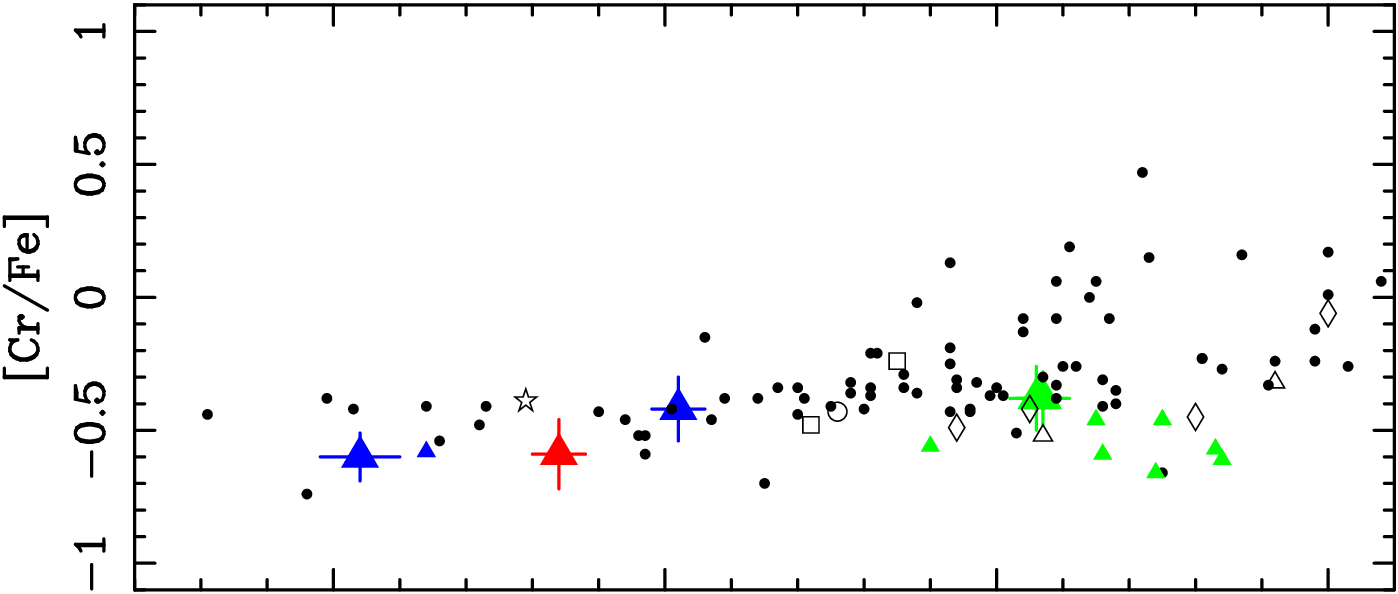}
\includegraphics[angle=-90, width=0.49\textwidth]{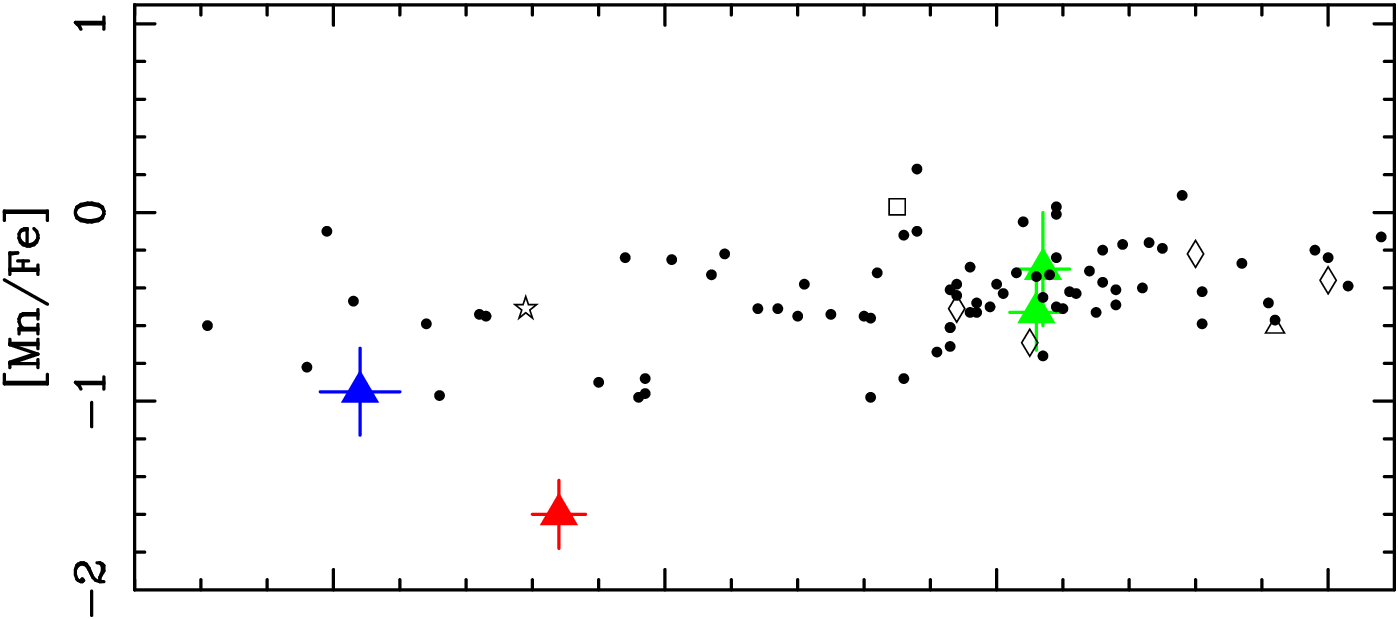}
\includegraphics[angle=-90, width=0.49\textwidth]{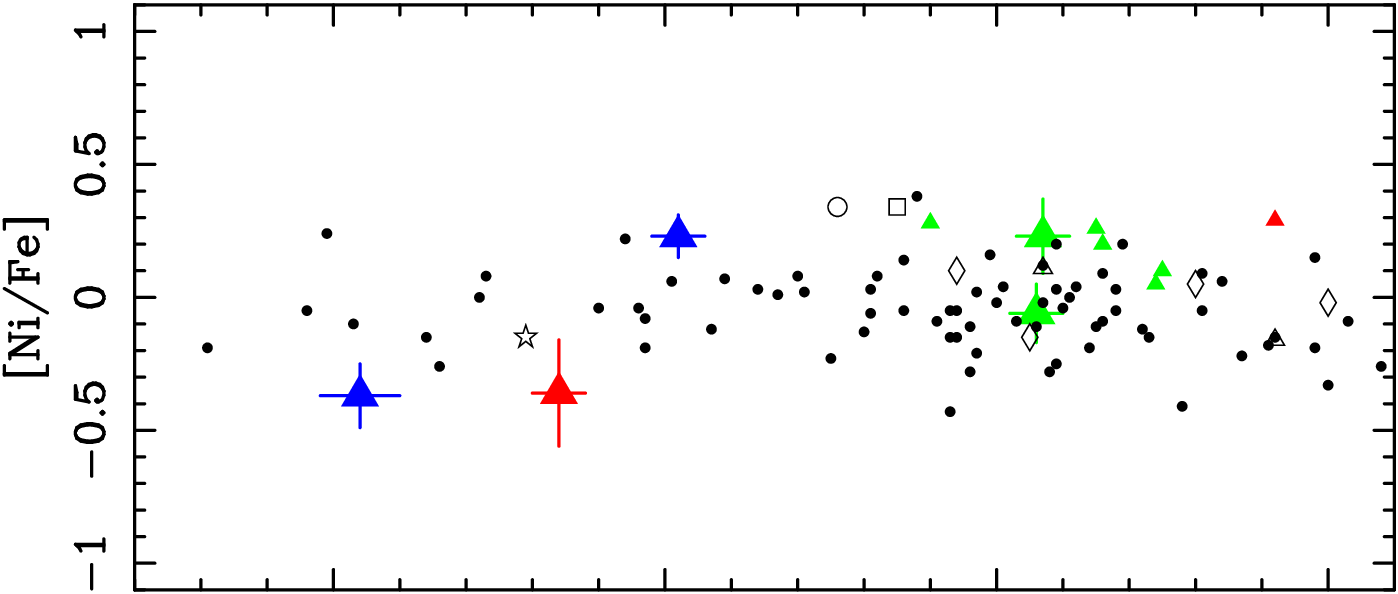}
\includegraphics[angle=-90, width=0.49\textwidth]{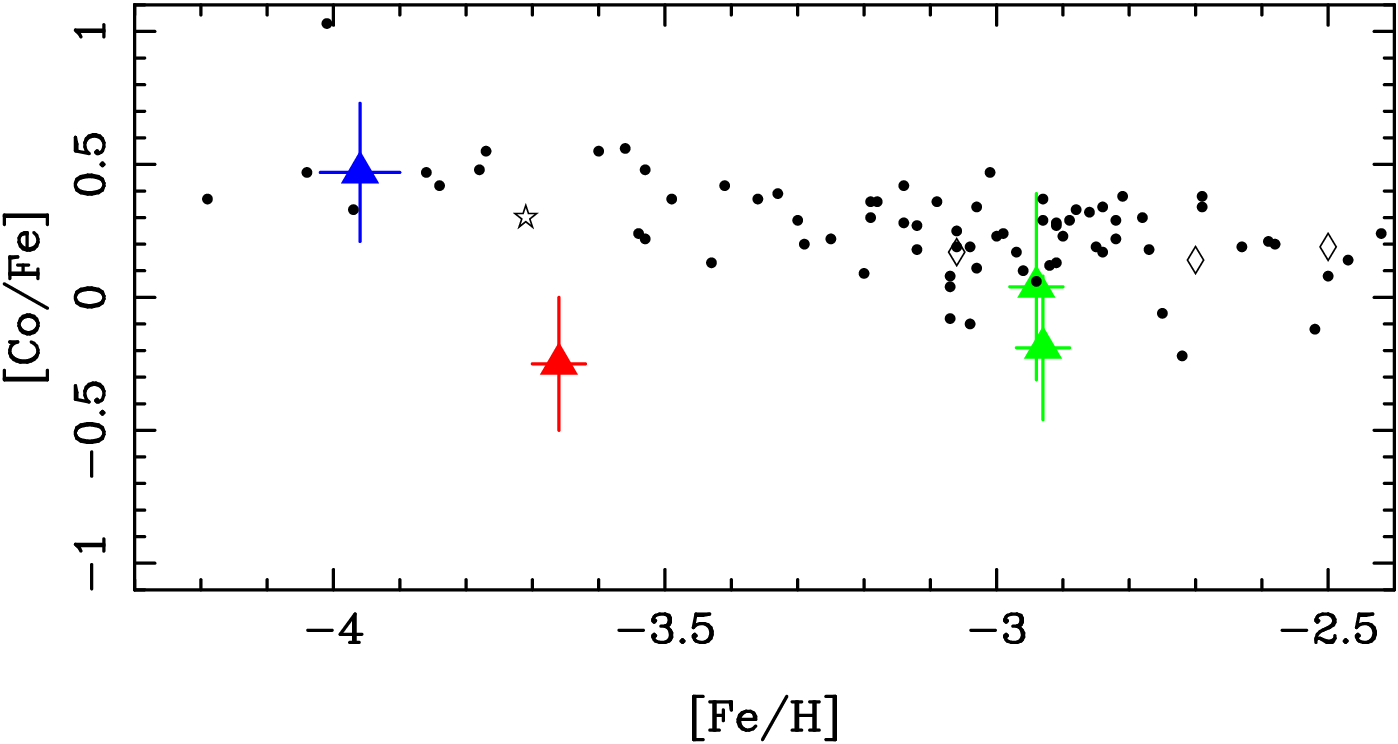}
\caption{Iron peak elements. Symbols as in Fig.\ref{fig:MgFe}.
\label{fig:CrMnNiCo}}
\end{figure}
\subsubsection{Abundances}

The Cr abundance of Scl07-050 was determined from three lines at 4254\AA,
4275\AA, and 4290\AA.  In the other stars, we used the line at
5208\AA\ from the red part of the spectrum and, for the two Sextans
stars, two additional lines at 5346\AA\ and 5410\AA. None of these
lines could be detected in Scl07-50. The abundance of the
5208\AA\ line was derived from spectral synthesis, in order to correct
for a blending line. 

The manganese abundance of the two Sextans stars was determined from
the line at 4824\AA, which was not detected in the other stars. In
Fnx05-42 we used two resonance lines of the triplet at
$\sim$4030\AA.  In Scl07-50, we could use the third line as well. We
did not use the resonance lines for the Sextans stars, since the
uncertainties of derived abundances were too high. 

The Co abundances were determined, when possible, from the three lines at
3995~\AA, 4119~\AA, and 4121~\AA.  

The nickel abundance of Scl07-50 was derived from two lines at
3807~\AA\ and 3858~\AA. These lines yielded very large errors
($>$0.5~dex) for all the other stars and were therefore not
considered. An exception was the line at 3858~\AA\ of Fnx05-042, which
had somewhat smaller errors ($\pm$0.2-0.3~dex).  The line at 5477~\AA\
was detected in all stars, except Scl07-50. For Fnx05-42 the
abundances of the blue and the red line agreed very well.  

\subsubsection{NLTE Effects}
\label{sec:FepeakNLTE}

Several recent publications showed clear evidence for the presence of
NLTE effects in the abundances of the iron peak elements. The most
important ones were an offset in Mn abundance when derived from the
resonance triplet at 4030\AA, \citep{cayrel04, lai08}, a trend in
[CrI/Fe] with metallicity and effective temperature \citep{lai08}, and
a deviation of Cr from ionization balance \citep{sobeck07}, as well as a difference of
[CrI/Fe] between turnoff  and giant stars \citep{bonifacio09}.

On theoretical grounds, \citet{bergemann08} and \citet{bergemann10}
calculated NLTE corrections for manganese and cobalt for a grid of
atmospheric parameters and applied their results to 17 stars with
metallicities between solar and [Fe/H]~=~$-$3.1. 
Their model atmospheres have higher
temperatures and gravities than the stars from our sample, however
the main parameter that controls the magnitude of NLTE effects is
the metallicity, in the sense that NLTE corrections increase with
decreasing [Fe/H]. Both [Co/Fe]$_{NLTE}$ and [Mn/Fe]$_{NLTE}$ are always 
higher compared to their LTE values. The differences can be up to
0.9~dex for cobalt and $\sim$0.5~dex for manganese. 

Whereas the LTE abundances show a  downturn of 
[Mn/Fe] toward lower metallicities, the NLTE abundances are flat with 
$\langle$[Mn/Fe]$\rangle$~$\sim$~$-$0.1 over the whole metallicity
range. Unfortunately, no calculations of NLTE effects
on [Cr/Fe] exist so far.

To summarize the above discussion, both observations and theory provide 
very strong evidence for the influence of NLTE effects on the 
abundances of Co, Cr, and Mn. These effects seem to be particularly
large at lower temperatures, for giant stars and when resonance lines
of neutral species are used, which are exactly the conditions we meet
with our EMP stars. Therefore it is hardly possible to draw any final
conclusions from our observed abundances, unless exact calculations
of NLTE corrections in the parameter range 4000~K~$\leq$~\Teff~
$\leq$~4800~K, 0.5~$\leq$~logg~$\leq$~2.0, and  $-$2.5~$\leq$~[Fe/H]
~$\leq$~$-$4.0 become available. 

With this in mind, the somewhat lower cobalt and manganese abundances
in Fnx05-042 might be explained considering that it has both very low
metallicity and very low gravity, whereas the other stars from our sample
are either slightly higher in metallicity (Sextans) or in surface 
gravity (Sculptor).

\subsubsection{Analysis}

In the early galaxy evolution, when only massive stars contribute to the
chemical enrichment, Co, Cr, and Mn are produced through
explosive nucleosynthesis during SN~II events. Co is produced in the
complete Si burning shell, Cr is mainly produced in the incomplete Si
burning shell, and by a small amount also in the complete Si burning
region, and Mn is produced exclusively in the incomplete Si burning
shell. Abundances of these elements in the galactic halo show a strong
odd-even effect (odd nuclei have lower abundances than the even
nuclei), which proves to be difficult to be reproduced all at once. 
Various attempts have been made with standard SN~II events
\citep[e.g.,][]{woosley95}, zero-metallicity SN~II events, including pair-instability SN
\citep[e.g.,][]{heger02}, varying the {\it
  mass-cut} (dividing the mass expelled from that falling back onto
  the remnant) \citep[e.g.,][]{nakamura99}, hypernovae (i.e. very
  energetic SNe, with E~$>$~10$^{52}$~erg) or varying the SN explosion
  energy \citep[e.g.,][]{nakamura01,umeda02,umeda05}.

The abundances of the iron peak elements are shown in
Fig.~\ref{fig:CrMnNiCo}. They generally confirm the results obtained
by most of the previous studies of metal-poor stars in the halo:
[Co/Fe] is increasing, whereas [Cr/Fe] and [Mn/Fe] are decreasing
toward lower metallicities. [Ni/Fe] is flat over the whole
metallicity range from $-$2.5 to $-$4.  The dispersion in [Cr/Fe] is
spectacularly small.

Within the limitation due to the NLTE effects, all very and extremely
metal-poor stars in systems as different as UFDs and the Milky Way,
and including the classical dSphs studied here, show similar abundance
ratios among their iron peak elements. This implies that the early
enrichment of the galaxy interstellar medium with iron peak elements
is independent of the environment.

\subsection{The heavy elements}

\subsubsection{Abundances}

The 6142\AA\ and 6497\AA\ barium lines are only detected in the two
Sextans stars and in Fnx05-42.   After
spectral synthesis, taking in account the hyperfine splitting, the
abundances derived from the equivalent widths were corrected by
$-$0.07~dex, $-$0.13~dex, and $-$0.09~dex, for Sex11-04, Sex24-72 and
Fnx05-42, respectively.  We could only estimate an upper limit on
[Ba/H] for Scl07-049. The barium abundance of Scl07-050 is determined
from the blue line at 4554.03\AA , which cannot be detected for the
other stars, since it is located in the gap between the blue and the
red lower CCD chips. Upper limits in [Eu/H] are derived from the
region around the 4129\AA\ line for all stars except Scl07-049. The
yttrium abundance can only be measured in Sex11-04, from its line at
4884\AA . For the other stars, only upper limits are estimated.  
The strontium abundance was determined from the two resonance lines at
4077.7\AA\ and 4215.5\AA. In Sex24-72 the latter was heavily
blended by a CN molecular bands and could not be used. The strongest
of all Sr lines (at 4077.7\AA\ in Fnx05-42) has an equivalent width of
231m\AA\ and is therefore in the domain where equivalent widths might
no longer be appropriate. Thus we did spectral synthesis in order to
derive the Sr abundance from this line. All other lines
are somewhat weaker, therefore we used their equivalent widths. For
the lines with equivalent widths $\sim$200m\AA\ however, we tested
what difference this would make and found that the abundances derived
from synthesis and equivalent widths differed by no more than 0.1~dex,
which is small compared to the errors.
NLTE effects play a negligible role in our results, because the corrections
are below 0.1dex for both Sr and Ba in our range of effective
temperatures \citep{short06,mashonkina08,andrievsky09}. 

\subsubsection{Analysis}

\cite{truran81} first demonstrated the $r$-process origin of barium in
very metal-poor stars. Since then two components have been identified: 
i) the main $r$-process produces the full range of neutron capture
elements; ii) another process called alternatively weak $r$-process
\citep{ishimaru05}, LEPP (light element primary process)
\citep{travaglio04}, or CPR (charged particle reactions) process
\citep{qian07} produces mainly the light ($Z< 56$ ) neutron capture
elements and little or no heavier ones, such as Ba. More
  recently, \citet{farouqi09} have proposed the superpositions of type
  II supernovae winds with different entropies to reproduce the full
  range of r-process elements. \citet{barklem05} and
\citet{francois07} demonstrated the sequential existence of these 
  (at least) two processes in the Milky Way halo stars, by showing
  that the main r-process dominates, once the heavy elements have been
  enriched beyond [Ba/H] $\sim -2.5$. Below this level, another
  process contributes to the enhancements of Sr as well as Y.

In Fig. \ref{fig:BaSrFe} we show the relation between [Ba/Fe] and
[Sr/Fe] as a function of the metallicity. From our initial comparison
sample, we only kept the $r$-process stars ([Ba/Eu]$<$0) in the [Ba/Fe] plot. 
This required available measurements of the Eu abundance \citep{lai08,
  francois07, honda04, aoki05}.  Similarly for the [Fe/H] $\ga -2.$
stars in Fornax \citep{letarte10} and Sculptor (Hill et al., in
prep), we required [Ba/Eu]$<$0. We also inserted the sample of Milky
Way halo stars of \citet{barklem05}, excluding carbon-rich and
[Ba/Eu]$>$0 stars as well.  There is no difference between
metal-poor dwarf and giant stars in Ba abundances \citep{bonifacio09}, thus
we kept both populations here.  We could not distinguish between $s$-
and $r$- process in Sextans, neither  for our present sample,
nor for the UFDs. Draco Ba abundances  of the stars at [Fe/H] $> -2.45$
are pure $r$-process
ones \citep{cohen09}. Finally, we included \citep{shetrone01} Ursa Minor and
\citep{shetrone03} Carina [Ba/Eu]$<$0 stars.

All dSph stars at [Fe/H] $<-3$ in Fig.\ref{fig:BaSrFe} are located on
top of or very close to the trend of [Ba/Fe] versus [Fe/H] defined by
the Galactic halo. Above this metallicity and below [Fe/H]$\sim -2.$,
the least massive dSph galaxies (here represented by stars in Coma
Berenices, Ursa Major, Draco) are preferentially found in the
[Ba/Fe] $<0$ regime. Sextans however, like its sibling higher
mass galaxies Scl and Fnx in the EMP regime, seems to
lie on the galactic halo trend (and dispersion).
At [Fe/H]$>-2.5$, the smallest dSphs keep very low Ba
abundances, while the Galactic halo stars have already reduced their
dispersion and cluster around [Ba/Fe] solar.  Carina, Sextans, Draco
and even Ursa Minor definitely reach [Ba/Fe]$\sim$0 at [Fe/H]$\sim
-2.$, with little dispersion, while fainter galaxies, corresponding to
M$_V$$< -8$ \citep{simon07}, do not.  This means that among the
galaxies that are fainter than Sextans, some will finally increase
their Ba abundance sufficiently to catch up with the solar Galactic
halo level.

Galactic mass is a crucial parameter to sustain the explosions of
supernovae \citep{ferrara00}, but also to shape the homogeneity of the
interstellar medium and the frequency at which gas can cool and form
stars \citep{revaz09}. In low-mass systems gas is expelled easily,
while it is retained in more massive ones, leading to a more complete
chemical evolution and homogeneity. Barium is the first chemical
element, which seems to provide clear evidence for differential early
evolution of the dSphs depending on their mass. Moreover,
Fig.\ref{fig:BaSrFe} provides clues on the site of production of the
main $r$-process, in particular by comparison with the
$\alpha$-elements which are produced in a universal way below [Fe/H]$\sim -2.5$
(Fig. \ref{fig:alphaFe} and \ref{fig:MgFe}).

Models should be generated to
investigate whether low-mass systems could
resist the explosions of a few high-mass supernovae but not multiple
lower mass ones, thereby loosing the
main $r$-process elements produced in O-Ne-Mg core collapse of 8-10
$\msol$ stars \citep{wanajo03}.  Alternatively, it is not
straightforward to reconcile that the neutron winds of core-collapse SNe
II \citep{qian07,farouqi09} would be expelled from the galaxy bodies,
while leaving a normal enrichment in the $\alpha$ elements but a
depletion in $r$-process ones.  The work of \citet{cescutti06}
provides another way to tackle the question of the enrichment in Ba
with time or [Fe/H]. The dispersion in [Ba/Fe] observed at [Fe/H]$\la
-3.5$ or below [Fe/H]$\le -2$ for the faint dSphs is of the order
1dex. This corresponds to the intrinsic spread predicted by
\citet{cescutti06}'s model 1, in which barium is produced by
15-30$\msol$ massive stars. Their stellar yields in barium is
increasing with decreasing stellar masses, by a factor $\sim$1000
between 30$\msol$ and 12$\msol$, but only by a factor 10 between
30$\msol$ and 15$\msol$. At [Fe/H]=$-3.5$ their lightest stellar mass
star dying is $\sim$ 15$\msol$. It is likely the minimum stellar
mass that can enrich the faintest dSphs. For the more massive ones,
the full range of 12-30$\msol$ provides a good fit to the data.

For [Sr/Fe], the distinction between low- and high- mass
galaxies is clearer. Our comparison sample is that of Sect.
\ref{sec:comp} to which we added the measurements of \citet{barklem05}
with no restriction on [Ba/Eu].  Galaxies similar to Draco or
  fainter have a constant [Sr/Fe] with a mean [Sr/Fe]=$-1.23$ and a
  dispersion of 0.29 dex, corresponding to the abundance ratio of the
  Milky Way halo stars at [Fe/H] $\le -3.5$, but up to more than one
  dex below the level of the Milky Way at [Fe/H] $> -2.6$. Conversely,
Scl07-50
and the two Sextans stars in our sample perfectly follow the galactic
halo trend in [Sr/Fe], while Fnx05-42 displays a higher [Sr/Fe] value
[Sr/Fe]$\sim +0.87$, yet still within the 2$\sigma$ dispersion of
\citet{barklem05} and echoing its relatively high barium content.

While different star formation histories lead to different evolution
in the strontium and barium enrichements, it is remarkable that in the
earliest stages of the galaxy evolution, here characterized by [Fe/H]
below $-3.5$ or so, all galaxies, from the UFDs to the massive Milky
Way, produce the same amount of r-process elements.

Figure \ref{fig:SrBa} sheds light on the relation between primary and
secondary $r$-process. \citet{francois07} precisely described that
although Milky Way halo stars show a decrease in [Sr/Fe] and increased
scatter below [Fe/H] of $-3$ (similarly to [Ba/Fe]), a strong
relationship of [Sr/Ba] as a function of [Ba/H] is hidden in this
dispersion. Indeed, below [Ba/H]=$-$3 [Sr/Ba] increases from a solar
ratio reaching up to +1 dex at [Ba/H]=$-$4. This is the strongest
evidence to date for the weak r-process to arise earlier in the Milky
Way evolution than the main r-process. Below this, [Sr/Ba]
scatters down back to solar values again around [Ba/H]=$-$5.  All four
stars for which we could measure Sr and Ba follow the [Sr/Ba]
Galactic trend (Fig. \ref{fig:SrBa}), suggesting a constant and
  universal ratio between the processes delivering heavy and light
  r-process elements. Nevertheless, further investigation is clearly
needed to clarify why some of the ultra faint dwarfs, such as Coma
Berenices and Ursa Major II seem to follow the same relation between
strontium and barium as the Milky Way halo and the classical dSph
stars, while others, such as Bo\"otes or Draco, are clear outsiders,
showing lower [Sr/H] than expected from their barium content.

\begin{figure}
\includegraphics[angle=-90, width=\hsize]{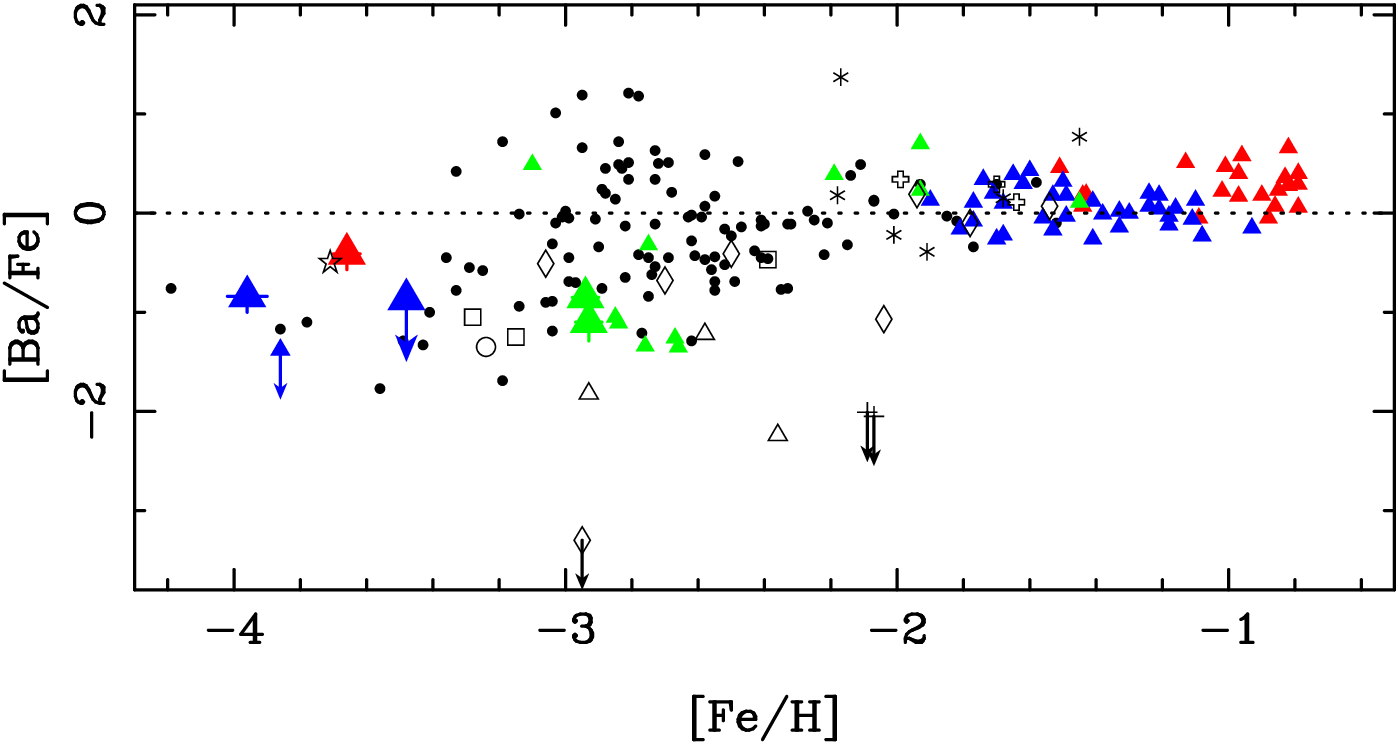}
\includegraphics[angle=-90, width=\hsize]{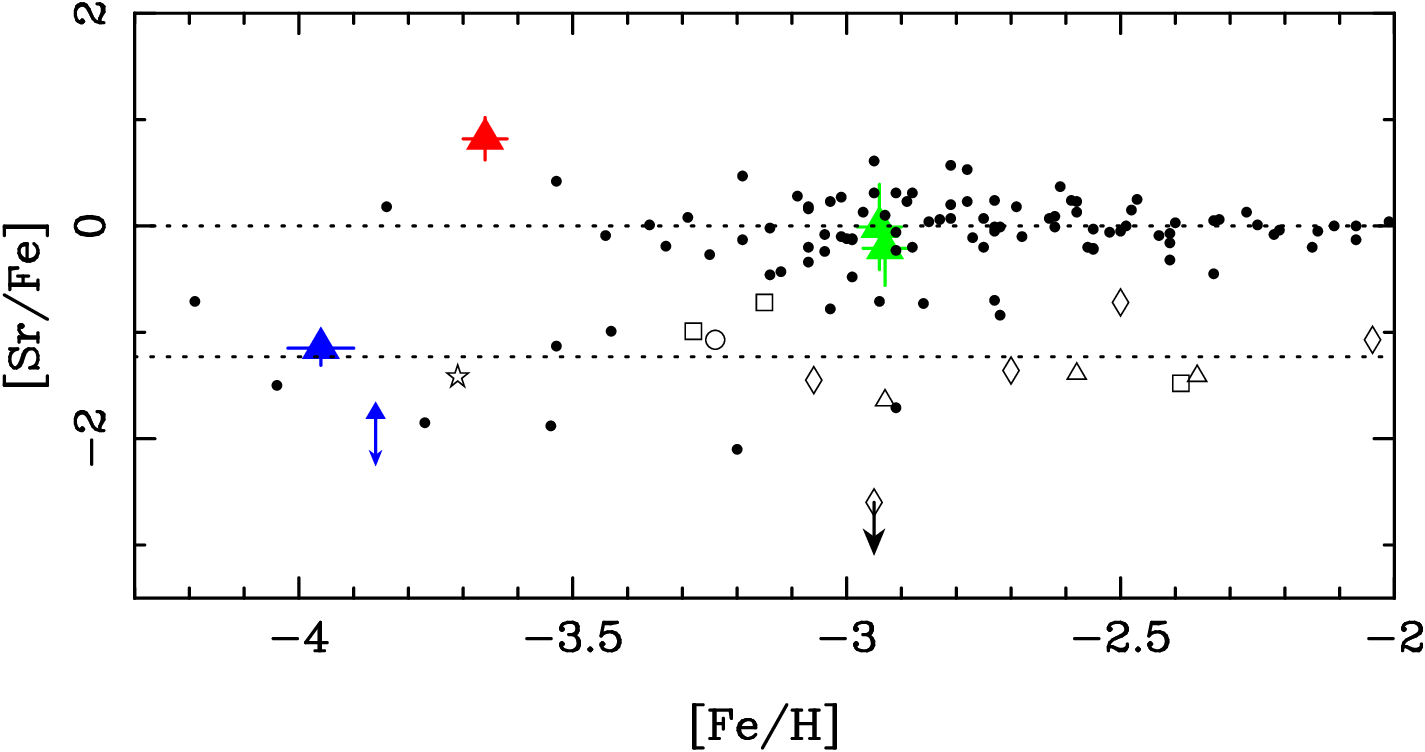}
\caption{Abundance ratios [Ba/Fe] and [Sr/Fe] as a function of
  [Fe/H].  General symbols are as in Fig.\ref{fig:MgFe}.  From our
  initial comparison sample, we only keep the $r$-process only stars
  in the [Ba/Fe] plot. This requires to have measurements of the Eu
  abundance. This was possible with \citet{lai08, francois07, honda04,
    aoki05}. We also inserted the sample of Milky Way halo stars of
  \citet{barklem05}, similarly excluding carbon-rich and [Ba/Eu]$>$0
  ($s$-process enriched stars). As to the UFDs, we use the analysis
  and upper limits in [Ba/Fe] of \citet{koch08b} on Hercules (plus
  sign).  In the case of Scl07-49, we could only estimate an upper
  limit for [Ba/H] in that we indicate with an arrow. Missing the blue
  part of its spectrum we could not measure its abundance in SrII.
  Data from \citet{shetrone01} in Ursa Minor are identified with small
  six-branches stars and. The Carina stars (crosses) at [Fe/H] $>
  -2.5$ are taken from \citet{shetrone03}. The dotted line at
    [Sr/Fe]=$-1.23$ corresponds to the mean abundance ratio of the
    smallest dwarfs (fainter than Draco) and to the level of the
    Galatic halo stars for [Fe/H] $\le -3.7$.
\label{fig:BaSrFe}}
\end{figure}

\begin{figure}
\includegraphics[angle=-90, width=\hsize]{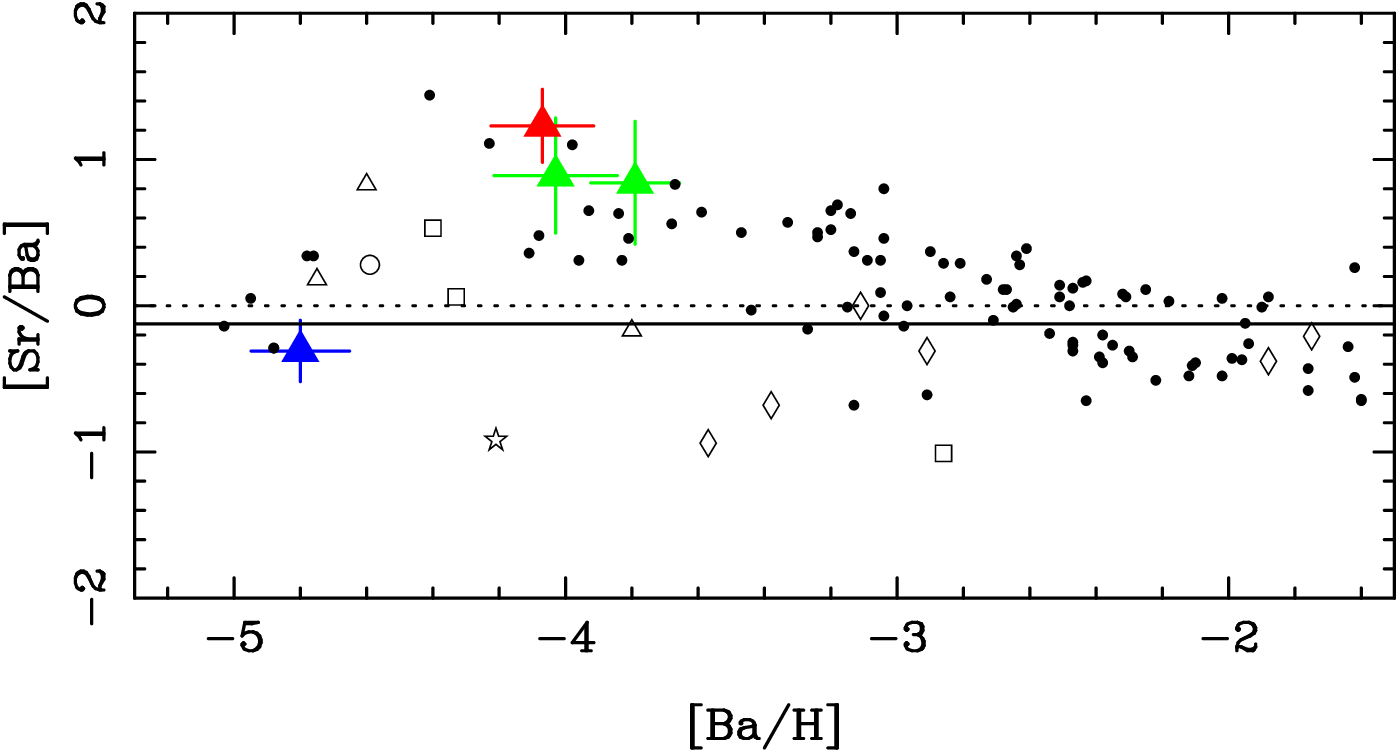}
\caption{[Sr/Ba] versus [Ba/H]  for the comparison sample (see Fig. \ref{fig:BaSrFe} and for the EMP stars of this work.  Symbols as in Fig.\ref{fig:MgFe}. The plain line indicates the solar r-process  [Sr/Ba], as derived by \citet{simmerer04}.
\label{fig:SrBa}}
\end{figure}

\section{Summary and concluding remarks}
\label{sect:conclusion}

We analyzed high-resolution VLT/UVES spectra of five metal-poor
candidates in three different dwarf spheroidal galaxies, Sextans,
Fornax, and Sculptor. The abundances of 15 chemical elements were
derived.

All stars from our sample have metallicities around or below
$-$3. This significantly increases the total number of extremely
metal-poor stars observed in dwarf galaxies, placing their metallicity
floor at a level comparable with the Milky Way Halo. Moreover our
sample contains the most metal poor star ever observed in a dwarf
galaxy, Scl07-50 with [Fe/H]~=~$-$3.96\plm0.06.

The analysis of the chemical composition of our sample stars and the
comparison with the Milky Way halo and other classical and ultra-faint
dwarf galaxies, allowed us unprecedented insights in the earliest
chemical enrichment of these galaxies. Our main results are the following:

\begin{itemize}

\item We showed that all our stars have undergone some degree of
  internal mixing, consistent with their advanced evolutionary
  stages. Taking into consideration mixing-induced carbon depletion, one Sextans
  star in our sample, Sex24-72, must be considered as carbon-enhanced.
  This carbon enhancement most likely reflects the pristine property
  of the ISM rather than mass transfer from an AGB
  companion. Similar moderately carbon-rich high-luminosity stars
  had been previously identified in Draco and Ursa MajorII, while they
  do not exist in the Milky Way halo. Interestingly, we found evidence
  for inhomogeneities in the early ISM in Sextans: the carbon
  abundances of our two Sextans stars are very different, despite their
  similar  metallicities, temperatures, gravities, and abundance
  ratios for all the other elements.

\item We showed that below [Fe/H]~=~$-$3, the abundances of the
  $\alpha$-elements show no major difference between the various
  galaxies in our comparison sample, from small to massive
  systems. Similarly, the abundances of the iron peak elements, 
  aluminum and sodium in dwarf galaxies follow the trends with
  metallicity seen in the Milky Way halo. This suggests that the
  conditions of nucleosynthesis and of early enrichment of the
  interstellar medium are universal for these elements, i.e.,
  independent on the properties of the host galaxy.

\item Below [Fe/H] $\sim -3.5$ all galaxies have similar low barium and
   strontium contents. Above this metallicity, galaxies fainter than
   Draco have [Sr/Fe]$\sim -1.23$ and [Ba/Fe]$\sim -1$, while the
   more massive ones increase their r-process chemical abundances,
   eventually reaching the solar level observed in the Milky
   Way. Despite this variation with galaxy mass, there is some
   evidence for a constant ratio between the "weak" and "main"
   r-processes, although in some cases, such as Bo\"otes, the Sr
   content is lower than expected from the abundance in Ba. This
   issue definitely deserves follow-up.

\end{itemize}

This study definitely strongly suggests the presence of EMP stars 
in all classical dSphs. Their analysis yields unique constraints on the
conditions of onset of star formation as well as on the sites of
nucleosynthesis. Future surveys of EMPS in dSphs will significantly
improve our understanding on how and where the first generations of
stars arise and assemble, firmly establishing trends and level of
homogeneities in the early interstellar media.

\begin{acknowledgements}
This publication makes use of data products from the Two Micron All
Sky Survey, which is a joint project of the University of
Massachusetts and the Infrared Processing and Analysis
Center/California Institute of Technology, funded by the National
Aeronautics and Space Administration and the National Science
Foundation.

We thank the VISTA commissioning team for providing $J$, $Ks$ photometry of
this object which was observed during the commissioning process.

The Hobby-Eberly Telescope (HET) is a joint project of the University
of Texas at Austin, the Pennsylvania State University, Stanford
University, Ludwig-Maximilians-Universitat Munchen, and
Georg-August-Universitat Gottingen. The HET is named in honor of its
principal benefactors, William P. Hobby and Robert E. Eberly.  MS
would like to thank the NSF for support for the HET observations
through AST-0306884.

\end{acknowledgements}

\bibliographystyle{aa}
\bibliography{bibtex}

\begin{thebibliography}{107}
\expandafter\ifx\csname natexlab\endcsname\relax\def\natexlab#1{#1}\fi

\bibitem[{{Alonso} {et~al.}(1999){Alonso}, {Arribas}, \&
  {Mart{\'{\i}}nez-Roger}}]{alonso99}
{Alonso}, A., {Arribas}, S., \& {Mart{\'{\i}}nez-Roger}, C. 1999, \aaps, 140,
  261

\bibitem[{{Alvarez} \& {Plez}(1998)}]{alvarez98}
{Alvarez}, R. \& {Plez}, B. 1998, \aap, 330, 1109

\bibitem[{{Anders} \& {Grevesse}(1989)}]{anders89}
{Anders}, E. \& {Grevesse}, N. 1989, \gca, 53, 197

\bibitem[{{Andrievsky} {et~al.}(2007){Andrievsky}, {Spite}, {Korotin}, {Spite},
  {Bonifacio}, {Cayrel}, {Hill}, \& {Fran{\c c}ois}}]{andrievsky07}
{Andrievsky}, S.~M., {Spite}, M., {Korotin}, S.~A., {et~al.} 2007, \aap, 464,
  1081

\bibitem[{{Andrievsky} {et~al.}(2008){Andrievsky}, {Spite}, {Korotin}, {Spite},
  {Bonifacio}, {Cayrel}, {Hill}, \& {Fran{\c c}ois}}]{andrievsky08}
{Andrievsky}, S.~M., {Spite}, M., {Korotin}, S.~A., {et~al.} 2008, \aap, 481,
  481

\bibitem[{{Andrievsky} {et~al.}(2009){Andrievsky}, {Spite}, {Korotin}, {Spite},
  {Fran{\c c}ois}, {Bonifacio}, {Cayrel}, \& {Hill}}]{andrievsky09}
{Andrievsky}, S.~M., {Spite}, M., {Korotin}, S.~A., {et~al.} 2009, \aap, 494,
  1083

\bibitem[{{Aoki} {et~al.}(2009){Aoki}, {Arimoto}, {Sadakane}, {Tolstoy},
  {Battaglia}, {Jablonka}, {Shetrone}, {Letarte}, {Irwin}, {Hill}, {Francois},
  {Venn}, {Primas}, {Helmi}, {Kaufer}, {Tafelmeyer}, {Szeifert}, \&
  {Babusiaux}}]{aoki09}
{Aoki}, W., {Arimoto}, N., {Sadakane}, K., {et~al.} 2009, \aap, 502, 569

\bibitem[{{Aoki} {et~al.}(2007){Aoki}, {Beers}, {Christlieb}, {Norris}, {Ryan},
  \& {Tsangarides}}]{aoki07}
{Aoki}, W., {Beers}, T.~C., {Christlieb}, N., {et~al.} 2007, \apj, 655, 492

\bibitem[{{Aoki} {et~al.}(2005){Aoki}, {Honda}, {Beers}, {Kajino}, {Ando},
  {Norris}, {Ryan}, {Izumiura}, {Sadakane}, \& {Takada-Hidai}}]{aoki05}
{Aoki}, W., {Honda}, S., {Beers}, T.~C., {et~al.} 2005, \apj, 632, 611

\bibitem[{{Asplund}(2005)}]{asplund05}
{Asplund}, M. 2005, \araa, 43, 481

\bibitem[{{Barklem} {et~al.}(2005){Barklem}, {Christlieb}, {Beers}, {Hill},
  {Bessell}, {Holmberg}, {Marsteller}, {Rossi}, {Zickgraf}, \&
  {Reimers}}]{barklem05}
{Barklem}, P.~S., {Christlieb}, N., {Beers}, T.~C., {et~al.} 2005, \aap, 439,
  129

\bibitem[{{Battaglia} {et~al.}(2008){Battaglia}, {Irwin}, {Tolstoy}, {Hill},
  {Helmi}, {Letarte}, \& {Jablonka}}]{battaglia08}
{Battaglia}, G., {Irwin}, M., {Tolstoy}, E., {et~al.} 2008, \mnras, 383, 183

\bibitem[{{Battaglia} {et~al.}(2006){Battaglia}, {Tolstoy}, {Helmi}, {Irwin},
  {Letarte}, {Jablonka}, {Hill}, {Venn}, {Shetrone}, {Arimoto}, {Primas},
  {Kaufer}, {Francois}, {Szeifert}, {Abel}, \& {Sadakane}}]{battaglia06}
{Battaglia}, G., {Tolstoy}, E., {Helmi}, A., {et~al.} 2006, \aap, 459, 423

\bibitem[{{Baumueller} {et~al.}(1998){Baumueller}, {Butler}, \&
  {Gehren}}]{baumueller98}
{Baumueller}, D., {Butler}, K., \& {Gehren}, T. 1998, \aap, 338, 637

\bibitem[{{Beers} \& {Christlieb}(2005)}]{beers05}
{Beers}, T.~C. \& {Christlieb}, N. 2005, \araa, 43, 531

\bibitem[{{Bergemann} \& {Gehren}(2008)}]{bergemann08}
{Bergemann}, M. \& {Gehren}, T. 2008, \aap, 492, 823

\bibitem[{{Bergemann} {et~al.}(2010){Bergemann}, {Pickering}, \&
  {Gehren}}]{bergemann10}
{Bergemann}, M., {Pickering}, J.~C., \& {Gehren}, T. 2010, \mnras, 401, 1334

\bibitem[{{Bersier} \& {Wood}(2002)}]{bersier02}
{Bersier}, D. \& {Wood}, P.~R. 2002, \aj, 123, 840

\bibitem[{{Bisterzo} {et~al.}(2010){Bisterzo}, {Gallino}, {Straniero},
  {Cristallo}, \& {K{\"a}ppeler}}]{bisterzo10}
{Bisterzo}, S., {Gallino}, R., {Straniero}, O., {Cristallo}, S., \&
  {K{\"a}ppeler}, F. 2010, \mnras, 404, 1529

\bibitem[{{Bonifacio} {et~al.}(2009){Bonifacio}, {Spite}, {Cayrel}, {Hill},
  {Spite}, {Fran{\c c}ois}, {Plez}, {Ludwig}, {Caffau}, {Molaro}, {Depagne},
  {Andersen}, {Barbuy}, {Beers}, {Nordstr{\"o}m}, \& {Primas}}]{bonifacio09}
{Bonifacio}, P., {Spite}, M., {Cayrel}, R., {et~al.} 2009, \aap, 501, 519

\bibitem[{{Bonifacio} {et~al.}(2006){Bonifacio}, {Zaggia}, {Sbordone},
  {Santin}, {Monaco}, {Monai}, {Molaro}, {Marconi}, {Girardi}, {Ferraro}, {di
  Marcantonio}, {Caffau}, \& {Bellazzini}}]{bonifacio06}
{Bonifacio}, P., {Zaggia}, S., {Sbordone}, L., {et~al.} 2006, {Abundances in
  Sagittarius Stars}, ed. {Randich, S.~\& Pasquini, L.}, 232--+

\bibitem[{{Buonanno} {et~al.}(1999){Buonanno}, {Corsi}, {Castellani},
  {Marconi}, {Fusi Pecci}, \& {Zinn}}]{buonanno99}
{Buonanno}, R., {Corsi}, C.~E., {Castellani}, M., {et~al.} 1999, \aj, 118, 1671

\bibitem[{{Carretta} {et~al.}(2002){Carretta}, {Gratton}, {Cohen}, {Beers}, \&
  {Christlieb}}]{carretta02}
{Carretta}, E., {Gratton}, R., {Cohen}, J.~G., {Beers}, T.~C., \& {Christlieb},
  N. 2002, \aj, 124, 481

\bibitem[{{Cayrel}(1988)}]{cayrel88}
{Cayrel}, R. 1988, in IAU Symposium, Vol. 132, The Impact of Very High S/N
  Spectroscopy on Stellar Physics, ed. {G.~Cayrel de Strobel \& M.~Spite},
  345--+

\bibitem[{{Cayrel} {et~al.}(2004){Cayrel}, {Depagne}, {Spite}, {Hill}, {Spite},
  {Fran{\c c}ois}, {Plez}, {Beers}, {Primas}, {Andersen}, {Barbuy},
  {Bonifacio}, {Molaro}, \& {Nordstr{\"o}m}}]{cayrel04}
{Cayrel}, R., {Depagne}, E., {Spite}, M., {et~al.} 2004, \aap, 416, 1117

\bibitem[{{Cayrel} {et~al.}(1991){Cayrel}, {Perrin}, {Barbuy}, \&
  {Buser}}]{cayrel91}
{Cayrel}, R., {Perrin}, M., {Barbuy}, B., \& {Buser}, R. 1991, \aap, 247, 108

\bibitem[{{Cescutti} {et~al.}(2006){Cescutti}, {Fran{\c c}ois}, {Matteucci},
  {Cayrel}, \& {Spite}}]{cescutti06}
{Cescutti}, G., {Fran{\c c}ois}, P., {Matteucci}, F., {Cayrel}, R., \& {Spite},
  M. 2006, \aap, 448, 557

\bibitem[{{Christlieb} {et~al.}(2002){Christlieb}, {Bessell}, {Beers},
  {Gustafsson}, {Korn}, {Barklem}, {Karlsson}, {Mizuno-Wiedner}, \&
  {Rossi}}]{christlieb02}
{Christlieb}, N., {Bessell}, M.~S., {Beers}, T.~C., {et~al.} 2002, \nat, 419,
  904

\bibitem[{{Cohen} {et~al.}(2008){Cohen}, {Christlieb}, {McWilliam}, {Shectman},
  {Thompson}, {Melendez}, {Wisotzki}, \& {Reimers}}]{cohen08}
{Cohen}, J.~G., {Christlieb}, N., {McWilliam}, A., {et~al.} 2008, \apj, 672,
  320

\bibitem[{{Cohen} {et~al.}(2004){Cohen}, {Christlieb}, {McWilliam}, {Shectman},
  {Thompson}, {Wasserburg}, {Ivans}, {Dehn}, {Karlsson}, \&
  {Melendez}}]{cohen04}
{Cohen}, J.~G., {Christlieb}, N., {McWilliam}, A., {et~al.} 2004, \apj, 612,
  1107

\bibitem[{{Cohen} \& {Huang}(2009)}]{cohen09}
{Cohen}, J.~G. \& {Huang}, W. 2009, \apj, 701, 1053

\bibitem[{{Cohen} {et~al.}(2006){Cohen}, {McWilliam}, {Shectman}, {Thompson},
  {Christlieb}, {Melendez}, {Ramirez}, {Swensson}, \& {Zickgraf}}]{cohen06}
{Cohen}, J.~G., {McWilliam}, A., {Shectman}, S., {et~al.} 2006, \aj, 132, 137

\bibitem[{{Coleman} \& {de Jong}(2008)}]{coleman08}
{Coleman}, M.~G. \& {de Jong}, J.~T.~A. 2008, \apj, 685, 933

\bibitem[{{Cristallo} {et~al.}(2009){Cristallo}, {Straniero}, {Gallino},
  {Piersanti}, {Dom{\'{\i}}nguez}, \& {Lederer}}]{cristallo09}
{Cristallo}, S., {Straniero}, O., {Gallino}, R., {et~al.} 2009, \apj, 696, 797

\bibitem[{{Denissenkov} \& {Pinsonneault}(2008)}]{denissenkov08}
{Denissenkov}, P.~A. \& {Pinsonneault}, M. 2008, \apj, 679, 1541

\bibitem[{{Eggleton} {et~al.}(2008){Eggleton}, {Dearborn}, \&
  {Lattanzio}}]{eggleton08}
{Eggleton}, P.~P., {Dearborn}, D.~S.~P., \& {Lattanzio}, J.~C. 2008, \apj, 677,
  581

\bibitem[{{Farouqi} {et~al.}(2009){Farouqi}, {Kratz}, {Mashonkina}, {Pfeiffer},
  {Cowan}, {Thielemann}, \& {Truran}}]{farouqi09}
{Farouqi}, K., {Kratz}, K., {Mashonkina}, L.~I., {et~al.} 2009, \apjl, 694, L49

\bibitem[{{Ferrara} \& {Tolstoy}(2000)}]{ferrara00}
{Ferrara}, A. \& {Tolstoy}, E. 2000, \mnras, 313, 291

\bibitem[{{Fran{\c c}ois} {et~al.}(2007){Fran{\c c}ois}, {Depagne}, {Hill},
  {Spite}, {Spite}, {Plez}, {Beers}, {Andersen}, {James}, {Barbuy}, {Cayrel},
  {Bonifacio}, {Molaro}, {Nordstr{\"o}m}, \& {Primas}}]{francois07}
{Fran{\c c}ois}, P., {Depagne}, E., {Hill}, V., {et~al.} 2007, \aap, 476, 935

\bibitem[{{Frebel} {et~al.}(2010{\natexlab{a}}){Frebel}, {Kirby}, \&
  {Simon}}]{frebel10a}
{Frebel}, A., {Kirby}, E.~N., \& {Simon}, J.~D. 2010{\natexlab{a}}, \nat, 464,
  72

\bibitem[{{Frebel} {et~al.}(2010{\natexlab{b}}){Frebel}, {Simon}, {Geha}, \&
  {Willman}}]{frebel10}
{Frebel}, A., {Simon}, J.~D., {Geha}, M., \& {Willman}, B. 2010{\natexlab{b}},
  \apj, 708, 560

\bibitem[{{Fulbright} {et~al.}(2004){Fulbright}, {Rich}, \&
  {Castro}}]{fulbright04}
{Fulbright}, J.~P., {Rich}, R.~M., \& {Castro}, S. 2004, \apj, 612, 447

\bibitem[{{Fusi Pecci} {et~al.}(1990){Fusi Pecci}, {Ferraro}, {Crocker},
  {Rood}, \& {Buonanno}}]{fusipecci90}
{Fusi Pecci}, F., {Ferraro}, F.~R., {Crocker}, D.~A., {Rood}, R.~T., \&
  {Buonanno}, R. 1990, \aap, 238, 95

\bibitem[{{Gehren} {et~al.}(2006){Gehren}, {Shi}, {Zhang}, {Zhao}, \&
  {Korn}}]{gehren06}
{Gehren}, T., {Shi}, J.~R., {Zhang}, H.~W., {Zhao}, G., \& {Korn}, A.~J. 2006,
  \aap, 451, 1065

\bibitem[{{Gratton} {et~al.}(2000){Gratton}, {Sneden}, {Carretta}, \&
  {Bragaglia}}]{gratton00}
{Gratton}, R.~G., {Sneden}, C., {Carretta}, E., \& {Bragaglia}, A. 2000, \aap,
  354, 169

\bibitem[{{Grevesse} \& {Sauval}(1998)}]{grevesse98}
{Grevesse}, N. \& {Sauval}, A.~J. 1998, Space Science Reviews, 85, 161

\bibitem[{{Gustafsson} {et~al.}(1975){Gustafsson}, {Bell}, {Eriksson}, \&
  {Nordlund}}]{gustafsson75}
{Gustafsson}, B., {Bell}, R.~A., {Eriksson}, K., \& {Nordlund}, A. 1975, \aap,
  42, 407

\bibitem[{{Gustafsson} {et~al.}(2008){Gustafsson}, {Edvardsson}, {Eriksson},
  {J{\o}rgensen}, {Nordlund}, \& {Plez}}]{gustafsson08}
{Gustafsson}, B., {Edvardsson}, B., {Eriksson}, K., {et~al.} 2008, \aap, 486,
  951

\bibitem[{{Gustafsson} {et~al.}(2003){Gustafsson}, {Edvardsson}, {Eriksson},
  {Mizuno-Wiedner}, {J{\o}rgensen}, \& {Plez}}]{gustafsson03}
{Gustafsson}, B., {Edvardsson}, B., {Eriksson}, K., {et~al.} 2003, in
  Astronomical Society of the Pacific Conference Series, Vol. 288, Stellar
  Atmosphere Modeling, ed. I.~{Hubeny}, D.~{Mihalas}, \& K.~{Werner}, 331--+

\bibitem[{{Heger} \& {Woosley}(2002)}]{heger02}
{Heger}, A. \& {Woosley}, S.~E. 2002, \apj, 567, 532

\bibitem[{{Heiter} \& {Eriksson}(2006)}]{heiter06}
{Heiter}, U. \& {Eriksson}, K. 2006, \aap, 452, 1039

\bibitem[{{Helmi} {et~al.}(2006){Helmi}, {Irwin}, {Tolstoy}, {Battaglia},
  {Hill}, {Jablonka}, {Venn}, {Shetrone}, {Letarte}, {Arimoto}, {Abel},
  {Francois}, {Kaufer}, {Primas}, {Sadakane}, \& {Szeifert}}]{helmi06}
{Helmi}, A., {Irwin}, M.~J., {Tolstoy}, E., {et~al.} 2006, \apjl, 651, L121

\bibitem[{{Hill}(2010)}]{hill10a}
{Hill}, V. 2010, in IAU Symposium, Vol. 265, IAU Symposium, ed. {K.~Cunha,
  M.~Spite, \& B.~Barbuy}, 219--226

\bibitem[{{Honda} {et~al.}(2004){Honda}, {Aoki}, {Kajino}, {Ando}, {Beers},
  {Izumiura}, {Sadakane}, \& {Takada-Hidai}}]{honda04}
{Honda}, S., {Aoki}, W., {Kajino}, T., {et~al.} 2004, \apj, 607, 474

\bibitem[{{Hurley-Keller} {et~al.}(1999){Hurley-Keller}, {Mateo}, \&
  {Grebel}}]{hurley-keller99}
{Hurley-Keller}, D., {Mateo}, M., \& {Grebel}, E.~K. 1999, \apjl, 523, L25

\bibitem[{{Ishimaru} {et~al.}(2005){Ishimaru}, {Wanajo}, {Aoki}, {Ryan}, \&
  {Prantzos}}]{ishimaru05}
{Ishimaru}, Y., {Wanajo}, S., {Aoki}, W., {Ryan}, S.~G., \& {Prantzos}, N.
  2005, Nuclear Physics A, 758, 603

\bibitem[{{Karachentsev} {et~al.}(2004){Karachentsev}, {Karachentseva},
  {Huchtmeier}, \& {Makarov}}]{karachentsev04}
{Karachentsev}, I.~D., {Karachentseva}, V.~E., {Huchtmeier}, W.~K., \&
  {Makarov}, D.~I. 2004, \aj, 127, 2031

\bibitem[{{Karakas}(2010)}]{karakas10}
{Karakas}, A.~I. 2010, \mnras, 114

\bibitem[{{Kirby} {et~al.}(2008){Kirby}, {Simon}, {Geha}, {Guhathakurta}, \&
  {Frebel}}]{kirby08}
{Kirby}, E.~N., {Simon}, J.~D., {Geha}, M., {Guhathakurta}, P., \& {Frebel}, A.
  2008, \apjl, 685, L43

\bibitem[{{Koch} {et~al.}(2008{\natexlab{a}}){Koch}, {Grebel}, {Gilmore},
  {Wyse}, {Kleyna}, {Harbeck}, {Wilkinson}, \& {Wyn Evans}}]{koch08}
{Koch}, A., {Grebel}, E.~K., {Gilmore}, G.~F., {et~al.} 2008{\natexlab{a}},
  \aj, 135, 1580

\bibitem[{{Koch} {et~al.}(2008{\natexlab{b}}){Koch}, {McWilliam}, {Grebel},
  {Zucker}, \& {Belokurov}}]{koch08b}
{Koch}, A., {McWilliam}, A., {Grebel}, E.~K., {Zucker}, D.~B., \& {Belokurov},
  V. 2008{\natexlab{b}}, \apjl, 688, L13

\bibitem[{{Lai} {et~al.}(2008){Lai}, {Bolte}, {Johnson}, {Lucatello}, {Heger},
  \& {Woosley}}]{lai08}
{Lai}, D.~K., {Bolte}, M., {Johnson}, J.~A., {et~al.} 2008, \apj, 681, 1524

\bibitem[{{Lee} {et~al.}(2003){Lee}, {Park}, {Park}, {Sohn}, {Oh}, {Yuk},
  {Rey}, {Lee}, {Lee}, {Kim}, {Han}, {Park}, {Lee}, {Jeon}, \& {Kim}}]{lee03}
{Lee}, M.~G., {Park}, H.~S., {Park}, J.-H., {et~al.} 2003, \aj, 126, 2840

\bibitem[{{Letarte} {et~al.}(2010){Letarte}, {Hill}, {Tolstoy}, {Jablonka},
  {Spite}, {Shetrone}, \& {Venn}}]{letarte10}
{Letarte}, B., {Hill}, V., {Tolstoy}, E., {et~al.} 2010, submitted to A\&A

\bibitem[{{Magain}(1984)}]{magain84}
{Magain}, P. 1984, \aap, 134, 189

\bibitem[{{Majewski} {et~al.}(1999){Majewski}, {Siegel}, {Patterson}, \&
  {Rood}}]{majewski99}
{Majewski}, S.~R., {Siegel}, M.~H., {Patterson}, R.~J., \& {Rood}, R.~T. 1999,
  \apjl, 520, L33

\bibitem[{{Mashonkina} {et~al.}(2008){Mashonkina}, {Zhao}, {Gehren}, {Aoki},
  {Bergemann}, {Noguchi}, {Shi}, {Takada-Hidai}, \& {Zhang}}]{mashonkina08}
{Mashonkina}, L., {Zhao}, G., {Gehren}, T., {et~al.} 2008, \aap, 478, 529

\bibitem[{{McWilliam} {et~al.}(1995){McWilliam}, {Preston}, {Sneden}, \&
  {Shectman}}]{mcwilliam95}
{McWilliam}, A., {Preston}, G.~W., {Sneden}, C., \& {Shectman}, S. 1995, \aj,
  109, 2736

\bibitem[{{Meynet} {et~al.}(2006){Meynet}, {Ekstr{\"o}m}, \&
  {Maeder}}]{meynet06}
{Meynet}, G., {Ekstr{\"o}m}, S., \& {Maeder}, A. 2006, \aap, 447, 623

\bibitem[{{Monkiewicz} {et~al.}(1999){Monkiewicz}, {Mould}, {Gallagher},
  {Clarke}, {Trauger}, {Grillmair}, {Ballester}, {Burrows}, {Crisp}, {Evans},
  {Griffiths}, {Hester}, {Hoessel}, {Holtzman}, {Krist}, {Meadows}, {Scowen},
  {Stapelfeldt}, {Sahai}, \& {Watson}}]{monkiewicz99}
{Monkiewicz}, J., {Mould}, J.~R., {Gallagher}, III, J.~S., {et~al.} 1999,
  \pasp, 111, 1392

\bibitem[{{Nakamura} {et~al.}(2001){Nakamura}, {Umeda}, {Iwamoto}, {Nomoto},
  {Hashimoto}, {Hix}, \& {Thielemann}}]{nakamura01}
{Nakamura}, T., {Umeda}, H., {Iwamoto}, K., {et~al.} 2001, \apj, 555, 880

\bibitem[{{Nakamura} {et~al.}(1999){Nakamura}, {Umeda}, {Nomoto}, {Thielemann},
  \& {Burrows}}]{nakamura99}
{Nakamura}, T., {Umeda}, H., {Nomoto}, K., {Thielemann}, F., \& {Burrows}, A.
  1999, \apj, 517, 193

\bibitem[{{Norris} {et~al.}(2010){Norris}, {Yong}, {Gilmore}, \&
  {Wyse}}]{norris10}
{Norris}, J.~E., {Yong}, D., {Gilmore}, G., \& {Wyse}, R.~F.~G. 2010, \apj,
  711, 350

\bibitem[{{Prochaska} {et~al.}(2000){Prochaska}, {Naumov}, {Carney},
  {McWilliam}, \& {Wolfe}}]{prochaska00}
{Prochaska}, J.~X., {Naumov}, S.~O., {Carney}, B.~W., {McWilliam}, A., \&
  {Wolfe}, A.~M. 2000, \aj, 120, 2513

\bibitem[{{Qian} \& {Wasserburg}(2007)}]{qian07}
{Qian}, Y. \& {Wasserburg}, G.~J. 2007, \physrep, 442, 237

\bibitem[{{Ram{\'{\i}}rez} \& {Mel{\'e}ndez}(2005)}]{ramirez05}
{Ram{\'{\i}}rez}, I. \& {Mel{\'e}ndez}, J. 2005, \apj, 626, 465

\bibitem[{{Ramsey} {et~al.}(1998){Ramsey}, {Adams}, {Barnes}, {Booth},
  {Cornell}, {Fowler}, {Gaffney}, {Glaspey}, {Good}, {Hill}, {Kelton},
  {Krabbendam}, {Long}, {MacQueen}, {Ray}, {Ricklefs}, {Sage}, {Sebring},
  {Spiesman}, \& {Steiner}}]{Ramsey98}
{Ramsey}, L.~W., {Adams}, M.~T., {Barnes}, T.~G., {et~al.} 1998, in Society of
  Photo-Optical Instrumentation Engineers (SPIE) Conference Series, Vol. 3352,
  Society of Photo-Optical Instrumentation Engineers (SPIE) Conference Series,
  ed. {L.~M.~Stepp}, 34--42

\bibitem[{{Revaz} {et~al.}(2009){Revaz}, {Jablonka}, {Sawala}, {Hill},
  {Letarte}, {Irwin}, {Battaglia}, {Helmi}, {Shetrone}, {Tolstoy}, \&
  {Venn}}]{revaz09}
{Revaz}, Y., {Jablonka}, P., {Sawala}, T., {et~al.} 2009, \aap, 501, 189

\bibitem[{{Ryan} {et~al.}(1996){Ryan}, {Norris}, \& {Beers}}]{ryan96}
{Ryan}, S.~G., {Norris}, J.~E., \& {Beers}, T.~C. 1996, \apj, 471, 254

\bibitem[{{Schlegel} {et~al.}(1998){Schlegel}, {Finkbeiner}, \&
  {Davis}}]{schlegel98}
{Schlegel}, D.~J., {Finkbeiner}, D.~P., \& {Davis}, M. 1998, \apj, 500, 525

\bibitem[{{Shetrone} {et~al.}(2003){Shetrone}, {Venn}, {Tolstoy}, {Primas},
  {Hill}, \& {Kaufer}}]{shetrone03}
{Shetrone}, M., {Venn}, K.~A., {Tolstoy}, E., {et~al.} 2003, \aj, 125, 684

\bibitem[{{Shetrone} {et~al.}(2001){Shetrone}, {C{\^o}t{\'e}}, \&
  {Sargent}}]{shetrone01}
{Shetrone}, M.~D., {C{\^o}t{\'e}}, P., \& {Sargent}, W.~L.~W. 2001, \apj, 548,
  592

\bibitem[{{Short} \& {Hauschildt}(2006)}]{short06}
{Short}, C.~I. \& {Hauschildt}, P.~H. 2006, \apj, 641, 494

\bibitem[{{Simmerer} {et~al.}(2004){Simmerer}, {Sneden}, {Cowan}, {Collier},
  {Woolf}, \& {Lawler}}]{simmerer04}
{Simmerer}, J., {Sneden}, C., {Cowan}, J.~J., {et~al.} 2004, \apj, 617, 1091

\bibitem[{{Simon} {et~al.}(2010){Simon}, {Frebel}, {McWilliam}, {Kirby}, \&
  {Thompson}}]{simon10}
{Simon}, J.~D., {Frebel}, A., {McWilliam}, A., {Kirby}, E.~N., \& {Thompson},
  I.~B. 2010, \apj, 716, 446

\bibitem[{{Simon} \& {Geha}(2007)}]{simon07}
{Simon}, J.~D. \& {Geha}, M. 2007, \apj, 670, 313

\bibitem[{{Skrutskie} {et~al.}(2006){Skrutskie}, {Cutri}, {Stiening},
  {Weinberg}, {Schneider}, {Carpenter}, {Beichman}, {Capps}, {Chester},
  {Elias}, {Huchra}, {Liebert}, {Lonsdale}, {Monet}, {Price}, {Seitzer},
  {Jarrett}, {Kirkpatrick}, {Gizis}, {Howard}, {Evans}, {Fowler}, {Fullmer},
  {Hurt}, {Light}, {Kopan}, {Marsh}, {McCallon}, {Tam}, {Van Dyk}, \&
  {Wheelock}}]{skrutskie06}
{Skrutskie}, M.~F., {Cutri}, R.~M., {Stiening}, R., {et~al.} 2006, \aj, 131,
  1163

\bibitem[{{Sobeck} {et~al.}(2007){Sobeck}, {Lawler}, \& {Sneden}}]{sobeck07}
{Sobeck}, J.~S., {Lawler}, J.~E., \& {Sneden}, C. 2007, \apj, 667, 1267

\bibitem[{{Spite}(1967)}]{spite67}
{Spite}, M. 1967, Annales d'Astrophysique, 30, 211

\bibitem[{{Spite} {et~al.}(2006){Spite}, {Cayrel}, {Hill}, {Spite}, {Fran{\c
  c}ois}, {Plez}, {Bonifacio}, {Molaro}, {Depagne}, {Andersen}, {Barbuy},
  {Beers}, {Nordstr{\"o}m}, \& {Primas}}]{spite06}
{Spite}, M., {Cayrel}, R., {Hill}, V., {et~al.} 2006, \aap, 455, 291

\bibitem[{{Spite} {et~al.}(2005){Spite}, {Cayrel}, {Plez}, {Hill}, {Spite},
  {Depagne}, {Fran{\c c}ois}, {Bonifacio}, {Barbuy}, {Beers}, {Andersen},
  {Molaro}, {Nordstr{\"o}m}, \& {Primas}}]{spite05}
{Spite}, M., {Cayrel}, R., {Plez}, B., {et~al.} 2005, \aap, 430, 655

\bibitem[{{Stancliffe} {et~al.}(2009){Stancliffe}, {Church}, {Angelou}, \&
  {Lattanzio}}]{stancliffe09}
{Stancliffe}, R.~J., {Church}, R.~P., {Angelou}, G.~C., \& {Lattanzio}, J.~C.
  2009, \mnras, 396, 2313

\bibitem[{{Starkenburg} {et~al.}(2010){Starkenburg}, {Hill}, {Tolstoy},
  {Gonz{\'a}lez Hern{\'a}ndez}, {Irwin}, {Helmi}, {Battaglia}, {Jablonka},
  {Tafelmeyer}, {Shetrone}, {Venn}, \& {de Boer}}]{starkenburg10}
{Starkenburg}, E., {Hill}, V., {Tolstoy}, E., {et~al.} 2010, \aap, 513, A34+

\bibitem[{{Stetson} {et~al.}(1998){Stetson}, {Hesser}, \&
  {Smecker-Hane}}]{stetson98}
{Stetson}, P.~B., {Hesser}, J.~E., \& {Smecker-Hane}, T.~A. 1998, \pasp, 110,
  533

\bibitem[{{Suda} \& {Fujimoto}(2010)}]{suda10}
{Suda}, T. \& {Fujimoto}, M.~Y. 2010, \mnras, 405, 177

\bibitem[{{Tolstoy}(2010)}]{tolstoy10}
{Tolstoy}, E. 2010, in IAU Symposium, Vol. 262, IAU Symposium, ed. {G.~Bruzual
  \& S.~Charlot}, 119--126

\bibitem[{{Tolstoy} {et~al.}(2009){Tolstoy}, {Hill}, \& {Tosi}}]{tolstoy09}
{Tolstoy}, E., {Hill}, V., \& {Tosi}, M. 2009, \araa, 47, 371

\bibitem[{{Tolstoy} {et~al.}(2004){Tolstoy}, {Irwin}, {Helmi}, {Battaglia},
  {Jablonka}, {Hill}, {Venn}, {Shetrone}, {Letarte}, {Cole}, {Primas},
  {Francois}, {Arimoto}, {Sadakane}, {Kaufer}, {Szeifert}, \&
  {Abel}}]{tolstoy04}
{Tolstoy}, E., {Irwin}, M.~J., {Helmi}, A., {et~al.} 2004, \apjl, 617, L119

\bibitem[{{Travaglio} {et~al.}(2004){Travaglio}, {Gallino}, {Arnone}, {Cowan},
  {Jordan}, \& {Sneden}}]{travaglio04}
{Travaglio}, C., {Gallino}, R., {Arnone}, E., {et~al.} 2004, \apj, 601, 864

\bibitem[{{Truran}(1981)}]{truran81}
{Truran}, J.~W. 1981, \aap, 97, 391

\bibitem[{{Tull}(1998)}]{tull98}
{Tull}, R.~G. 1998, in Presented at the Society of Photo-Optical
  Instrumentation Engineers (SPIE) Conference, Vol. 3355, Proc. SPIE Vol. 3355,
  p. 387-398, Optical Astronomical Instrumentation, Sandro D'Odorico; Ed., ed.
  S.~{D'Odorico}, 387--398

\bibitem[{{Umeda} \& {Nomoto}(2002)}]{umeda02}
{Umeda}, H. \& {Nomoto}, K. 2002, \apj, 565, 385

\bibitem[{{Umeda} \& {Nomoto}(2005)}]{umeda05}
{Umeda}, H. \& {Nomoto}, K. 2005, \apj, 619, 427

\bibitem[{{Venn} {et~al.}(2004){Venn}, {Irwin}, {Shetrone}, {Tout}, {Hill}, \&
  {Tolstoy}}]{Venn04}
{Venn}, K.~A., {Irwin}, M., {Shetrone}, M.~D., {et~al.} 2004, \aj, 128, 1177

\bibitem[{{Wanajo} {et~al.}(2003){Wanajo}, {Tamamura}, {Itoh}, {Nomoto},
  {Ishimaru}, {Beers}, \& {Nozawa}}]{wanajo03}
{Wanajo}, S., {Tamamura}, M., {Itoh}, N., {et~al.} 2003, \apj, 593, 968

\bibitem[{{Woosley} \& {Weaver}(1986)}]{woosley86}
{Woosley}, S.~E. \& {Weaver}, T.~A. 1986, \araa, 24, 205

\bibitem[{{Woosley} \& {Weaver}(1995)}]{woosley95}
{Woosley}, S.~E. \& {Weaver}, T.~A. 1995, \apjs, 101, 181

\end{thebibliography}

\begin{center}
\scriptsize
\begin{longtable}{cccc|cc cc cc cc cc}
\caption{Linelist, equivalent widths and scattering corrections. These corrections are subtracted from
the original abundances calculated with \texttt{calrai}. The exponents have the following meanings: (1) Equivalent widths are not used, abundances are determined from synthesis instead; (2) Hyperfine
structure was taken into account.  (3) Lines are not used, due to very
high uncertainties (e.g. blendlines, low S/N etc.)}\\
\hline
\hline
$\lambda$ & EL  & $\chi _{ex}$ & log(gf) &&   &&  &EW $\pm \Delta$EW &$\Delta_{scattering}$  &&  &&\\
&&&&&&&&&&&&& \\
          &     &              &         & S24-72 && S11-04 && Fnx05-42 && Scl07-49 &&Scl07-50& \\
\hline
\hline
\endfirsthead
\hline
\hline
$\lambda$ & EL  & $\chi _{ex}$ & log(gf) &&   &&  &EW $\pm \Delta$EW &$\Delta_{scattering}$  &&  &&\\
 & & & &&&&&& & & & &  \\
          &     &              &         & S24-72 && S11-04 && Fnx05-42 && Scl07-49 &&Scl07-50& \\
\hline
\hline
\endhead
6300.31& O1&0.00&$-$9.750& $<$11          &0.02& $<$20         &0.04& $<$9          &0.03&               &    &               &\\
\hline                                                                                                             
5889.97&NA1&0.00& 0.122  & 306\3          &    & 217.9 \plm5.7 &0.08& 160.6 \plm4.5 &0.07& 133.8 \plm6.1 &0.04&  83.9 \plm7.8 &0.02\\
5895.92&NA1&0.00&$-$0.190& 284\1          &0.05& 196.4 \plm3.0 &0.07& 143.2 \plm4.7 &0.07& 125.0 \plm2.0 &0.04&  66.2 \plm5.0 &0.02\\
\hline                                                                                                                             
3829.35&MG1&2.71&$-$0.210&                &    &               &    &               &    &               &    &  87.1\1       &0.16\\
3832.30&MG1&2.71& 0.150  &                &    &               &    &               &    &               &    & 111.2\1       &0.19\\
5172.70&MG1&2.71&$-$0.390& 213.0 \plm5.9  &0.12&248.6 \plm9.9  &0.17&188.0 \plm4.4  &0.15& 178.5 \plm5.2 &0.09& 94.1  \plm4.5 &0.04\\
5183.60&MG1&2.72&$-$0.160& 240.2 \plm7.0  &0.11&278.3 \plm8.8  &0.17&216.1 \plm5.8  &0.15& 200.2 \plm7.1 &0.08& 117.6 \plm4.8 &0.05\\
5528.41&MG1&4.35&$-$0.357&  55.5 \plm2.2  &0.08& 82.6 \plm2.2  &0.16& 44.9 \plm1.7  &0.11&               &    &               &    \\
\hline                                                                                                                             
3944.01&AL1&0.00&$-$0.640&                &    & 208.3 \plm15.7&0.54& 140.2 \plm10.1&0.45&               &    &  57.8 \plm5.1 &0.10\\
3961.52&AL1&0.01&$-$0.340&179.9\1         &    & 213.9 \plm22.6&0.54& 158.6 \plm20.5&0.45&               &    &  72.3 \plm4.8 &0.11\\
\hline                                                                                                                             
3905.52&SI1&1.91&$-$1.090&                &    &               &    &               &    &               &    &  99.8 \plm8.4 &0.16\\
4102.94&SI1&1.91&$-$2.920& 130\1          &    & 139.4\1       &0.56&  67.1 \plm6.7 &0.53&               &    &               &    \\
\hline                                                                                                                             
4226.73&CA1&0.00& 0.240  &                &    &               &    &               &    &               &    & 100.2 \plm3.0 &0.11\\
5588.75&CA1&2.52& 0.210  &  57.7 \plm5.9  &0.08&  46.8 \plm5.4 &0.12&  22.1 \plm1.8 &0.09&               &    &               &    \\
5857.45&CA1&2.93& 0.230  &                &    &  32.0 \plm2.7 &0.09&               &    &               &    &               &    \\
6102.73&CA1&1.88&$-$0.790&  35.3 \plm3.0  &0.04&  37.2 \plm1.4 &0.06&               &    &               &    &               &    \\
6122.23&CA1&1.89&$-$0.320&  68.7 \plm3.1  &0.05&  79.8 \plm2.8 &0.07&  65.0 \plm3.9 &0.06&  53.7 \plm3.5 &0.03&               &    \\
6162.17&CA1&1.90&$-$0.090&                &    & 108.9 \plm3.0 &0.07&  56.8 \plm3.1 &0.05&               &    &               &    \\
6439.08&CA1&2.52& 0.390  &  47.9 \plm2.6  &0.04&  81.7 \plm3.7 &0.06&               &    &  48.0 \plm3.9 &0.02&               &    \\
\hline                                                                                                                             
4246.82&SC2&0.31& 0.240  &                &    &               &    & 128.6 \plm6.8 &0.41&               &    &  81.0 \plm3.5 &0.07\\
4314.08&SC2&0.62&$-$0.100&                &    &               &    & 109.2 \plm6.2 &0.42&               &    &  41.2 \plm3.7 &0.05\\
4400.39&SC2&0.61&$-$0.540&                &    &               &    &  82.6 \plm4.4 &0.38&               &    &               &    \\
5031.02&SC2&1.36&$-$0.400&  40.4\2        &0.10&  47.3\2       &0.18&  25.8\2       &0.15&  31.0\2       &0.06&               &    \\
5526.79&SC2&1.77& 0.030  &  48.5\2        &0.06&  56.9\2       &0.12&               &    &  40.0\2       &0.04&               &    \\
\hline                                                                                                                             
4981.73&TI1&0.84& 0.500  &  65.9 \plm2.8  &0.11&  78.0 \plm2.6 &0.14&  34.4 \plm2.8 &0.12&  39.4 \plm3.6 &0.06&               &    \\
4991.07&TI1&0.84& 0.380  &  50.7 \plm2.9  &0.10&  71.5 \plm3.5 &0.14&  30.1 \plm3.3 &0.12&  34.7 \plm5.3 &0.06&               &    \\
4999.50&TI1&0.83& 0.250  &  49.0 \plm2.2  &0.10&  61.9 \plm3.0 &0.13&  29.1 \plm4.0 &0.12&  26.0 \plm4.0 &0.06&               &    \\
5039.96&TI1&0.02&$-$1.130&  34.9 \plm2.2  &0.08&  49.9 \plm2.7 &0.09&               &    &  26.0 \plm5.0 &0.05&               &    \\
5064.65&TI1&0.05&$-$0.930&                &    &  59.9 \plm2.2 &0.09&               &    &  29.6 \plm3.2 &0.05&               &    \\
5173.74&TI1&0.00&$-$1.120&  29.1 \plm3.4  &0.07&  50.0 \plm2.6 &0.08&               &    &               &    &               &    \\
5192.97&TI1&0.02&$-$1.010&  32.3 \plm2.3  &0.07&  63.5 \plm2.7 &0.08&               &    &               &    &               &    \\
5210.39&TI1&0.05&$-$0.580&                &    &               &    &               &    &  30.1 \plm3.8 &0.04&               &    \\
\hline                                                                                                                             
3913.47&TI2&1.12&$-$0.530&                &    &               &    &               &    &               &    &  55.8 \plm2.6 &0.09\\
4012.39&TI2&0.57&$-$1.610& 100.0 \plm10.8 &0.40&               &    &               &    &               &    &               &    \\
4028.34&TI2&1.89&$-$1.000&                &    &               &    &  34.6 \plm9.4 &0.59&               &    &               &    \\
4290.22&TI2&1.16&$-$1.120&                &    &               &    &               &    &               &    &  40.0 \plm4.0 &0.05\\
4300.05&TI2&1.18&$-$0.770&                &    &               &    &               &    &               &    &  60.5 \plm4.3 &0.06\\
4443.79&TI2&1.08&$-$0.710&                &    &               &    & 116.3 \plm10.2&0.39&               &    &  51.6 \plm2.5 &0.05\\
4450.48&TI2&1.08&$-$1.450&                &    &               &    &  73.3 \plm8.8 &0.38&               &    &               &    \\
4468.51&TI2&1.13&$-$0.620&                &    &               &    &               &    &               &    &  57.8 \plm2.5 &0.05\\
4501.27&TI2&1.12&$-$0.750&                &    &               &    &               &    &               &    &  56.9 \plm2.7 &0.05\\
4533.97&TI2&1.24&$-$0.770&                &    &               &    &               &    &               &    &  56.3 \plm3.1 &0.05\\
4563.76&TI2&1.22&$-$0.960&                &    &               &    &               &    &               &    &  34.1 \plm1.9 &0.04\\
4571.97&TI2&1.57&$-$0.520&                &    &               &    &               &    &               &    &  35.8 \plm2.6 &0.04\\
4865.61&TI2&1.12&$-$2.590&  18.9 \plm3.7  &0.11&  28.5 \plm5.5 &0.20&               &    &               &    &               &    \\
5129.16&TI2&1.89&$-$1.390&  44.8 \plm4.5  &0.10&  53.7 \plm2.2 &0.19&  24.0 \plm2.3 &0.14&               &    &               &    \\
5154.07&TI2&1.57&$-$1.520&                &    &  58.9 \plm2.8 &0.17&               &    &               &    &               &    \\
5185.91&TI2&1.89&$-$1.350&  46.8 \plm3.1  &0.09&  49.3 \plm2.4 &0.17&               &    &               &    &               &    \\
5188.68&TI2&1.58&$-$1.220&  89.5 \plm4.3  &0.12&               &    &  63.8 \plm2.5 &0.16&  58.8         &0.06&               &    \\
5336.77&TI2&1.58&$-$1.700&  62.7 \plm5.0  &0.08&  68.1 \plm2.5 &0.16&  29.0 \plm4.0 &0.11&  38.1         &0.05&               &    \\
5381.01&TI2&1.57&$-$1.780&  36.9 \plm4.7  &0.07&  54.4 \plm5.2 &0.14&               &    &               &    &               &    \\
5418.77&TI2&1.58&$-$2.110&                &    &  42.2 \plm1.8 &0.13&               &    &  23.1         &0.04&               &    \\
\hline                                                                                                                             
4254.33&CR1&0.00&$-$0.110&                &    &               &    &               &    &               &    &  53.0 \plm3.1 &0.06\\
4274.80&CR1&0.00&$-$0.230&                &    &               &    &               &    &               &    &  39.8 \plm2.8 &0.06\\
4289.72&CR1&0.00&$-$0.360&                &    &               &    &               &    &               &    &  42.8 \plm6.2 &0.06\\
5206.04&CR1&0.94& 0.019  & 110.8 \plm3.1  &0.10& 126.0 \plm2.1 &0.12&  64.2 \plm1.8 &0.10&  66.8 \plm2.8 &0.06&               &    \\
5208.42&CR1&0.94& 0.160  & 141.4\1        &0.10&162.2\1        &0.12& 90.0\1        &0.11& 93.9\1        &0.07&               &    \\
5345.80&CR1&1.00&$-$0.980&  33.2 \plm3.8  &0.07&  44.4 \plm3.2 &0.10&               &    &               &    &               &    \\
5409.80&CR1&1.03&$-$0.720&  52.4 \plm2.9  &0.07&  77.0 \plm3.6 &0.10&               &    &               &    &               &    \\
\hline                                                                                                                             
4030.75&MN1&0.00&$-$0.480& 214\3          &    &               &    &               &    &               &    &  48.8\2       &0.08\\
4033.06&MN1&0.00&$-$0.620& 190\3          &    & 180\3         &    &  95.8\2       &0.27&               &    &  33.3\2       &0.07\\
4034.48&MN1&0.00&$-$0.810&                &    & 164\3         &    &  80.4\2       &0.27&               &    &  27.9\2       &0.07\\
4823.52&MN1&2.32& 0.140  &  38.6\2        &0.13&  44.3\2       &0.21&               &    &               &    &               &    \\
\hline                                                                                                                             
3920.26&FE1&0.12&$-$1.750&                &    &  213.7\3      &    &               &    &               &    & 120.7 \plm3.3 &0.16\\
3922.91&FE1&0.05&$-$1.650&                &    &               &    &               &    &               &    & 117.1 \plm5.5 &0.16\\
4005.24&FE1&1.56&$-$0.610&                &    &  205.2\3      &    &  148.2\3      &    &               &    &  71.1 \plm3.7 &0.11\\
4045.81&FE1&1.48& 0.280  &                &    &               &    &  198.1\3      &    &               &    & 117.8 \plm6.1 &0.15\\
4063.59&FE1&1.56& 0.070  &                &    &               &    &  159.1\3      &    &               &    & 110.7 \plm4.4 &0.14\\
4071.74&FE1&1.61&$-$0.020&                &    &  251.7\3      &    &  184.5\3      &    &               &    & 109.2 \plm6.2 &0.14\\
4132.06&FE1&1.61&$-$0.670&                &    &  228.3\3      &    &  132.4\3      &    &               &    &  81.1 \plm4.9 &0.11\\
4143.87&FE1&1.56&$-$0.460&                &    &               &    &               &    &               &    &  97.2 \plm5.2 &0.12\\
4202.03&FE1&1.48&$-$0.700&                &    &  185.6\3      &    &  145.5\3      &    &               &    &  86.9 \plm3.7 &0.11\\
4260.47&FE1&2.40&$-$0.020&                &    &  158.3\3      &    &               &    &               &    &  63.0 \plm6.2 &0.08\\
4271.15&FE1&2.45&$-$0.350&                &    &               &    &   93.3\3      &    &               &    &  41.2 \plm3.4 &0.06\\
4271.76&FE1&1.48&$-$0.160&                &    &  247.7\3      &    &  145.4\3      &    &               &    & 109.1 \plm3.5 &0.11\\
4325.76&FE1&1.61&$-$0.010&                &    &  234.3\3      &    &  168.4\3      &    &               &    & 105.8 \plm4.3 &0.10\\
4375.93&FE1&0.00&$-$3.030&                &    &  201.8\3      &    &  183.0\3      &    &               &    &  94.4 \plm5.4 &0.08\\
4383.55&FE1&1.48& 0.200  &                &    &  268.0\3      &    &  180.6\3      &    &               &    & 124.3 \plm3.5 &0.10\\
4404.75&FE1&1.56&$-$0.140&                &    &               &    &  189.6\3      &    &               &    &  93.1 \plm4.9 &0.09\\
4415.13&FE1&1.61&$-$0.610&                &    &               &    &               &    &               &    &  85.8 \plm4.9 &0.08\\
4461.65&FE1&0.09&$-$3.200&                &    &  209.0\3      &    &  117.4\3      &    &               &    &  70.1 \plm6.0 &0.06\\
4871.32&FE1&2.87&$-$0.360&  94.8 \plm3.0  &0.18& 111.3 \plm5.0 &0.26&  49.3 \plm2.3 &0.19&  59.4 \plm2.1 &0.09&               &    \\
4872.14&FE1&2.88&$-$0.570&  79.4 \plm4.0  &0.17&  80.3 \plm5.7 &0.25&  46.1 \plm2.6 &0.19&  50.1 \plm4.3 &0.08&               &    \\
4891.49&FE1&2.85&$-$0.110&                &    & 121.7 \plm4.5 &0.25&  67.0 \plm2.4 &0.21&  76.8 \plm5.2 &0.10&  32.2 \plm3.2 &0.03\\
4918.99&FE1&2.87&$-$0.340&  95.2 \plm3.4  &0.17& 110.3 \plm2.6 &0.24&  65.0 \plm4.4 &0.20&  65.1 \plm2.4 &0.09&               &    \\
4920.50&FE1&2.83& 0.070  & 114.4 \plm5.5  &0.17& 130.2 \plm4.4 &0.24&  77.1 \plm3.0 &0.21&  96.2 \plm3.1 &0.11&  36.7 \plm2.2 &0.03\\
4939.69&FE1&0.86&$-$3.340&                &    &               &    &  68.1 \plm4.7 &0.12&               &    &               &    \\
4966.10&FE1&3.33&$-$0.890&  41.4 \plm3.6  &0.11&  41.5 \plm2.6 &0.17&               &    &               &    &               &    \\
4994.13&FE1&0.92&$-$3.080& 110.5 \plm2.6  &0.10& 126.3 \plm3.2 &0.14&  72.2 \plm2.8 &0.12&               &    &               &    \\
5001.86&FE1&3.88& 0.010  &  29.1 \plm2.7  &0.11&  32.8 \plm3.2 &0.19&               &    &               &    &               &    \\
5006.12&FE1&2.83&$-$0.628&  88.5 \plm3.3  &0.15& 101.4 \plm4.0 &0.22&  41.7 \plm2.3 &0.15&  55.6 \plm4.9 &0.08&               &    \\
5041.07&FE1&0.96&$-$3.090& 106.4 \plm4.2  &0.10& 135.8 \plm4.0 &0.13&  72.2 \plm1.8 &0.12&               &    &               &    \\
5041.76&FE1&1.49&$-$2.200& 106.3 \plm3.3  &0.11& 131.2 \plm4.3 &0.15&               &    &  70.3 \plm4.1 &0.07&               &    \\
5049.82&FE1&2.28&$-$1.360&  83.0 \plm3.7  &0.13& 107.1 \plm2.5 &0.19&  44.4 \plm2.5 &0.14&  60.5 \plm2.6 &0.07&               &    \\
5051.64&FE1&0.92&$-$2.800& 125.3 \plm3.7  &0.09&148.4 \plm3.8  &0.13&102.9 \plm3.7  &0.12&               &    &               &    \\
5068.77&FE1&2.94&$-$1.040&  50.6 \plm2.7  &0.11& 55.1 \plm2.5  &0.18&               &    &               &    &               &    \\
5079.74&FE1&0.99&$-$3.220&  96.4 \plm3.6  &0.09&               &    & 67.8 \plm3.8  &0.11&               &    &               &    \\
5083.34&FE1&0.96&$-$2.960& 111.7 \plm4.0  &0.09&130.2 \plm3.4  &0.13& 73.7 \plm3.4  &0.11&               &    &               &    \\
5110.41&FE1&0.00&$-$3.760& 156.8 \plm5.2  &0.07&165.5 \plm4.5  &0.10&119.2 \plm3.6  &0.08&               &    &               &    \\
5123.72&FE1&1.01&$-$3.070&  91.9 \plm3.3  &0.09&118.3 \plm3.3  &0.12& 48.7 \plm2.8  &0.10&               &    &               &    \\
5127.36&FE1&0.92&$-$3.310&  96.7 \plm4.0  &0.09&114.1 \plm5.2  &0.12& 54.0 \plm2.4  &0.10&               &    &               &    \\
5150.84&FE1&0.99&$-$3.040&  90.5 \plm3.3  &0.09&117.9 \plm3.2  &0.12& 61.7 \plm2.5  &0.10&               &    &               &    \\
5151.91&FE1&1.01&$-$3.320&  87.1 \plm3.6  &0.09& 95.6 \plm3.8  &0.12& 48.9 \plm3.1  &0.10&               &    &               &    \\
5162.29&FE1&4.18& 0.020  &  31.2 \plm3.2  &0.10& 33.9 \plm2.0  &0.17&               &    &               &    &               &    \\
5166.28&FE1&0.00&$-$4.200& 135.4 \plm3.5  &0.06&155.4 \plm3.4  &0.09& 93.8 \plm3.7  &0.07&               &    &               &    \\
5171.61&FE1&1.48&$-$1.751& 132.5 \plm2.4  &0.09&141.3 \plm3.5  &0.13& 97.4 \plm3.3  &0.12&  97.4 \plm3.6 &0.07&  30.6 \plm2.3 &0.02\\
5191.46&FE1&3.04&$-$0.550&  61.7 \plm2.7  &0.11& 75.6 \plm2.9  &0.18& 32.0 \plm2.2  &0.12&  38.1 \plm2.3 &0.06&               &    \\
5192.34&FE1&3.00&$-$0.520&  68.2 \plm2.5  &0.11& 80.1 \plm3.6  &0.18& 40.1 \plm3.4  &0.13&  51.7 \plm3.1 &0.06&               &    \\
5194.94&FE1&1.56&$-$2.090& 116.9 \plm3.7  &0.10&128.9 \plm2.3  &0.13& 71.3 \plm2.8  &0.11&  79.9 \plm2.9 &0.06&               &    \\
5215.19&FE1&3.27&$-$0.930&  33.5 \plm1.5  &0.10& 40.9 \plm2.9  &0.13&               &    &               &    &               &    \\
5216.28&FE1&1.61&$-$2.102&  97.3 \plm3.0  &0.10&124.6 \plm2.8  &0.13& 53.6 \plm4.3  &0.10&  67.1 \plm2.9 &0.05&               &    \\
5225.53&FE1&0.11&$-$4.790&  79.6 \plm3.5  &0.06&101.3 \plm3.9  &0.08& 31.3 \plm2.9  &0.07&               &    &               &    \\
5254.96&FE1&0.11&$-$4.760&  75.0 \plm3.5  &0.06& 91.5 \plm3.6  &0.08&               &    &               &    &               &    \\
5266.56&FE1&3.00&$-$0.390&  86.7 \plm2.5  &0.11& 95.4 \plm2.8  &0.17& 43.3 \plm2.3  &0.12&  53.9 \plm2.7 &0.06&               &    \\
5269.54&FE1&0.86&$-$1.320& 206.9 \plm4.6  &0.08&213.1 \plm6.0  &0.12&183.3 \plm9.4  &0.10&               &    &               &    \\
5281.79&FE1&3.04&$-$0.830&  46.8 \plm2.6  &0.08&               &    &               &    &               &    &               &    \\
5283.62&FE1&3.24&$-$0.520&  61.6 \plm3.4  &0.10& 66.3 \plm2.2  &0.16&               &    &               &    &               &    \\
5302.30&FE1&3.28&$-$0.880&  34.0 \plm2.7  &0.08& 51.9 \plm2.3  &0.14&               &    &               &    &               &    \\
5307.37&FE1&1.61&$-$2.812&  42.5 \plm2.5  &0.07& 69.9 \plm4.0  &0.11&               &    &               &    &               &    \\
5324.19&FE1&3.21&$-$0.100&  78.7 \plm2.5  &0.11& 98.8 \plm3.3  &0.17& 43.6 \plm2.5  &0.12&  50.0 \plm3.9 &0.05&               &    \\
5328.04&FE1&0.92&$-$1.470&                &    &               &    &169.2 \plm4.9  &0.09&               &    &               &    \\
5339.93&FE1&3.27&$-$0.680&  47.0 \plm2.9  &0.08& 57.0 \plm1.9  &0.14&               &    &               &    &               &    \\
5367.48&FE1&4.42& 0.550  &                &    & 33.8 \plm1.6  &0.10&               &    &               &    &               &    \\
5369.96&FE1&4.37& 0.540  &  33.8 \plm3.1  &0.08&               &    &               &    &               &    &               &    \\
5371.50&FE1&0.96&$-$1.644& 193.3 \plm4.2  &0.07&205.1 \plm4.5  &0.11&157.2 \plm4.7  &0.09&               &    &  85.1 \plm5.5 &0.03\\
5383.37&FE1&4.31& 0.500  &  35.3 \plm3.4  &0.08&               &    &               &    &               &    &               &    \\
5393.17&FE1&3.24&$-$0.920&  50.2 \plm2.6  &0.08& 53.8 \plm2.4  &0.13&               &    &               &    &               &    \\
5397.14&FE1&0.91&$-$1.992& 173.0 \plm5.8  &0.07&190.1 \plm5.3  &0.10&140.2 \plm4.2  &0.08&               &    &  71.9 \plm4.1 &0.02\\
5405.79&FE1&0.99&$-$1.852& 173.9 \plm4.6  &0.07&201.9 \plm3.3  &0.10&140.2 \plm2.9  &0.08&               &    &  75.3 \plm2.8 &0.02\\
5415.19&FE1&4.39& 0.510  &  46.7 \plm3.1  &0.08& 33.4 \plm2.1  &0.12&               &    &               &    &               &    \\
5424.07&FE1&4.32& 0.520  &  41.4 \plm3.2  &0.08& 54.2 \plm2.8  &0.15&               &    &               &    &               &    \\
5429.70&FE1&0.96&$-$1.880& 179.0 \plm4.1  &0.07&185.7 \plm4.2  &0.10&157.1 \plm4.3  &0.08&               &    &               &    \\
5434.52&FE1&1.01&$-$2.120& 164.9 \plm4.4  &0.06&178.2 \plm3.1  &0.09&130.9 \plm3.6  &0.08&               &    &               &    \\
5446.92&FE1&0.99&$-$1.910& 172.3 \plm4.1  &0.06&202.4 \plm4.6  &0.10&138.7 \plm4.3  &0.08&               &    &               &    \\
5455.61&FE1&1.01&$-$2.090&                &    &206.2 \plm4.3  &0.10&131.5 \plm4.7  &0.08&               &    &               &    \\
5455.61&FE1&1.01&$-$2.090& 174.2 \plm3.8  &0.06&               &    &               &    &               &    &               &    \\
5497.52&FE1&1.01&$-$2.850& 110.9 \plm4.6  &0.06&138.9 \plm3.3  &0.08& 82.6 \plm2.1  &0.07&               &    &               &    \\
5501.48&FE1&0.96&$-$3.050& 107.7 \plm4.1  &0.06&135.4 \plm3.4  &0.08& 75.8 \plm2.4  &0.07&               &    &               &    \\
5506.79&FE1&0.99&$-$2.790& 122.8 \plm3.9  &0.06&150.0 \plm4.5  &0.08& 97.2 \plm3.9  &0.07&               &    & 23.5 \plm1.8  &0.01\\
5569.62&FE1&3.42&$-$0.540&  40.9 \plm2.1  &0.07& 50.8 \plm2.6  &0.11&               &    &               &    &               &    \\
5572.84&FE1&3.40&$-$0.310&                &    &               &    & 27.1 \plm1.6  &0.09&               &    &               &    \\
5576.09&FE1&3.43&$-$1.000&                &    & 38.8 \plm2.4  &0.10&               &    &               &    &               &    \\
5586.76&FE1&3.37&$-$0.140&  79.9 \plm2.9  &0.08& 86.5 \plm4.2  &0.13&               &    &               &    &               &    \\
5615.66&FE1&3.33& 0.050  &  75.6 \plm3.0  &0.08&107.0 \plm1.8  &0.13& 57.4 \plm2.6  &0.09&59.7 \plm2.7   &0.04&               &    \\
5956.70&FE1&0.86&$-$4.570&                &    & 30.7 \plm3.1  &0.11&               &    &               &    &               &    \\
6136.62&FE1&2.45&$-$1.500&  70.4 \plm3.6  &0.04& 93.8 \plm3.5  &0.06&               &    &  42.4 \plm4.23&0.02&               &    \\
6137.70&FE1&2.59&$-$1.366&  66.3 \plm1.3  &0.04& 79.7 \plm2.1  &0.06&               &    &               &    &               &    \\
6191.57&FE1&2.43&$-$1.416&  80.3 \plm2.7  &0.04&107.3 \plm2.6  &0.06&               &    &               &    &               &    \\
6213.43&FE1&2.22&$-$2.660&  30.3 \plm2.2  &0.03& 42.4 \plm2.4  &0.04&               &    &               &    &               &    \\
6219.29&FE1&2.20&$-$2.438&  41.4 \plm2.0  &0.03& 52.1 \plm2.5  &0.04&               &    &               &    &               &    \\
6230.74&FE1&2.56&$-$1.276&  84.4 \plm2.3  &0.04& 89.4 \plm3.5  &0.06& 43.0 \plm1.5  &0.04&  60.6 \plm8.9 &0.02&               &    \\
6252.57&FE1&2.40&$-$1.757&  65.3 \plm2.7  &0.03& 89.6 \plm2.2  &0.05& 44.6 \plm2.9  &0.04&  30.5 \plm4.3 &0.02&               &    \\
6290.97&FE1&4.73&$-$0.760&                &    &               &    & 28.8 \plm2.9  &0.04&               &    &               &    \\
6297.80&FE1&2.22&$-$2.740&  53.1 \plm3.6  &0.03& 34.0 \plm3.3  &0.05&               &    &               &    &               &    \\
6393.61&FE1&2.43&$-$1.630&  80.5 \plm2.0  &0.03& 93.8 \plm1.9  &0.05&               &    &  42.3 \plm4.0 &0.02&               &    \\
6421.36&FE1&2.28&$-$2.014&  56.2 \plm2.4  &0.02& 80.5 \plm2.7  &0.04&               &    &               &    &               &    \\
6430.86&FE1&2.18&$-$1.946&  67.3 \plm2.9  &0.03& 95.2 \plm2.6  &0.04& 39.1 \plm2.7  &0.03&  35.7 \plm4.1 &0.01&               &    \\
6494.98&FE1&2.40&$-$1.270&  94.2 \plm3.3  &0.03&122.9 \plm3.3  &0.04&               &    &  55.5 \plm2.8 &0.02&               &    \\
6593.88&FE1&2.43&$-$2.390&  22.8 \plm3.5  &0.02&               &    &               &    &               &    &               &    \\
\hline                                                                                                                             
5018.43&FE2&2.89&$-$1.220& 126.4 \plm4.4  &0.18&140.4 \plm5.6  &0.24& 99.7 \plm4.9  &0.24& 100.8 \plm3.8 &0.13&  30.1         &0.03\\
4923.92&FE2&2.89&$-$1.320& 115.3 \plm5.3  &0.20&124.3 \plm5.5  &0.27& 86.5 \plm3.4  &0.26&  82.3 \plm3.1 &0.12&  35.8         &0.03\\
5276.00&FE2&3.20&$-$1.950&  58.2 \plm3.2  &0.11& 71.0 \plm3.6  &0.18&               &    &  39.3 \plm2.6 &0.04&               &    \\
6247.56&FE2&3.89&$-$2.360&                &    & 18.4 \plm3.0  &0.07&               &    &               &    &               &    \\
5197.57&FE2&3.23&$-$2.100&                &    &               &    &               &    &  34.2 \plm5.3 &0.07&               &    \\
\hline                                                                                                                             
3995.30&CO1&0.92&$-$0.220& 101.9\2        &0.34& 113.7\2       &0.45&               &    &               &    &  44.8\2       &0.08\\
4118.77&CO1&1.05&$-$0.490&                &    &               &    &  57.0\2       &0.38&               &    &               &    \\
4121.31&CO1&0.92&$-$0.320& 105.1\2        &0.29& 127.5\2       &0.40&  69.4\2       &0.38&               &    &  38.6\2       &0.07\\
\hline                                                                                                                             
3807.13&NI1&0.42&$-$1.180& 123\3          &    & 214\3         &    & 118\3         &    &               &    &  51.4 \plm4.3 &0.11\\
3858.29&NI1&0.42&$-$0.970&                &    & 198\3         &    & 114.3 \plm7.7 &0.48&               &    &  62.8 \plm10.2&0.11\\
5476.92&NI1&1.83&$-$0.890& 112.2 \plm3.2  &0.08& 109.2 \plm4.3 &0.12&  50.4 \plm5.5 &0.08& 78.2  \plm3.2 &0.05&               &\\
\hline                                                                                                         

4077.71&SR2&0.00& 0.167  & 173            &0.31& 204.6         &0.41& 231\1         &0.40&               &    &  65.0         &0.08\\
4215.52&SR2&0.00&-0.145  &                &0.30& 176.3         &0.40& 197.6         &0.40&               &    &  46.5         &0.08\\
\hline                                                                                                         
4883.69& Y2&1.08& 0.070  & $<$21          &0.11& 22.6\1        &0.15& $<$37         &0.15&  $<$14        &0.07&  $<$15        &0.03\\
\hline                                                                                                         
4554.03&BA2&0.00& 0.170  &                &    &               &    &               &    &               &    &  16.8\2       &0.04\\
6141.73&BA2&0.70&$-$0.077& 41.0\2         &0.03& 68.6\2        &0.05& 53.0\2        &0.04& $<$13\2       &0.02&               &\\
6496.91&BA2&0.60&$-$0.380& 46.7\2         &0.03& 63.7\2        &0.05& 39.0\2        &0.02& $<$10\2       &0.01&               &\\
\hline                                                                                                         
4129.70&EU2&0.00& 0.204  & $<$25          &0.42& $<$21         &0.30& $<$29         &0.34&               &    & $<$20         &0.06\\
\hline
\hline
\label{tab:linelist}
\end{longtable}
\end{center}

\begin{landscape}
\begin{table}
\begin{center}
\caption{Abundances}
\tiny
\tabcolsep 4pt
\label{tab:abundances}
\begin{tabular}{lrrrrrrrrrrrrrrrrrrrrr}

\hline
                  &  FeI    &  FeII   &   C     &    OI     &  NaI    &  MgI      &  AlI    &  SiI     &  CaI    & ScII    &  TiI    &  TiII   &  CrI    &   MnI   &  CoI    &  NiI    & SrII    &   YII     &  BaII     &  EuII      \\

                  &         &         &         &           &         &           &         &          &         &         &         &         &         &         &         &         &         &           &           &            \\      

\textbf{log$\epsilon$(X)$_{\odot}$} &   7.50  &  7.50  &  8.55   &   8.87     &    6.33  &   7.58   &  6.47    &   7.55  &   6.36  &  3.17  &   5.02  &  5.02    &  5.67   & 5.39    & 4.92    & 6.25    &  2.97  &    2.24    &  2.13     &    0.51        \\              
\hline                                                                                                                                                                                                                               
	&         &         &         &           &         &           &         &          &         &         &         &         &         &         &         &         &         &           &           &            \\              
                                                                                                                                                                                                                             
\textbf{Sex24-72} &         &         &         &           &         &           &         &          &         &         &         &         &         &         &         &         &         &           &           &            \\
                  &         &         &         &           &         &           &         &          &         &         &         &         &         &         &         &         &         &           &           &            \\
                                                                                                                                                                                                                                     
Nb Lines          &   68    &    3    &         &           &   1     &   3       &   1     &   1      &   4     &   2     &   7     &   6     &   4     &   1     &   2     &   1     &   1     &           &   2       &            \\     
 $[$X/H]          & $-$2.93 & $-$2.86 & $-$2.53 & $<$$-$2.29& $-$2.60 & $-$2.83   & $-$3.43 & $-$2.75  & $-$2.81 & $-$3.01 & $-$3.12 & $-$2.70 & $-$3.31 & $-$3.23 & $-$3.12 & $-$2.70 & $-$3.14 & $<$$-$3.45& $-$4.03   & $<$$-$3.09 \\
 $[$X/Fe]         &         &         &  0.40   & $<$ 0.64  &  0.33   &  0.10     & $-$0.5  & 0.18     &  0.12   & $-$0.08 & $-$0.19 &  0.23   & $-$0.38 & $-$0.30 & $-$0.19 &    0.23 & $-$0.21 & $<$$-$0.52& $-$1.10   & $<$$-$0.16 \\
Error             &  0.04   &   0.08  &  0.19   &           &  0.30   &  0.06     &   0.5   &  0.35    &  0.09   &  0.10   &  0.07   &  0.09   &  0.10   &  0.30   &  0.27   &  0.14   &  0.35   &     -     &  0.19     &            \\   
                                                                                                                                                                                                                                     
\hline                                                                                                                                                                                                                               
                  &         &         &         &           &         &           &         &          &         &         &                             &         &         &         &         &           &           &            \\      
                                                                                                                                                                                                                                     
\textbf{Sex11-04} &         &         &         &           &         &           &         &          &         &         &                             &         &         &         &         &           &           &            \\
                  &         &         &         &           &         &           &         &          &         &         &                             &         &         &         &         &           &           &            \\                                                                                                                   
Nb Lines          &   66    &   4     &         &           &   2     &   3       &   2     &   1      &   6     &   2     &   7     &   7     &   4     &   1     &   2     &   1     &   1     &   1       &   2       &            \\
 $[$X/H]          & $-$2.94 & $-$2.70 & $-$3.94 & $<$$-$2.08& $-$3.00 & $-$2.63   & $-$2.65 & $-$2.47  & $-$2.79 & $-$3.00 & $-$3.15 & $-$2.72 & $-$3.32 & $-$3.47 & $-$2.90 & $-$3.00 & $-$2.95 & $-$3.38   & $-$3.79   & $<$$-$3.27 \\
$[$X/Fe]          &         &         &  $-$1.00&  $<$ 0.86 & $-$0.06 &  0.31     &  0.29   &  0.47    &  0.15   & -0.06   & $-$0.21 &  0.22   & $-$0.38 & $-$0.53 &  0.04   & $-$0.06 & -0.01   & $-$0.44   & $-$0.85   & $<$$-$0.33 \\
Error             &  0.04   &  0.09   &  0.32   &           &  0.08   &  0.08     &  0.15   &  0.30    &  0.07   &  0.11   &  0.05   &  0.07   &  0.12   &  0.20   &  0.35   &  0.11   &  0.40   &  0.16     &  0.14     &            \\
\hline                                                                                                                                                                                                                               
                  &         &         &         &           &         &           &         &          &         &         &                             &         &         &         &         &           &           &            \\      
                                                                                                                                                                                                                                     
\textbf{Fnx05-42} &         &         &         &           &         &           &         &          &         &         &                             &         &         &         &         &           &           &            \\   
                  &         &         &         &           &         &           &         &          &         &         &                             &         &         &         &         &           &           &            \\
Nb Lines          &  40     &   2     &         &           &   2     &   3       &   2     &   1      &   3     &   4     &   4     &   7     &   2     &   2     &   2     &   2     &    1    &           &   2       &            \\
$[$X/H]           & $-$3.66 & $-$3.42 & $-$4.46 & $<$$-$2.45& $-$3.69 & $-$3.20   & $-$3.88 & $-$3.52  & $-$3.15 & $-$3.38 & $-$3.71 & $-$3.27 & $-$4.25 & $-$5.26 & $-$3.91 & $-$4.02 & $-$2.84 & $<$$-$3.18& $-$4.07   & $<$$-$3.11 \\
$[$X/Fe]          &         &         & $-$0.80 & $<$ 1.21  & $-$0.05 &  0.41     & $-$0.22 &  0.14    &  0.53   &  0.28   & $-$0.05 &  0.39   & $-$0.59 & $-$1.60 & $-$0.25 & $-$0.36 &  0.82   & $<$ 0.48  & $-$0.41   & $<$ 0.55   \\
Error             &  0.04   &  0.08   &  0.23   &           &  0.07   &  0.08     &  0.21   &  0.14    &  0.12   &  0.09   &  0.07   &  0.08   &  0.13   &  0.18   &  0.25   &  0.20   &  0.20   &           &  0.16     &            \\
\hline                                                                                                                                                                                                                               
                  &         &         &         &           &         &           &         &          &         &         &                             &         &         &         &         &           &           &            \\      
                                                                                                                                                                                                                                     
\textbf{Scl07-49} &         &         &         &           &         &           &         &          &         &         &                             &         &         &         &         &           &           &            \\   
                  &         &         &         &           &         &           &         &          &         &         &                             &         &         &         &         &           &           &            \\                   
Nb Lines          &   21    &   4     &         &           &   2     &   2       &         &          &   2     &   2     &   6     &   3     &   2     &         &         &   1     &         &           &           &            \\
$[$X/H]           & $-$3.48 & $-$3.22 &         &           & $-$3.72 & $-$3.26   &         &$<$$-$2.05& $-$2.98 &$-$3.08  &$-$3.33  & $-$3.03 & $-$3.90 &         &         & $-$3.25 &         & $<$$-$3.51& $<$$-$4.44&            \\
$[$X/Fe]          &         &         &         &           & $-$0.24 &  0.22     &         &$<$1.43   &  0.50   &  0.40   &  0.15   &  0.55   & $-$0.42 &         &         &  0.23   &         & $<$0.05   & $<$$-$0.88&            \\
Error             &  0.04   &  0.10   &         &           &  0.11   &  0.08     &         &          &  0.06   &  0.11   &  0.08   &  0.09   &  0.12   &         &         &  0.08   &         &           &           &            \\
                                                                                                                                                                                                                                     
\hline                                                                                                                                                                                                                               
                  &         &         &         &           &         &           &         &          &         &         &                             &         &         &         &         &           &           &            \\
                                                                                                                                                                                                                                     
\textbf{Scl07-50} &         &         &         &           &         &           &         &          &         &         &                             &         &         &         &         &           &           &            \\
                  &         &         &         &           &         &           &         &          &         &         &                             &         &         &         &         &           &           &            \\           
Lines             &   21    &   2     &         &           &   2     &   4       &   2     &   1      &   1     &   2     &         &   9     &   3     &   3     &   2     &   2     &   1     &           &   1       &            \\
$[$X/H]           & $-$3.96 & $-$4.08 & $-$3.96 &           & $-$4.03 & $-$3.97   & $-$4.63 & $-$3.97  & $-$4.45 & $-$3.75 &         & $-$3.75 & $-$4.56 & $-$4.91 & $-$3.49 & $-$4.33 & $-$5.11 & $<-$3.22  & $-$4.80   & $<$$-$2.57 \\
$[$X/Fe]          &         &         &  0.00   &           & $-$0.07 & $-$0.01   & $-$0.67 & $-$0.01  & $-$0.49 &  0.21   &         &  0.21   & $-$0.60 & $-$0.95 &  0.47   & $-$0.37 & $-$1.15 & $<$ 0.74  & $-$0.84   & $<$ 1.39   \\
Error             &  0.06   & 0.11    &  0.50   &           &  0.10   &  0.13     &  0.10   &  0.20    &  0.13   &  0.10   &         &  0.08   &  0.09   &  0.23   &  0.26   &  0.12   &  0.16   &           &  0.16     &            \\
\hline       
\hline
\end{tabular}			
\end{center}
\end{table}
\end{landscape}

\end{document}